%% file: main.tex
\documentclass[fleqn,10pt]{olplainarticle}

\usepackage{comment}
\usepackage{hyperref}
\usepackage{hyperref}
 \hypersetup{
     colorlinks=true,
     linkcolor=darkgray,
     filecolor=blue,
     citecolor=black,      
     urlcolor=purple,
}
\usepackage{booktabs}

\usepackage{todonotes}
\reversemarginpar
\usepackage{graphicx,lipsum}
\usepackage{sidecap}
\usepackage{caption}
\usepackage{subcaption}
\usepackage{wrapfig}
\usepackage[T1]{fontenc}
\usepackage{soul}
\usepackage[inline]{enumitem}
\newlist{inlist}{enumerate*}{1}
\setlist[inlist]{label=\textit{\roman*)}}
\usepackage[table,xcdraw]{xcolor}
\usepackage{pdflscape} 
\usepackage{longtable}
\usepackage{textcomp}
\usepackage{transparent}

\usetikzlibrary{arrows.meta,angles,quotes}
\colorlet{linecol}{black!75}
\usepackage{forest}
\usepackage{adjustbox}

\usepackage{svg}

\newlist{todolist}{itemize}{3}
\setlist[todolist]{label=$\square$}
\usepackage{pifont}
\makeatletter
\patchcmd{\pgfutilsolvetwotwoleqfloat}
  { \noexpand\pgfmathfloatdivide@}
  {\noexpand\pgfmathfloatdivide@}
  {}{}
\makeatother

\usepackage{array}
\newcolumntype{L}[1]{>{\raggedright\let\newline\\\arraybackslash\hspace{0pt}}m{#1}}
\newcolumntype{C}[1]{>{\centering\let\newline\\\arraybackslash\hspace{0pt}}m{#1}}
\newcolumntype{R}[1]{>{\raggedleft\let\newline\\\arraybackslash\hspace{0pt}}m{#1}}
\usepackage{booktabs, makecell, caption, threeparttable, tabularx, boldline}

\newcommand\semrel[1]{\texttt{#1}}
\newcommand\topic[1]{``#1''}
\newcommand\concept[1]{``#1''}
\newcommand\kos[1]{\textit{#1}}
\newcommand\triple[3]{<\textit{#1}, \textit{#2}, \textit{#3}>}

\newcommand\maintableurl{\url{https://doi.org/10.48366/R732033}} 
\newcommand\githubrepo{\url{https://github.com/angelosalatino/kos-rf}}

\usepackage{rotating}
\def\Rot#1{ \multicolumn{1}{l}{\rlap{\rotatebox{45}{#1}~}}}
\usepackage{changepage}

\title{A Survey on Knowledge Organization Systems of Research Fields: Resources and Challenges}

\author[1,*, \href{https://orcid.org/0000-0002-4763-3943}{[0000-0002-4763-3943]}]{Angelo Salatino}
\author[1, \href{https://orcid.org/0009-0009-9477-7112}{[0009-0009-9477-7112]}]{Tanay Aggarwal}
\author[2, \href{https://orcid.org/0000-0002-5193-7851}{[0000-0002-5193-7851]}]{Andrea Mannocci}
\author[1,3, \href{https://orcid.org/0000-0001-6557-3131}{[0000-0001-6557-3131]}]{Francesco Osborne}
\author[1, \href{https://orcid.org/0000-0003-0015-1952}{[0000-0003-0015-1952]}]{Enrico Motta}
\affil[1]{Knowledge Media Institute, The Open University, Milton Keynes, United Kingdom}
\affil[2]{CNR-ISTI --- National Research Council, Institute of Information Science and Technologies ``Alessandro Faedo'', 56124 Pisa, Italy}
\affil[3]{Department of Business and Law, University of Milano Bicocca, Milan, Italy}
\affil[*]{Corresponding author: angelo.salatino@open.ac.uk}

\begin{abstract} 
Knowledge Organization Systems (KOSs), such as term lists, thesauri, taxonomies, and ontologies, play a fundamental role in categorising, managing, and retrieving information. In the academic domain, KOSs are often adopted for representing research areas and their relationships, primarily aiming to classify research articles, academic courses, patents, books, scientific venues, domain experts, grants, software, experiment materials, and several other relevant products and agents. These structured representations of research areas, widely embraced by many academic fields, have proven effective in empowering AI-based systems to \textit{i)} enhance the retrievability of relevant documents, \textit{ii)} enable advanced analytic solutions to quantify the impact of academic research, and \textit{iii)} analyse and forecast research dynamics. 
This paper aims to present a comprehensive survey of the current KOS for academic disciplines. We analysed and compared 45 KOSs according to five main dimensions: scope, structure, curation, usage, and links to other KOSs. Our results reveal a very heterogeneous scenario in terms of scope, scale, quality, and usage, highlighting the need for more integrated solutions for representing research knowledge across academic fields. We conclude by discussing the main challenges and the most promising future directions.
\end{abstract}


\begin{document}\sloppy

\maketitle

\section{Introduction}
\label{introduction}

Knowledge Organization Systems (KOSs), such as term lists, thesauri, taxonomies, and ontologies, play a fundamental role in categorising, managing, and retrieving information~\citep{mazzocchi2018}.
Specifically, 
they ``model the underlying semantic structure of a domain and provide semantics, navigation,
and translation through labels, definitions, typing, relationships, and properties for concepts''~\citep{zeng2008knowledge}. 



In the academic domain, KOSs are often adopted for representing research areas and their relationships~\citep{sugimoto2018}, with the primary aim of classifying research articles, books, courses, patents, grants, software, experiment materials, scientific venues, domain experts, organisations, and several other relevant items and agents. 
These structured representation of knowledge have been adopted by most academic fields and proved very effective in 
\begin{inlist}
\item improving the retrievability of relevant documents
~\citep{newman2010evaluating,salatino2019improving}, 
\item enabling advanced analytic solutions to quantify the impact of research~\citep{ding2014,qiu2017,sugimoto2018,salatino2023}, and 
\item understanding and forecasting research dynamics~\citep{scharnhorst2012,qiu2017}.
\end{inlist}

More recently, these KOSs have become even more instrumental given the fast-growing number of publications, the rise of Open Science and Open Access articles, the thriving role of interdisciplinary research, and the emergence of vast online repositories of articles, academic courses, and other academic materials~\citep{auer2018towards}. 
This transformation poses new opportunities but also new challenges. For example, in the recent COVID-19 pandemic, there was a lot of discussion on how the scientific community was being ``drowning in COVID-19 papers'' and had to resort to new tools based on robust representations of research concepts~\citep{brainard2020scientists}. 
To address these issues, KOSs have been increasingly incorporated into various AI systems to assist researchers in navigating literature~\citep{dai2020fullmesh} and semi-automating systematic reviews~\citep{bolanos2024artificialintelligenceliteraturereviews}. Despite the emergence of new AI systems based on Large Language Models (LLMs) in the last two years, structured and machine-readable representations of domain knowledge continue to be invaluable as they aid in formulating precise queries to identify relevant publications, reduce hallucinations, and enhance interpretability~\citep{bechard2024reducinghallucinationstructuredoutputs, gnoli2024library}.


KOSs of research areas are very heterogeneous in terms of scope, scale, quality, and usage. 
Some fields (e.g., \topic{Biomedical}) are well covered by a variety of KOSs (e.g., \kos{MeSH}, \kos{UMLS}, \kos{NLM}) that are used to categorise research products and are routinely adopted by libraries, online repositories, researchers, and organisations. 
Other fields (e.g., \topic{Mathematics}) prevalently rely on one widely accepted KOSs (e.g., \kos{Mathematics Subject Classification}). 
A few research areas (e.g., \topic{Geography}, \topic{History}, \topic{Material Science}, \topic{Political Science}, and \topic{Sociology}) do not even have their own specific KOSs. 
Several large KOSs cover multiple academic disciplines, but often with coarse-grained representation that is not sufficient for the needs of the specific fields. 
To the best of our knowledge, we still lack a systematic and in-depth analysis of these knowledge organization systems and their characteristics.

The objective of this paper is to present a comprehensive survey of KOSs for academic fields. 
We defined formal inclusion and exclusion criteria that led to the identification of 45 candidates. 
We analysed them according to five main aspects.
The first is the \textbf{scope} of a KOS in terms of its coverage of academic disciplines.  
The second aspect is the \textbf{structure}, which includes features such as the number of concepts, the maximum depth, the type of hierarchy, and the presence of synonyms.
The third aspect is the \textbf{curation}, which includes information about the formats, the license, the frequency of the updates, the procedure used for the generation, and the languages in which the KOS is available.
The fourth aspect deals with the \textbf{links to other KOSs}, which allow users and tools to interconnect and adopt multiple KOSs for a richer characterisation of research areas. The final aspect regards their \textbf{usage} in digital libraries, repositories, and research communities.

We conclude the paper by discussing the main challenges and opportunities in this field, highlighting the most promising future directions. 
We also analyse how current solutions could be integrated and interlinked to produce a more comprehensive and granular representation of all academic disciplines. 

In adherence to Open Science principles, we release the table that describes all the identified KOSs according to the 15 features on Open Research Knowledge Graph\footnote{Full table describing the analysed KOSs according to the 15 features --- \maintableurl}, as well as the code\footnote{The code for processing the analysed KOSs is available on GitHub --- \githubrepo} we developed for processing them.

The rest of this manuscript is organised as follows. Section~\ref{background} introduces KOSs and discusses their applications. Section~\ref{methodology} describes the methodology that we employed to identify the 45 KOSs and introduces the full set of features used for the analysis.
Section~\ref{results} presents the results of our analysis, offering a thorough, feature-by-feature assessment of the current landscape, and illustrating significant patterns. 
Section~\ref{challenges} discusses the ongoing challenges and possible future directions. 
Section~\ref{sec:ttv} outlines the threats to the validity of our analysis.
Finally, Section~\ref{conclusions} 
concludes the paper by summarising the contributions
and the main findings.

\section{Background}\label{background}

Knowledge organization systems play a crucial role in research by providing structured frameworks for organising complex information, allowing researchers to establish clear categories, discern relationships, and navigate large datasets with increased efficiency. 
To underscore their importance, we will draw upon the literature to describe their usage in the research domain, specifically focusing on two main angles: digital libraries (Section~\ref{digitallibraries}) and information science (Section~\ref{informationscience}).

\subsection{Knowledge Organization Systems in Digital Libraries}\label{digitallibraries}


Knowledge organization systems form the backbone of effective search and retrieval in digital libraries, providing a systematic means for categorising and organising knowledge, retrieving information, facilitating preservation, and ensuring interoperability~\citep{hodge2000systems}. Annotating research products with appropriate research concepts facilitates semantic searches, leading to more effective information retrieval.





In the literature, we can find different types of KOSs, such as: taxonomies, glossaries, dictionaries, synonym rings, gazetteers, authority files, subject headings, thesauri, classification schemes, semantic networks, and ontologies~\citep{hodge2000systems, zeng2008knowledge}. \cite{zeng2008knowledge} comprehensively emphasises the interplay between the complexity of their structures and their expected functions. The complexity of their structure can range from simple ``flat'' structures (e.g., \textit{pick lists}, \textit{dictionaries}), to two-dimensional hierarchical structures (e.g., \textit{taxonomies}), and finally, to multidimensional structures, creating networks according to diverse semantic types and relationship (e.g., \textit{ontologies}).
Generally, KOSs with higher structural complexity exhibit greater capacity to suit various functions, including i) disambiguation of terms, ii) management of synonyms or equivalent terms, iii) establishment of semantic relationships, particularly hierarchical and associative links, and iv) representation of both conceptual relationships and attributes within knowledge models. 
We refer the interested reader to Hodge's book~\citep{hodge2000systems} and \citep{zeng2008knowledge} for additional details on the various types of KOSs.

Because KOSs are means for organising information, they are at the heart of every digital library~\citep{hodge2000systems}.
Indeed, well-known publishers like Elsevier, Springer Nature, the Institute of Electrical and Electronics Engineers (IEEE), and the Association for Computing Machinery (ACM) have developed their own system to provide full text of documents linked to bibliographic records.
Major bibliographic databases like Web of Science, Scopus, Dimensions, and OpenAlex also employ KOSs to organise their vast collections of bibliographic records.

The annotation of documents based on the concepts within KOSs can be either performed manually or automatically. Manual annotation tasks are undertaken by human experts, typically experienced curators or editors, who leverage their domain knowledge to critically assess document content and assign the most pertinent concepts. In contrast, automatic annotation employs 
a range of computational tools that often incorporate advanced artificial intelligence techniques~\citep{salatino2019improving}. 
For instance, OpenAlex, which is a major bibliographic catalogue of scientific papers, employs a deep learning model that, based on research papers' title, abstract, citations, and journal name, can define the appropriate topics drawn from the \kos{OpenAlex Topics} vocabulary~\citep{openalex2024}.

In addition to the automatic classification of content, KOSs also support additional tasks, such as augmented retrieval~\citep{shiri2002thesaurus}, recommender systems~\citep{cleverley2015best}, integration and interoperability~\citep{zeng2019knowledge}, and knowledge management and preservation~\citep{chowdhury2010digital,Ganguly_2017}. 
For augmented retrieval, KOSs support precision search and query expansion by allowing users to execute highly specific queries using controlled vocabulary and discover more relevant content. In particular, by leveraging the related terms or broader concepts within KOSs, users can either manually expand their searches through user interfaces~\citep{shiri2002thesaurus}, or rely on the search engine to automatically expand their queries~\citep{mu2014explicitly}.

With regard to recommender systems, KOSs enable the development of applications that enhance content discovery by suggesting related content based on subject matter and providing personalised recommendations derived from user search patterns, fostering serendipitous discovery and richer user engagement~\citep{cleverley2015best,thanapalasingam2018}.


In the context of integration and interoperability, KOSs establish a framework for the semantic enrichment of data, facilitating seamless integration of research products across different digital libraries or repositories and consequently enhancing their interoperability~\citep{zeng2019knowledge}.

Finally, KOSs can also contribute to the long-term preservation of information by ensuring it is organised logically and can be easily retrieved and understood in the future~\citep{chowdhury2010digital}. In this regard, \cite{Ganguly_2017} argue that adding standardised values as
metadata, selecting them from pre-defined controlled vocabularies rather than guessing keywords, improves the long-term preservation of digital objects.

In conclusion, digital libraries often employ KOSs to improve document organisation and provide a wide range of advanced features.


\subsection{Knowledge Organization Systems in Information Science}\label{informationscience}


The research community has utilised KOSs of research topics to enable and support a variety of tasks in this domain, such as analysis of the scientific landscape~\citep{reymond2020patents,ijerph15061113,angioni2022leveraging}, trend analysis and forecasting~\citep{yan2014research, salatino2018augur}, analyse the composition of a research team~\citep{salatino2023}, and assessing impact~\citep{sjogaarde2022association}. Here we outline a small sample of approaches employed for these tasks.

In the context of analysing the scientific landscape, \cite{angioni2022leveraging} developed the AIDA Dashboard, a tool that facilitates the analysis of conferences and journals in Computer Science, providing valuable insights into main authors, organisations, and countries. The dashboard leverages the \kos{Computer Science Ontology}~\citep{salatino2018} to provide a very granular representation of the venues' research topics, as well as to rank venues within topics using various metrics. Further contributions include \cite{reymond2020patents},  who examined patents in the Humanities using the \kos{UNESCO Thesaurus}. Their findings provided useful insights into potential research questions and unexplored research avenues. Furthermore, the work of \cite{ijerph15061113} introduced a tool that assists users in analysing research trends by expanding initial queries using \kos{MeSH} terms, facilitating a more comprehensive exploration of the research landscape.


For research trends analysis, \cite{ilgisonis2022catch} performed a systematic retrospective analysis of the frequencies of \kos{MeSH} concepts across twelve years. Their analysis revealed potential shifts in scientific priorities, and they employed the same patterns to predict emerging trends within a five-year timeframe.
In addition, \cite{ovalle2013influence} studied whether the European Framework Programmes shaped the scientific output in \topic{nanotechnology} of its member states. Their study compared this output to global trends as well as patterns of international collaboration. The authors relied on the representations of \topic{nanotechnology} within \kos{EuroVoc}, \kos{MeSH}, and three additional KOSs to construct a refined search query for retrieving relevant papers from Web of Science. 

Within the analysis of research team composition, \cite{salatino2023} investigated how the diversity of expertise of a research team can influence their scientific impact. In this experiment, research topics from the \kos{Computer Science Ontology}~\citep{salatino2018} were employed to model the researcher's expertise. Specifically, they characterise the expertise of an author at the time of collaboration as the distribution of research topics of their paper over the preceding five years. Additionally, ~\cite{Kang2015} mapped researchers' expertise according to \kos{MeSH} terms and developed a matching algorithm to find potential interdisciplinary collaborators. 

Regarding citation impact, \cite{sjogaarde2022association} performed an analysis of the correlation between topic growth and article citation. Their methodology leverages an in-house, automatically generated KOS~\citep{Sjogarde2020} to systematically categorise topics and disciplines within their dataset. Their topic-based analysis yields compelling insights that could shape future research policy decisions. Moreover, \cite{artsim2020} introduced a method for estimating the impact of recently published papers,  based on the premise that similar papers often experience comparable popularity trajectories. In this context, they leveraged citation networks and metadata, including the \kos{Computer Science Ontology}, to assess similarity.

These analyses and approaches underscore the value and importance of KOSs in research, as they provide the essential structure and organisation needed for developing various downstream applications~\citep{salatino2020ontology}.


\section{Survey methodology}\label{methodology}
In this section, we describe the process we followed to identify and analyse KOSs of academic disciplines.  
We first define the types of KOSs that are objects of this analysis (Sec.~\ref{definitions}) and describe the inclusion and exclusion criteria adopted in the survey (Sec.~\ref{inclusion_criteria}). We then describe the strategy we adopted to find the KOSs (Sec.~\ref{finding}) and discuss the set of features that we will use for describing and comparing them (Sec.~\ref{features}).

To ensure clarity and consistency, we adopt the following typographical conventions: KOSs are indicated in italics, e.g., \kos{Medical Subject Headings}, while research concepts (topics and disciplines) are enclosed in double quotation marks, e.g., \concept{Medicine}. These conventions are designed to help readers easily distinguish between different types of entities.


\subsection{Concepts}
\label{definitions}

In this survey, we focus on KOSs of academic fields. KOSs are commonly used to categorise items into specific classes. In this context, classes correspond to research topics within various disciplines. The items categorised can include a wide range of artefacts, such as documents (e.g., research papers, patents, project reports), videos (e.g., university courses), datasets, software, projects, and more.
As proposed in~\cite{salatino2019early}, we define a \textbf{research topic} as the subject of study or issue that is of interest to the academic community, and it is explicitly addressed by research papers. 
Examples of research topics are \topic{Acoustics}\footnote{Acoustics --- \url{https://physh.aps.org/concepts/40a5bd01-6544-4502-8321-458c33878bf3}} in the \kos{PhySH}\footnote{PhySH --- \url{https://physh.aps.org}} taxonomy, \topic{Hydrocodone}\footnote{Hydrocodone --- \url{https://meshb.nlm.nih.gov/record/ui?ui=D006853}} in \kos{MeSH}\footnote{MeSH --- \url{https://meshb.nlm.nih.gov}}, or \topic{Web Ontology Language (OWL)}\footnote{Web Ontology Language (OWL) --- \url{https://dl.acm.org/topic/ccs2012/10002951.10003260.10003309.10003315.10003316}} in \kos{ACM CCS}\footnote{ACM CCS --- \url{https://www.acm.org/publications/class-2012}}. On the other hand, we define \textbf{research field} or \textbf{discipline} as a broad area of knowledge within academia, which consists of several research topics. For instance, \topic{Mathematics} and \topic{Medicine} are disciplines encompassing various and more specific research topics such as \topic{algebra}, \topic{calculus}, \topic{oncology}, and \topic{cardiology}.

KOSs in this space primarily fall into four possible \textbf{types of system}: term lists, hierarchical taxonomies, thesauri, and ontologies. In the following, we describe each one of them.









A \textbf{term list} is a flat list of subject headings or descriptors which support the organisation of a collection of documents~\citep{hedden2010taxonomies,Zaharee2013}. The ANSI/NISO Z39.19-2005 standard refers to it also as ``pick list''~\citep{niso2005}, whereas within the ISO 25964-2, it is referred to as ``terminology''~\citep{ISO25964}.  The most important distinction from other types of KOS is that term lists do not define any relationships between the subjects.

\begin{figure}[!t]
\begin{minipage}[b]{.5\textwidth}
\centering
\begin{subfigure}[b]{\textwidth}
  \centering
  \includegraphics[width=\linewidth]{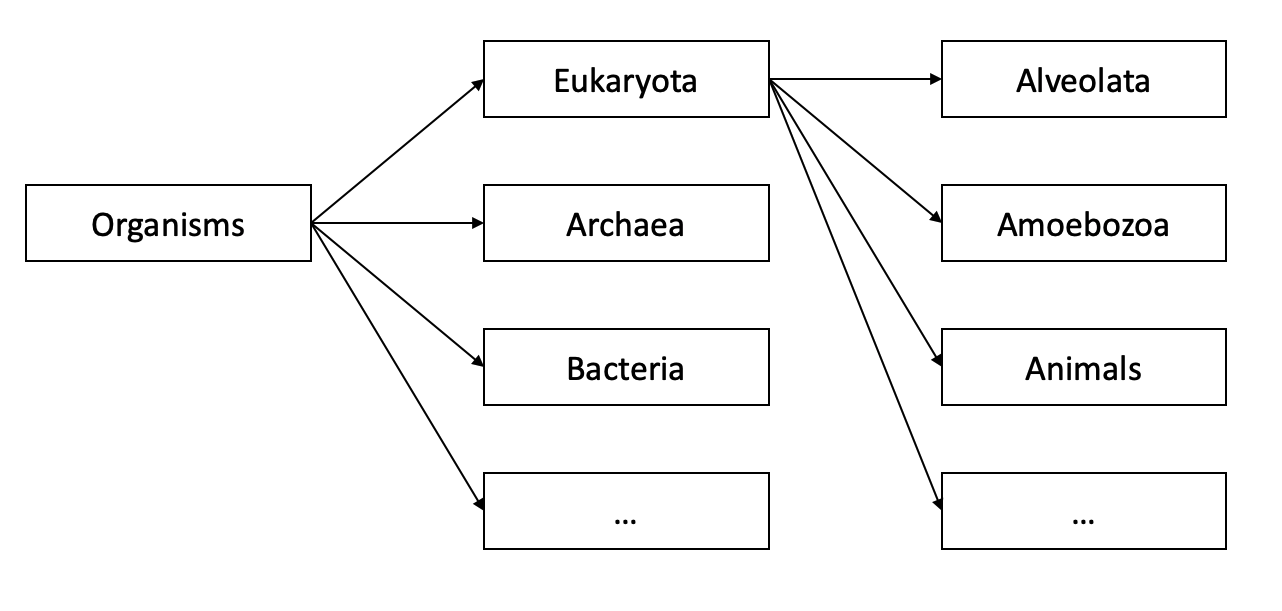}
  \caption{Taxonomy}
  \label{fig:taxonomy}
\end{subfigure}%
\begin{subfigure}[b]{\textwidth}
  \centering
  \includegraphics[width=\linewidth]{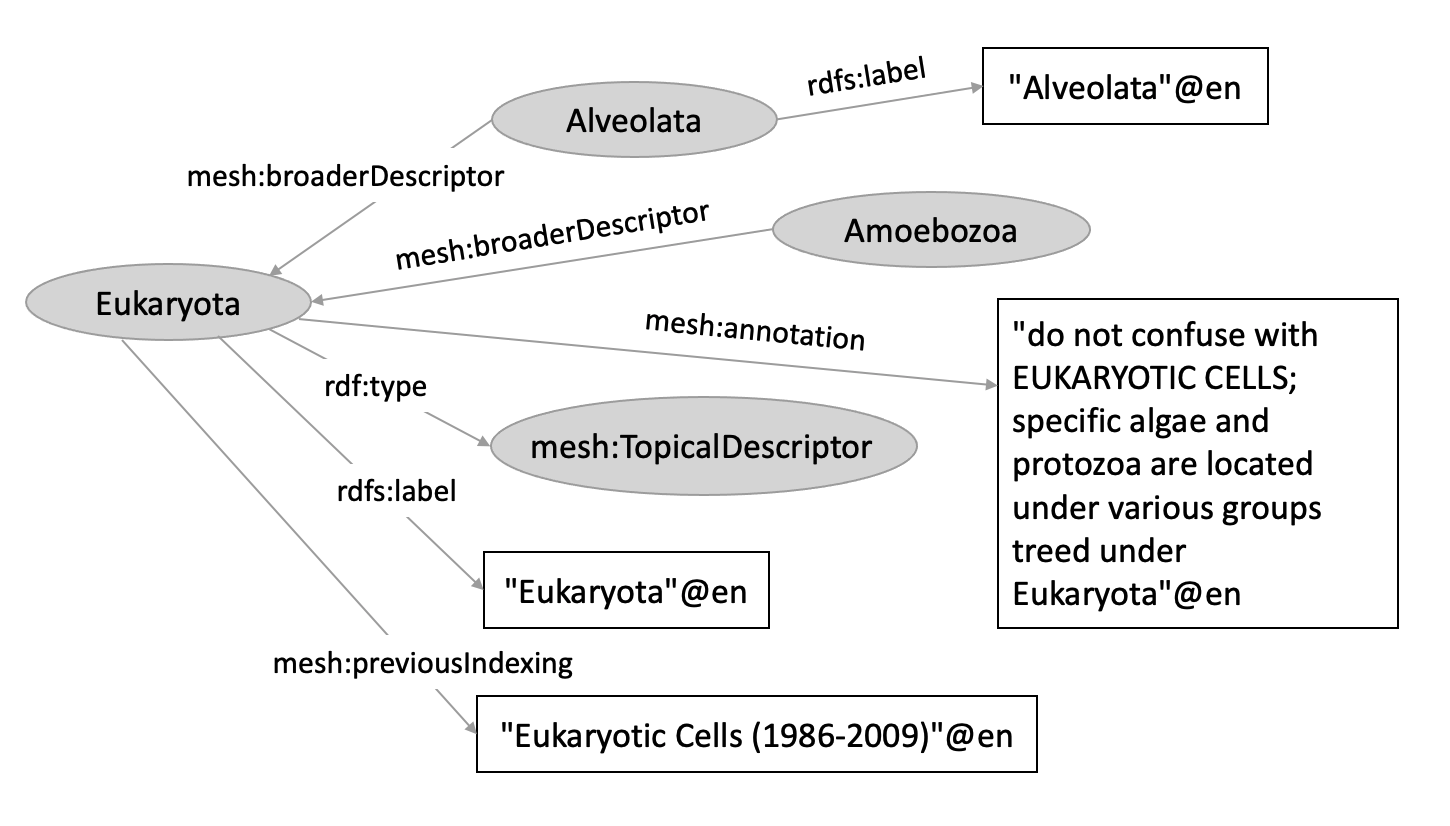}
  \caption{Ontology}
  \label{fig:ontology}
\end{subfigure}
\end{minipage}

\caption{A small portion of the Medical Subject Headings represented both through a taxonomy (a) and an ontology (b). 
The taxonomy consists solely of a set of terms connected by hierarchical relationships. The ontology is instead a rather complex structure as it contains entities (grey ovals) and literals (white boxes), connected through semantic relationships.}
\end{figure}

A \textbf{hierarchical taxonomy} (from here on just \textit{taxonomy}) organises the classes in a hierarchical structure with parent-child relationships~\citep{rasch1987nature}.
In practical terms, as shown in Figure~\ref{fig:taxonomy}, a taxonomy is typically organised in a tree structure, with a root node at the top 
that unfolds in several sub-branches. 
An important characteristic is that all items classified according to a class can also be considered under all its super-classes. For instance, in Figure~\ref{fig:taxonomy}, all \topic{Bacteria} are also \topic{Organisms}. 

A \textbf{thesaurus} can be seen as an extension of the taxonomy. Subjects are organised in a hierarchical structure and they can also be described according to additional properties, such as a description, related terms, and synonyms~\citep{niso2005}.

Finally, an \textbf{ontology}, from the \topic{Information Technology} perspective, is a formal and explicit description of knowledge within a domain, categorising things according to their essential or relevant qualities~\citep{gruber1993}. 
Ontologies conceptualise a subject area according to an abstract model of how domain experts rationalise knowledge in the domain. 
In practice, ontologies consist of a set of concepts, objects, and other entities and the relationships between them~\citep{genesereth2012}. 
Compared to thesauri, ontologies are more expressive since they can also express axioms and restrictions, so to provide local constraints on properties. 
As an example, Figure~\ref{fig:ontology} shows a portion of the \kos{MeSH} ontology. In particular, 
the two \semrel{mesh:broaderDescriptor} relationships define the super-class of both \topic{Alveolata} and \topic{Amoebozoa} which is \topic{Eukaryota}. Next, \semrel{rdfs:label} provides a human-readable version of a given class. Moreover, the \semrel{mesh:previousIndexing} points to the same element in a previous version of the KOS, and the \semrel{mesh:annotation} provides a human-readable description of the class.

In certain circumstances, as we will see from the analysis, the naming conventions are somewhat disorganised. 
Some KOSs are named after a particular category but, upon closer examination, actually represent a different category based on the definitions discussed above. \color{black}
For instance, the \kos{UNESCO Thesaurus} and the \kos{STW Thesaurus for Economics}, despite their names, have evolved away from traditional thesaurus structures, aligning with the expressiveness of ontologies~\citep{8873649}.

\subsection{Selection criteria}\label{inclusion_criteria}

In this section, we define standard inclusion (IC) and exclusion (EC) criteria for the survey. 

We selected all KOSs which match the following inclusion criteria:
\begin{enumerate}[label={\textbf{IC \arabic*.}},align=left]
\item they describe academic research topics as defined in Sec.~\ref{definitions};
\item they cover at least one of the following 19 broad research fields: \topic{Art}, \topic{Biology}, \topic{Business}, \topic{Chemistry}, \topic{Computer science}, \topic{Economics}, \topic{Engineering}, \topic{Environmental science}, \topic{Geography}, \topic{Geology}, \topic{History}, \topic{Materials science}, \topic{Mathematics}, \topic{Medicine}, \topic{Philosophy}, \topic{Physics}, \topic{Political science}, \topic{Psychology}, and \topic{Sociology};  
\item they are adopted by the main bibliographic databases or regularly used by the scientific community.
\end{enumerate}

For the \textbf{IC 2}, we adopted the top-level fields of the Microsoft's Fields of Study~\citep{sinha2015} due to their comprehensive coverage.

In contrast, we excluded KOSs that match the following criteria:
\begin{enumerate}[label={\textbf{EC \arabic*.}},align=left]
\item do not offer an English version;
\item are exclusively tailored to the content of a specific library and therefore are not adopted by a community of users.
\end{enumerate}

We established the \textbf{EC 2} criterion because there is an abundance of KOSs created by specific libraries (e.g., the \kos{Aarhus University Library Classification System}\footnote{Aarhus University Library Classification System --- \url{http://web.archive.org/web/20210721074627/https://library.au.dk/en/subject-areas/political-science/classification-system}}) for internal needs, but not available or (re)used by the scientific community. 


\subsection{Methodology for the retrieval of relevant KOSs}\label{finding}

Identifying all the relevant KOSs has proven to be a challenging task.
A common approach, when pursuing a survey or a review of the literature,  is to rely on a search engine, e.g., Scopus, Web of Science, Google Scholar, and construct a particular query that would return all research papers reporting the objects under review~\citep{publications9010012}.
However, a good number of systems organising research areas are not well described or documented in research papers. 
Therefore, relying on this typical approach would have produced limited results. To this end, we designed and performed the following systematic strategy.

\begin{enumerate}[label={\textbf{Phase \arabic*:}},align=left]

\item We started by querying Google Scholar for potential candidates using the following query  
``\textit{(controlled vocabulary OR taxonomy OR thesaurus OR ontology OR subject headings OR subject classification) AND (research OR science)}''. This step identified 7 KOSs.

\item We analysed the websites of academic publishers, pre-print archives, and academic search engines to identify the KOSs they use to organise their content. Specifically, we considered Scopus, Web of Science, Dimensions.ai, PubMed, the Springer Nature portal, ACM, IEEE, OpenAlex, ArXiv, and the last instance of Microsoft Academic Graph. This led to the identification of 9 additional KOSs.

\item We adopted the Google search engine to identify additional KOSs, using the query ``\textit{("controlled vocabulary" OR "term list" OR "taxonomy" OR "thesaurus" OR "ontology" OR "subject headings" OR "subject classification")}'' in combination with the 19 broad fields listed in \textbf{IC 2}. As a result, we identified 12 additional KOSs.

\item We contacted researchers working in the field of \topic{Digital Libraries} and asked them whether they could point us toward any additional effort. They suggested 2 more KOSs.

\item We expanded the resulting KOSs by analysing their links on Wikipedia.
Specifically, we relied on the ``See also'' section, typically listing online databases and other related KOSs, which allowed us to discover 5 additional ones.

\item We retrieved and considered all the KOSs that are explicitly connected, or otherwise referenced into the KOSs identified at previous stages. As a result, we gathered 8 additional KOSs. 


\item Whenever we could not find at least one KOS within one of the 19 disciplines defined in IC2 (e.g., \topic{History}), we reached out to professors in the missing disciplines, asking which KOS they employ, if any. 
Their response directed us toward 2 additional multi-field KOSs.
\end{enumerate}


\subsection{Methodology for the analysis of KOSs}\label{features}

We analysed each KOS according to the five main aspects summarised in Figure~\ref{fig:categories}: \textit{Scope}, \textit{Structure}, \textit{Curation}, \textit{External Links}, and \textit{Usage}.

\input{tables/features}

\textbf{Scope.}
The scope is the set of research fields covered by a KOS. Some KOSs focus on one specific field, such as \kos{PhySH} for \topic{Physics} or \kos{Mathematics Subject Classification} for \topic{Mathematics}. 
These may also include elements from other complementary fields. This is the case of \kos{PhilPapers Taxonomy} that mainly focuses on \topic{Philosophy}, but also describes some concepts from other fields, including \topic{Mathematics}. 
Another category of KOSs covers multiple fields by design, typically because they aim to offer good coverage of an extended set of scientific or academic disciplines. 
A good example is the \kos{UNESCO Thesaurus}, a well-known KOS which aims to cover all academic disciplines and is adopted by several libraries.

\textbf{Structure.}
The structural characteristics of a KOS include many aspects based on its topology and the way subjects are arranged. 
First, we classified each KOS according to the four types defined in Sec.~\ref{definitions}: term lists, taxonomies, thesauri, and ontologies. For example, the \kos{Web of Science Categories} is a flat list of terms, and thus, we characterised it as a term list. The \kos{Mathematics Subject Classification} and ANZSRC's \kos{Fields of Research} are taxonomies since they arrange topics in a hierarchy. The \kos{IEEE Thesaurus} is a thesaurus as it offers hierarchical information as well as synonyms. Finally, \kos{TheSoz} is an ontology because, in addition to the hierarchical structure and synonyms, it adds axioms and constraints. 


We then considered the number of concepts and the maximum depth of the hierarchical tree (which is one in the case of term lists). The depth was computed as the number of levels from the most generic concept (root) to the most specific concepts (leaves). Depth can be generally used as an indicator of specificity, since deep taxonomies tend to include a granular representation of very detailed (narrow) topics and may allow for more fine-grained content organisation.

If the KOS is hierarchical, we also characterised it as either mono-hierarchical or poly-hierarchical according to the type of employed hierarchy. In mono-hierarchical KOSs, each concept is assigned to a single-parent category. In contrast, poly-hierarchical KOS enable concepts to belong to multiple parent categories. As we will see, some KOSs attempt to organise their concepts within a strict mono-hierarchical structure (e.g., the \kos{Mathematics Subject Classification}), while others (e.g., \kos{STW Thesaurus for Economics}, \kos{Medical Subject Headings}) are organised in a poly-hierarchical structure. 

We also considered if the KOS contains synonyms that can be used to refer to the same concept or related terms. For instance, the \kos{Computer Science Ontology} uses \topic{Ontology Matching} and \topic{Ontology Mapping} as different labels for the same research topics. The \kos{IEEE Thesaurus} offers instead a set of related terms that are not necessarily synonyms but are semantically similar. For instance, \topic{4G mobile communication} is a related term of \topic{5G mobile communication}. The presence of different labels typically facilitates the classification of documents by both human annotators and automatic approaches. 

We used Python scripts for automatically extracting the structural features from the KOSs published in standard formats (e.g., RDF). The scripts are publicly available as Jupyter notebooks on a GitHub repository\footnote{Jupyter notebooks for analysing KOSs --- \githubrepo}.
For the knowledge organization systems that are available only as HTML or PDF, we manually extracted these features.

\textbf{Curation.}
The curatorial aspect takes into consideration many features regarding the creation, update, and publication of the KOS.
First, we considered how the KOS was generated according to four categories: manual, automatic, semi-automatic, and community effort. The first and more common category includes all the KOSs that were crafted manually by domain experts, such as the \kos{ACM Computing Classification Scheme}. The \textit{automatic} category covers the KOSs that were generated automatically using NLP or Machine Learning approaches, such as the \kos{Computer Science Ontology}. The \textit{semi-automatic} category covers KOSs generated by blending domain experts and automatic tools. For instance, the \kos{OpenAlex Topics} uses a manual approach for the first three levels and an automatic one for the last level. The \textit{community effort} category includes the KOSs that were crowdsourced with the support of the community, such as \kos{Physics Subject Headings}.

We also considered the format used for publishing the KOSs. A good number of KOSs are published according to W3C standard formats such as Ontology Web Language (OWL), Terse RDF Triple Language (TTL), N-Triples (NT) for expressing data in the Resource Description Framework (RDF)\footnote{Resource Description Framework (RDF) --- \url{https://www.w3.org/RDF}} data model, as well as JSON, and CSV.
An alternative is the OBO format\footnote{OBO Format --- \url{http://purl.obolibrary.org/obo/oboformat/spec.html}}, which is still a working draft standard for representing ontologies, based on the principles of OWL and predominantly used in the field of \topic{Biology}. Other KOSs are instead published in semi-structured formats, for instance, as a list of terms in HTML pages or PDF files. These solutions are typically not machine-interpretable and offer very limited support to automatic classifiers. 

Next, we observed the distribution license. Some KOSs are open and released under Creative Commons licenses\footnote{Creative Commons --- \url{https://creativecommons.org}}, while some others are copyrighted and may require a subscription fee.
We also reported the frequency of updates and the date of the latest release, which is a good indication of whether the KOS is actively maintained.
In addition, we considered whether the KOS is distributed in languages other than English. This is crucial for the interoperability of digital libraries across the different languages.
Finally, we reported the current maintainers of those KOSs, as they are the first point of contact for interested readers.
All these curatorial features have been identified from the official websites and relevant literature.

\textbf{External Links.}
Links to other KOSs allow the integration of different knowledge bases, potentially offering a more comprehensive representation of scientific disciplines. Therefore, we documented whether a KOS includes links to other KOSs, including those referencing general knowledge graphs, such as DBpedia\footnote{DBpedia --- \url{https://wiki.dbpedia.org}} and Wikidata\footnote{Wikidata --- \url{https://www.wikidata.org}}.
For example, the \textit{STW Thesaurus for Economics} is mapped to Wikidata, DBpedia, and others.
We discuss the crucial implications of this mechanism in Section~\ref{challenges}.

\textbf{Usage.}
The final aspect that we considered regards the presence of collections of documents or other artefacts that are annotated or organised according to the KOS. Their existence attests to the fact that a portion of the community actively uses that KOS. Furthermore, annotated datasets can also be used for training machine learning classifiers able to categorise new documents according to the original KOS~\citep{kandimalla2020,salatino2021}. For instance, the \kos{Fields of Research} is currently adopted by institutional repositories in both Australia and New Zealand as well as Dimensions.ai\footnote{Dimension.ai --- \url{https://app.dimensions.ai/discover/publication}} to index their publications metadata. Another example is \kos{Physics Subject Headings} employed to classify the articles of the Physical Review Journal\footnote{Physical Review Journal --- \url{https://journals.aps.org}}.


\input{analysis}

\input{new_challenges}

\input{treats_to_validity}


\section{Conclusions}\label{conclusions}

Knowledge organization systems of academic fields (e.g., term lists, taxonomies, thesauri, ontologies) are an important part of the academic ecosystem and enable the categorisation, management, and retrieval of items and information.
These solutions have become particularly important in the last few years given the ever-growing number of publications, the rise of Open Science, and the emergence of vast online repositories of articles, courses, and other academic materials~\citep{auer2018towards}.

This article provides a systematic overview of 45 KOSs, with 23 focusing on a single field and 22 covering multiple fields.
We propose an analysis framework that characterises them according to five main dimensions: scope, structure, curation, usage, and links to other KOSs. 
The comparative table describing the 45 KOSs according to the 15 features is available on \maintableurl, as a living review. All the code produced and the data retrieved during this analysis are openly available at \githubrepo.
These resources can be freely reused by researchers to re-run the analysis in the future.


Our findings indicate that the current generation of KOSs requires substantial enhancements in scope, quality, and granularity. Notably, seven disciplines (\topic{History}, \topic{Political Science}, \topic{Environmental Science}, \topic{Material Science}, \topic{Geography}, \topic{Sociology}, and \topic{Business}) lack dedicated, field-specific KOSs. Additionally, among the existing multidisciplinary KOSs, only five provide comprehensive coverage of the 19 fields analysed in this study.

The analysed KOSs exhibit considerable diversity in both the number of concepts and structural depth. Some KOSs include as few as 48 concepts (e.g., \kos{Fields of research and development}) with a relatively shallow structure, while others encompass more than 3 million concepts and feature a very deep structure, extending beyond 30 levels (e.g., \kos{Open Biological and Biomedical Ontology} and \kos{Unified Medical Language System}). 
The majority of the KOSs analysed (24) are still traditional taxonomies. However, a growing number of more recent KOSs (18) are now based on ontologies. Regarding hierarchical structures, there is a fairly balanced distribution: 24 KOSs utilise a poly-hierarchical structure, while 20 employ a mono-hierarchical one.
Traditionally, KOSs have been created and maintained manually. However, in recent years, there has been a significant shift toward more automated methods. Consequently, eight KOSs, including \kos{OpenAlex Topics} and \kos{Microsoft’s Fields of Study}, have implemented modern automatic or semi-automatic pipelines for their generation and updating processes. 
Eighteen KOSs are updated annually, demonstrating a significant commitment to maintaining up-to-date resources in some disciplines (e.g., \topic{Medicine}, \topic{Engineering}, \topic{Agriculture} and \topic{Computer Science}). Finally, 26 KOSs are available under open licenses.

Our findings indicate that there is currently no existing multi-field KOS that simultaneously is comprehensive in topic coverage, granular, consistently updated, and openly accessible (see Table~\ref{tab:analysistables}). The creation of such a KOS could revolutionise the field by dramatically improving content organisation across digital libraries, simplifying the process of research information gathering, facilitating the monitoring of research and development activities, ensuring high data quality, and supporting evidence-based policy making.



We have identified key future research directions in this domain, along with the challenges that accompany them. These priorities include generating higher-quality multi-field KOSs, developing new automatic methods for integrating and updating KOSs, adopting standardised formats, expanding language coverage, and evaluating KOSs across diverse characteristics. Furthermore, there is a critical need for more precise and scalable methods for automatically classifying articles according to these systems. 

Overall, this is a promising area of research that has yet to fully capitalise on recent advances in generative AI, which hold significant potential to drive progress in the field.
To achieve this, we believe it is essential for the Open Science, Digital Libraries, and Artificial Intelligence communities to collaborate in developing a unified framework of interlinked resources. We hope that this survey paper serves as a meaningful first step toward this direction.

\section*{Acknowledgements}
The authors wish to thank the reviewers for their constructive feedback, which is reflected in
this version of the manuscript. They also gratefully acknowledge The Open University Library for providing access to the Dewey Decimal Classification.

\section*{Author Contributions}
\textbf{Angelo Salatino}: Conceptualization, Formal analysis, Methodology, Software, Visualization, Writing – original draft. 
\textbf{Tanay Aggarwal}: Formal analysis, Software. 
\textbf{Andrea Mannocci}: Conceptualization, Formal analysis, Methodology, Writing – review \& editing. 
\textbf{Francesco Osborne}: Conceptualization, Formal analysis, Methodology, Writing – review \& editing. 
\textbf{Enrico Motta}: Formal analysis, Writing – review \& editing.

\section*{Competing Interests}
No competing interests to declare. 

\section*{Funding Information}
No funding was received for this research.

\section*{Data Availability}

The code for processing the analysed KOSs is available on a GitHub repository: \githubrepo. A complete table of all KOSs and their analysed features can be found at \maintableurl.

\bibliography{taxonomies}





\end{document}

%% file: tables/features.tex
\tikzset{
  my rounded corners/.append style={rounded corners=2pt},
}
\forestset{parent color/.style args={#1}{
    {fill=#1},
    for tree={fill/.wrap pgfmath arg={#1!##1}{1/level()*80},draw=#1!80!darkgray}},
    root color/.style args={#1}{fill={{#1!60!gray!25},draw=#1!80!darkgray}}
}


\definecolor{mycolor1}{RGB}{255,255,255}
\definecolor{mycolor2}{RGB}{255,255,255}
\definecolor{mycolor3}{RGB}{255,255,255}
\definecolor{mycolor4}{RGB}{255,255,255}
\definecolor{mycolor5}{RGB}{255,255,255}
\definecolor{mycolor6}{RGB}{255,255,255}

\begin{figure*}[h]
\begin{adjustbox}{width=\textwidth}
    \begin{forest}
      for tree={
        line width=1pt,
        if={level()<2}{
          my rounded corners,
          draw=linecol,
        }{},
        edge={color=linecol, >={Triangle[]}, ->},
        if level=0{%
          l sep+=1.5cm,
          align=center,
          font=\bfseries,
          parent anchor=south,
          tikz={
            \path (!1.child anchor) coordinate (A) -- () coordinate (B) -- (!l.child anchor) coordinate (C) pic [draw, angle radius=15mm, every node/.append style={fill=white, opacity=0.6}
            , "based on"] {angle};
          },
        }{%
          if level=1{%
            parent anchor=south,
            child anchor=north,
            tier=parting ways,
            align=center,
            font=\bfseries,
            for descendants={
              child anchor=west,
              parent anchor=south,
              anchor=west,
              align=left,
            },
          }{
            if level=2{
              shape=coordinate,
              no edge,
              grow'=0,
              calign with current edge,
              xshift=20pt,
              for descendants={
                parent anchor=south west,
                l sep+=-20pt
              },
              for children={
                edge path={
                  \noexpand\path[\forestoption{edge}] (!to tier=parting ways.parent anchor) |- (.child anchor)\forestoption{edge label};
                },
                font=\bfseries,
                for descendants={
                  no edge,
                },
              },
            }{},
          },
        }%
      },
      [Analysis, fill={mycolor1}
        [Scope, fill={mycolor2}
          [
            [Discipline
              [1. Mathematics\\2. Physics\\3. Economics\\4. ...]
            ]
          ]
        ]
        [Structure, fill={mycolor3}
          [
            [Type of system
                [1. Terms list\\2. Taxonomy\\3. Thesaurus\\4. Ontology]
            ]
            [Number of Concepts]
            [Depth]
            [Kind of hierarchy 
                [1. Mono-hierarchic\\2. Poly-hierarchic]
            ]
            [Related terms
                [1. Synonyms\\2. Related terms]
            ]
          ]
        ]
        [Curation, fill={mycolor4}
          [
            [Generation 
                [1. Automatic\\2. Manual\\3. Semi-automatic\\4. Community effort]
            ]
            [Distribution formats]
            [License]
            [Frequency of update]
            [Last update/revision]
            [Languages
                [(in addition to English)]
            ]
            [Providers/Maintainers]
          ]
        ]
        [External Links, fill={mycolor5}
          [
            [Mapping\\to external\\KOSs]
          ]
        ]
        [Usage, fill={mycolor6}
          [
            [Usage in\\Digital\\Libraries]
          ]
        ]
      ]
      \end{forest}
\end{adjustbox}
\caption{\label{fig:categories} The aspects and features used for the analysis of Knowledge Organization Systems.}
\end{figure*}

%% file: analysis.tex
\section{Analysis of Knowledge Organization Systems}
\label{results}

This section analyses the  KOSs of research fields utilising the 15 features reported in Figure~\ref{fig:categories}. We identified  45 KOSs: 23 specialising in single disciplines and 22 covering multiple fields. 
Table~\ref{tab:matrix} summarises these KOSs according to some key characteristics, i.e., primary discipline, number of concepts, hierarchical depth, type of system, kind of hierarchy, and whether it contains related terms. An extended version of this table with all the analysed features is available at \maintableurl.

This section presents the outcomes of our analysis, structured according to the hierarchical structure illustrated in Figure~\ref{fig:categories}. The subsections correspond to the five key aspects introduced in Section~\ref{features} (Scope, Structure, Curation, External Links, and Usage) and are further divided into smaller subsections, each addressing a specific feature.

It is important to note that the results presented in this section are based on an analysis conducted up to April 2024. 
Due to the dynamic nature of some KOSs, minor discrepancies in values may have occurred since then; however, the overall insights and conclusions of this manuscript remain valid.

\input{tables/bigtable}

\subsection{Scope}\label{sec:scope}

Out of the 45 identified KOSs, 22 cover multiple fields, while 23 focus on a single field of science. 
However, only 12 of the 19 fields under analysis (see IC 2, Sec. \ref{inclusion_criteria}) are addressed by at least one single-field KOS, while the remaining ones rely exclusively on broader multi-field solutions.

Notably, 6 fields are covered by more than one KOS. In \topic{Medicine}, we identified four single-field KOSs: the \kos{Medical Subject Headings}, the  \kos{Unified Medical Language System}, the \kos{National Library of Medicine classification}, and the \kos{Biomedical Ontologies from BioPortal}. Although the last expands toward \topic{Biomedicine} and hence more into the biological basis of health and disease. 
\topic{Computer Science} can rely on three KOSs: the \kos{Computer Science Ontology}, the \kos{ACM Computing Classification Scheme}, and \kos{Computer Science Subject Headings from Wikipedia}.
In the field of \topic{Biology}, we identified two KOSs: the \kos{Open Biological and Biomedical Ontology} and \kos{EDAM}. However, \kos{EDAM} focuses more on \topic{Bioinformatics}, describing the applied side of \topic{Biology} with \topic{Computer Science} tools.  In the field of \topic{Geology}, we found the \kos{GeoRef Thesaurus} and the \kos{U.S. Geological Survey Library Classification System}. \topic{Economics} is also covered by two KOSs: the \kos{Journal of Economic Literature Classification System} and the \kos{STW Thesaurus for Economics}. \topic{Physics} can rely on the \kos{Physics and Astronomy Classification Scheme} and \kos{Physics Subject Headings}. 

On the other hand, four KOSs were not directly assigned to one of the standard 19 broad fields, as they focus on more specialised areas.
Specifically, we associated \kos{Agrovoc} to \topic{Agriculture}, which is a sub-area of the \topic{Environmental Science}; \kos{Biomedical Ontologies from BioPortal} to \topic{Biomedicine}, which covers both \topic{Biology} and \topic{Medicine}; \kos{EDAM} to \topic{Bioinformatics}, which includes both \topic{Biology} and \topic{Computer Science}; and \kos{TheSoz} to \topic{Social Science}, which is typically considered a super-area of \topic{Economics}, \topic{Sociology}, \topic{Psychology}, and \topic{Political Science}.

We were unable to identify a single-field KOS for the following seven fields: \topic{History}, \topic{Political Science}, \topic{Environmental Science}, \topic{Material Science}, \topic{Geography}, \topic{Sociology}, and \topic{Business}. 
We reached out to a number of professors and domain experts in such fields, who confirmed this finding and mentioned that they usually rely on generic multi-field KOSs, such as the \kos{Dewey Decimal Classification} and the \kos{Library of Congress Classification}. To the best of their knowledge, KOSs for such fields are yet to be developed.

In the category of multi-field KOSs, we observed a substantial diversity in topic coverage. Table~\ref{tab:topic-spread} outlines the various fields covered by each KOS. For each system, the green fields represent the primary areas of focus, the blue fields indicate secondary areas with only a limited number of research topics, and the orange fields are those that are merely mentioned without any specific sub-areas. Notably, only five of these KOSs, emphasised in bold within the table, consistently cover all the 19 top-level research areas presented in Section~\ref{methodology}.

\input{tables/topic-distro-table}

\subsection{Structure}
This section analyses the KOSs based on the previously defined structural features, focusing on their size, depth, type, hierarchical organisation, and the presence of related terms.

\subsubsection{Type of KOS}

Table~\ref{tab:matrix} (column `Sy') indicates the category of each system using the following designations: \textbf{T} (taxonomy), \textbf{O} (ontology), \textbf{U} (thesaurus), and \textbf{L} (term list). Taxonomies (23 KOSs) and ontologies (18 KOSs) demonstrated the highest prevalence.  
In contrast, term lists and thesauri were less represented, with only one (\kos{Web of Science Categories}) and three KOSs (\kos{UMLS}, \kos{IEEE Thesaurus}, and \kos{GeoRef Thesaurus}), respectively.

This scenario suggests a clear division between the two approaches to categorising research topics. On one side, some communities prefer to use straightforward hierarchical taxonomies, often encoded in very simple formats or structured documents. On the other side, other communities embrace the richer expressivity of ontologies, which enable detailed descriptions of research topics, the relationships between them (e.g., causal, contributory, part/whole, and ancestral), and constraints~\citep{kendall2019ontology}. Thesauri may be less common because they are more complex than hierarchical taxonomies but lack the full expressive power of ontologies.

\subsubsection{Number of concepts}

The number of concepts within KOSs varies widely, spanning from smaller systems like \kos{Fields of Research and Development} (48 concepts) to vast ontologies like the \kos{Biomedical Ontologies from BioPortal} (13 million concepts). Fourteen KOSs contain fewer than 1,000 concepts, seventeen have between 1,000 and 10,000 concepts, nine include 10,000 to 100,000 concepts, and five have more than 100,000 concepts.

The median number of concepts within the analysed KOSs is approximately 4,700.  
Sixteen single-field KOSs exceed this median in concept count, while sixteen of the multi-field KOSs contain fewer concepts. This pattern suggests a trend: single-field KOSs tend to be larger and more specialised, likely to capture the intricacies within their specific domains. In contrast, multi-field KOSs seem to be designed to offer a broader overview across various fields, often resulting in a smaller number of concepts.

The six multi-field KOSs that exceed the median number of concepts, thereby providing a more granular representation of topics, are \kos{OpenAlex Taxonomy} (4,798 concepts), \kos{EuroVoc} (7,423), \kos{OpenAIRE's Field of Science Taxonomy} (50K), \kos{Dewey Decimal Classification} (60K), \kos{Library of Congress Classification} (467K), and \kos{Microsoft's Fields of Study} (704K).

\subsubsection{Depth}

Similarly to the number of concepts, the depth of KOSs also displays considerable variety, ranging from 1 (\kos{Web of Science Categories}) to 39 (\kos{Open Biological and Biomedical Ontology}). 
Specifically, 15 KOSs have up to 3 levels, 6 feature between 4 and 5 levels, 9 employ exactly 6 levels, 3 use between 7 and 10 levels, and 11 extend beyond 10 levels, as illustrated in Figure~\ref{fig:depths}.

The median depth stands at 6, with 18 of the 22 multi-field KOSs falling below this threshold. This finding further supports the notion that the majority of multi-field systems aim to provide a broad overview across diverse research fields.

\begin{figure}
    \centering
    \includesvg[width=0.8\linewidth]{figures/distribution_depths.svg}
    \caption{Depth distribution for single-field and multi-field KOSs.}
    \label{fig:depths}
\end{figure}

\subsubsection{Kind of hierarchy}

Column `Hr' in Table~\ref{tab:matrix} reveals a near-equal distribution between poly-hierarchical (24) and mono-hierarchical (20) KOSs. Only one KOS (\kos{Web of Science Categories}) is considered non-hierarchical since it is a flat list of terms.
An analysis of the 23 single-field KOSs reveals that poly-hierarchical structures are more prevalent than mono-hierarchical ones (see Table~\ref{tab:kohcov}). In contrast, multi-field systems display a nearly balanced distribution, with 11 mono-hierarchical and 10 poly-hierarchical structures. This trend suggests that single-field KOSs tend to favour poly-hierarchical structures, potentially due to the need to represent finer-grained research topics that often stem from multiple parent topics.

\begin{table}[h!]
\centering
\scriptsize
\rowcolors{2}{orange!20}{white}
    \caption{Distribution of KOSs according to the type of hierarchy and their coverage (single or multi-field). The * value does not account for \kos{Web of Science Categories}, which is non-hierarchical.}
    \label{tab:kohcov}
    \begin{tabular}{r|c|c|c}
\cmidrule{2-4}
                  & \textbf{Mono-hierarchical} & \textbf{ Poly-hierarchical} & \textbf{Total} \\
\midrule
Single-field KOSs & 9                 & 14                & 23    \\
Multi-field KOSs  & 11                & 10                & 21*   \\
\midrule
Total             & 20                & 24                &      \\
\bottomrule
    \end{tabular}
\end{table}


\subsubsection{Related terms}


Table~\ref{tab:matrix} (column `RT') reveals that the 21 KOSs incorporating related terms are mainly ontologies and thesauri. This aligns with the inherent capacity of these structures to express associative relationships (e.g., related terms). Interestingly, 15 of them are single-field KOSs.


Noteworthy exceptions are the \kos{Mathematics Subject Classification} and the \kos{Dewey Decimal Classification}. Although the \kos{Mathematics Subject Classification} is traditionally considered a taxonomy~\citep{dunne2020}, it also includes related terms. This functionality is enabled on the zbMATH website\footnote{zbMATH --- \url{https://zbmath.org/classification}}, where the KOS is displayed with a ``see also'' anchor, without adherence to the ANSI/NISO standard~\citep{niso2005}. Similarly, the \kos{Dewey Decimal Classification} uses the ``see-also'' references for synonyms and related terms~\citep{mitchell2001relationships}.


\subsection{Curation}


This section explores the creation of KOSs, focusing on their generation methodologies, distribution formats, licensing, update frequency, available languages, and curators.

\subsubsection{Generation}
As previously discussed, the generation of KOSs can be categorised as manual, automated, semi-automated, or community-based.


The majority (32) of the analysed KOSs are developed manually. 
Developing these large knowledge bases manually is typically both time-consuming and very expensive. 

Six KOSs have already adopted semi-automatic methodologies: \kos{Biomedical Ontologies from BioPortal}, \kos{KNOWMAK}, \kos{Microsoft's Fields of Study}, \kos{OpenAIRE's Field of Science Taxonomy}, \kos{Subject Resource Application Ontology}, and \kos{OpenAlex Topics}.
For instance, \kos{OpenAlex Topics} combines manually curated concepts from the \kos{All Science Journal Classification Codes} with over 4,500 new research topics automatically identified through citation clustering. 
The approach first involved clustering papers based on citation patterns to form thematically related groups. These groups were then labelled using large language models and subsequently integrated with the existing concepts in the \kos{All Science Journal Classification Codes}~\citep{openalex2024}.


Only two KOSs are generated using a fully automated pipeline, both within the field of computer science: the \kos{Computer Science Ontology} and \kos{Computer Science Subject Headings from Wikipedia}. 
The \kos{Computer Science Ontology} was generated using Klink-2~\citep{osborne2015klink}, which processed 16 million scientific publications.
Klink-2 identifies relationships between topics by analysing various indicators, including co-occurrence patterns, temporal distributions, and label similarity. 
The \kos{Computer Science Subject Headings from Wikipedia} was created through an automated approach aimed at enhancing and refining the \topic{Computer Science} branch of the Wikipedia category system. This approach integrates community detection, machine learning, and manually developed heuristics to identify and incorporate additional topics from Wikipedia articles~\citep{han2020}.

Finally, five KOSs leverage direct community expertise for their construction: \kos{Open Biological and Biomedical Ontology}, \kos{Medical Subject Headings}, \kos{Unified Medical Language System}, \kos{PhilPapers Taxonomy}, and \kos{Physics Subject Headings}. They use various technologies to facilitate collaboration and enable researchers to suggest modifications to the KOSs. For instance, \kos{Physics Subject Headings} encourages researchers to propose changes through GitHub issues, while \kos{Medical Subject Headings} requests researchers to submit cases on their portal to suggest new terms, alterations, or corrections to the tree structure. \kos{PhilPapers Taxonomy} consults experts in relevant areas, gathers insights from forum discussions, and incorporates feedback provided to the editors.



\subsubsection{Distribution formats}

Table~\ref{tab:formats} details the diverse formats in which KOSs are released. These range from PDF and HTML to more machine-readable options like CSV (and Excel), TSV, MARC, and RDF\footnote{Resource Description Framework --- \url{https://www.w3.org/RDF}} (i.e., Resource Description Framework), an open standard established by the World Wide Web Consortium\footnote{World Wide Web Consortium --- \url{https://www.w3.org}} (W3C). Several KOSs are released in multiple formats to cover different use cases.

The majority of KOSs (27) are accessible in HTML format, primarily through their providers' websites. In this format, the KOS is essentially a webpage listing and connecting concepts. While this is human-readable, the underlying data lacks the structured organisation necessary for direct and automated processing by computer systems.
Several KOSs are available in machine-interpretable formats: 17 utilise the RDF standard, and 10 are available in CSV format. Finally, 6 KOSs are exclusively available in PDF format, which complicates their integration with other KOSs and hinders automatic analysis.




\input{tables/formats}

\subsubsection{License}

The majority of the KOSs under analysis (26) employ open licenses from Creative Commons (CC), Open Data Commons (ODC), Open Database License (ODbL), or MIT licenses. Their openness varies considerably, as systems like \kos{Physics Subject Headings} (CC0 1.0) are highly open, while others, such as \kos{TheSoz} (CC BY-NC-ND 3.0), impose slightly stricter terms. Nonetheless, the CC BY license remains the most prevalent.

Eleven KOSs are freely accessible online but maintain copyright restrictions, meaning that they can be browsed but cannot be downloaded, modified, or redistributed without explicit permission from the copyright holder. 

Two KOSs are free from copyright: \kos{Medical Subject Headings} and \kos{National Library of Medicine classification}.
Four KOSs are copyrighted and inaccessible online. For instance, both the \kos{Dewey Decimal Classification} and the \kos{Library of Congress Classification} necessitate licensing fees even for browsing purposes.

Finally, \kos{Open Biological and Biomedical Ontology} and \kos{Biomedical Ontologies from BioPortal} represent broader initiatives that integrate multiple ontologies. The licensing terms for these efforts are contingent on the individual ontologies they incorporate. For the former, its mission dictates that all incorporated ontologies must be openly accessible under licenses such as CC BY 3.0, CC BY 4.0, or CC0 1.0, ensuring unrestricted use. In contrast, BioPortal presents a more diverse licensing landscape. Indeed, while many ontologies are openly available, some have specific terms of use set by their providers.

\subsubsection{Frequency of updates}\label{sec:frequency}

Determining the frequency of updates proved to be a significant challenge, as providers rarely disclose their updating schedules explicitly. In some instances, we were able to infer updating patterns by analysing the dates of previous releases. 

Eight KOSs are continuously updated and maintained, with new revisions produced monthly or more frequently. These include the \kos{Library of Congress Subject Headings}, the \kos{Art \& Architecture Thesaurus}, the \kos{Unesco Thesaurus}, \kos{TheSoz}, the \kos{Dewey Decimal Classificaion}, the \kos{Open Biological and Biomedical Ontology}, the \kos{PhilPapers Taxonomy}, and the \kos{Agrovoc}. 

Three systems are updated regularly, although less frequently, with revisions taking place twice a year: \kos{Unified Medical Language System}, \kos{National Library of Medicine classification}, \kos{EuroVoc}.

Seven KOSs are updated once a year: the \kos{STW Thesaurus for Economics}, the \kos{Medical Subject Headings}, the \kos{Computer Science Ontology}, \kos{OpenAlex Topics}, the \kos{IEEE Thesaurus}, \kos{OpenAIRE's Field of Science Taxonomy}, and the \kos{Subject Resource Application Ontology}.  
Five KOSs receive less frequent updates (roughly every 10--15 years), these include the \kos{ACM Computing Classification System}, the \kos{Mathematics Subject Classification}, \kos{Science Metrix Classification}, \kos{Socio-Economic Objective}, and the \kos{Fields of Research}.

Finally, three KOSs are no longer actively maintained but continue to be used by their respective communities:  \kos{Research Fields, Courses and Disciplines}, \kos{Microsoft's Fields of Study} and the \kos{Physics and Astronomy Classification Scheme}.

There are many factors that can influence the frequency of updates of a KOS. One factor is related to the discipline and its evolution pace. For instance, the field of \topic{Medicine} is a fast-advancing field, and for this reason, there is a new version of \kos{Medical Subject Headings} released every year.

Another factor influencing the frequency of updates is the depth (i.e., specificity) of a KOS. For instance, the \kos{Agrovoc Thesaurus}, whose depth equals 14, requires more frequent updates to include new emerging topics and readjust the hierarchical structure due to epistemological changes. For this reason, \kos{Agrovoc Thesaurus} receives monthly updates.
On the contrary, the \kos{Fields of Research (ANZSRC)} consists of only three levels, and since the modelled concepts can be considered quite generic, it is reasonable to assume that their structure will uphold over a relatively longer timespan. 
Indeed, the \kos{Fields of Research (ANZSRC)} received the most recent update in 2020, after 12 years.

We also investigated whether the frequency of updates is related to the type of system, as shown in Figure~\ref{frequency_update}. Our analysis revealed that among the 18 KOSs updated within a year, the majority were ontologies (11), followed by taxonomies (5), and thesauri (2). In contrast, the 8 KOSs that were either updated every ten years or discontinued were taxonomies. This suggests that while taxonomies might be effective initially for organising subjects hierarchically, they may become increasingly challenging and expensive to maintain over time due to the growing complexity and unique characteristics of the subjects they represent. On the other hand, ontologies, often implemented in flexible formats such as OWL and RDF, may be easier to maintain.


\begin{figure}
    \centering
    \includesvg[width=0.8\linewidth]{figures/updatefrequency.svg}
    \caption{Frequency of update of KOSs in relation to their type.}
    \label{frequency_update}
\end{figure}

\subsubsection{Last Update}

Twenty systems have received their most recent update in 2023 or later, while thirteen were last updated between 2020 and 2022. This indicates that within the last five years, thirty-three KOSs have been revised to reflect the evolving nature of their respective fields. Of these, 17 are single-field systems, and 15 are multi-field.
The remaining twelve systems either received their last update before 2020 or their latest release date could not be determined.

Notably, key disciplines such as \topic{Chemistry} (with \kos{ChemOnt}, last updated in 2016) and \topic{Geology} (with the \kos{U.S. Geological Survey Library Classification System}, last updated in 2000, and the \kos{GeoRef Thesaurus}, last updated in 2008) can only rely on possibly outdated KOSs.
This is potentially a significant issue, as the outdated representations in these disciplines can hinder the dissemination of modern research efforts that address topics not fully covered by these KOSs.

\subsubsection{Languages}\label{sec:language}

Our exclusion criteria ensured that all KOSs included in this study had at least one version available in English. Specifically, thirty-five KOSs are exclusively available in English. In contrast, ten KOSs provide research topics in multiple languages, reflecting their intended applications and jurisdictional requirements. For instance, \kos{EuroVoc}, developed by the Publications Office of the European Union, supports 23 languages, enabling interoperability across European digital libraries. Table~\ref{tab:languages_kos} lists the 10 multilingual KOSs, detailing the number of languages they support and other relevant characteristics to provide a comprehensive overview of their features.


The 10 multilingual KOSs exhibit substantial variability in the number of supported languages. Some provide broad language support, such as the \kos{Science Metrix Classification} (26 languages) and the \kos{Agrovoc Thesaurus} (42 languages). Notably, the  \kos{Art and Architecture Thesaurus} demonstrates exceptional linguistic inclusivity with its impressive coverage of 167 languages. In contrast, other KOSs support only a few languages, such as the \kos{STW Thesaurus for Economics} (German and English), the \kos{Unesco Thesaurus} (5 languages), and the \kos{Fields of research and development} (6 languages).

Furthermore, KOSs often vary in the number of concepts available in different languages. Typically, non-English versions only provide a partial representation of the domain described by the English version.
For instance, as shown in Figure~\ref{fig:agrovoc-distro}, \kos{Agrovoc} provides good coverage in English, French, Turkish, Spanish, Arabic, and a few other languages. However, the number of available concepts is significantly limited in languages such as Estonian, Burmese, Khmer, and Greek. Naturally, achieving uniform coverage poses greater challenges for KOSs with both a high volume of concepts and extensive multilingual support. 
Only a few smaller KOSs, such as \kos{Eurovoc}, \kos{Unesco Thesaurus}, \kos{Fields of Research and Development}, \kos{Science Metrix Classification}, and \kos{STW Thesaurus for Economics}, have managed to maintain consistent representation across all supported languages, as reported by column `Uniform' in Table~\ref{tab:languages_kos}. Notably, \kos{The Soz} provides near-uniform coverage in English, German, and French, while its Russian counterpart remains under development.

Finally, an analysis of the 10 multilingual KOSs reveals that the majority (6) are ontologies, while 3 are taxonomies, and 1 is a thesaurus. This trend likely reflects the inherent capacity of ontologies to effectively manage concept labels across multiple languages~\citep{montiel2011enriching}.

\begin{table}[!h]
\centering
\scriptsize
\rowcolors{2}{white}{orange!20}
\caption{\label{tab:languages_kos}
This table presents a comparison of 10 multilingual KOSs, detailing the number of languages they cover, whether language coverage is uniform, their scope coverage (Single-field or Multi-field), type of system (\textbf{Sy} = {\textbf{O}ntology, \textbf{T}axonomy, and Thesa\textbf{U}rus}), number of concepts (\#Concepts), and depth.}
\begin{tabular}{lrclcrr}
\toprule
\textbf{Knowledge Organization System}             & \textbf{Languages} & \textbf{Uniform} & \textbf{Scope Coverage} & \textbf{Sy} & \textbf{\#Concepts} & \textbf{Depth} \\
\midrule
Art and Architecture Thesaurus     & 167       &         & Single-field    & O              & 58K         & 13    \\
Agrovoc Thesaurus                  & 42        &         & Single-field    & O              & 41K         & 14    \\
Dewey Decimal Classification       & 30        &         & Multi-field     & T              & 60K         & 14    \\
Unified Medical Language System    & 28        &         & Single-field    & U              & 3.3M        & 30    \\
Science Metrix Classification      & 26        & Yes       & Multi-field     & T              & 199         & 3     \\
EuroVoc                            & 23        & Yes       & Multi-field     & O              & 7439        & 6     \\
Fields of research and development & 6         & Yes       & Multi-field     & T              & 48          & 2     \\
Unesco Thesaurus                   & 5         & Yes       & Multi-field     & O              & 4482        & 6     \\
TheSoz                             & 4         &         & Single-field    & O              & 8223        & 6     \\
STW Thesaurus for Economics        & 2         & Yes       & Single-field    & O              & 14K         & 13   \\
\bottomrule
\end{tabular}
\end{table}

\begin{figure*}[!h]
    \centering
    \includesvg[width=\linewidth]{figures/agrovoc-lang-distro4.svg}
    \caption{Distribution of terms per language in the Agrovoc Thesaurus. Produced with term counts available on \url{https://agrovoc.fao.org/browse/agrovoc} (Agrovoc v.2024-04, Apr. 2024).}
    \label{fig:agrovoc-distro}
\end{figure*}

\subsubsection{Providers and Maintainers}

We categorised the providers of all knowledge organization systems into five distinct groups: 
\begin{inlist}
    \item Publishers \& Information Service Providers,	
    \item Funding Bodies \& Government Agencies \& Research Councils \& Policy Makers,
    \item Research Institutes \& Universities,	
    \item National Libraries, and 
    \item Open Initiatives \& Consortia.
\end{inlist}

Publishers \& Information Service Providers represent the most prevalent category with 16 KOSs, including notable members such as Elsevier, Springer Nature, the Institute of Electrical and Electronics Engineers (IEEE), the Association for Computing Machinery (ACM), Clarivate, and the American Psychological Association (APA). Indeed, KOSs are essential for publishers as they need to organise, manage, and deliver content effectively in order to enhance discoverability, accessibility, and user engagement~\citep {salatino2019improving}.

The second-largest group comprises Funding Bodies, Government Agencies, Research Councils \& Policy Makers. This category includes 12 organisations such as the European Commission, UNESCO, the Food and Agriculture Organization (FAO), the Australian Research Council, the New Zealand Ministry of Business, and the Organisation for Economic Co-operation and Development (OECD). In this context, KOSs are crucial for streamlining operations, promoting transparency, and enhancing decision-making.

Research Institutes and Universities, including The Open University (UK), University of Sheffield (UK), University of Illinois at Urbana-Champaign (USA), University of Alberta (CA), Athena Research Center (GR), and The Getty Research Institute (USA) 
produced 8 KOSs. 

National Libraries, including the US National Library of Medicine, the Library of Congress, the German National Library of Economics (ZBW), and the German National Library of Science and Technology (TIB) provided 7 KOSs.

Finally, only 2 KOSs are supported by Open Initiatives and Consortia: the Open Biological and Biomedical Ontology (OBO) Foundry (\kos{Open Biological and Biomedical Ontology}) and FAIRsharing (\kos{Subject Resource Application Ontology}).


Our analysis showed that Funding Bodies \& Government Agencies \& Research Councils \& Policy Makers primarily developed multi-field KOSs (9 out of 12), likely due to their interdisciplinary focus. In contrast, Research Institutes \& Universities predominantly created single-field KOSs (6 out of 8), possibly because research teams tend to specialise in a specific area. The remaining organisational categories demonstrated a balanced approach in developing both single and multi-field KOSs.

\subsection{External Links}

Nineteen KOSs provide links to external knowledge bases. This is typically done by defining mappings that indicate that two entities in different representations refer to the same real-world object or concept. They typically rely on well-know semantic relationships such as \texttt{owl:same\_as}, \texttt{skos:exactMatch}, and \texttt{skos:closeMatch}.
As an example, the concept \topic{sunflowers} available in \kos{Agrovoc Thesarus}\footnote{Sunflowers (plural) in Agrovoc Thesarus --- \url{http://aims.fao.org/aos/agrovoc/c_aad037e4}} has a \texttt{skos:closeMatch} with \topic{sunflower} in \kos{Eurovoc}\footnote{Sunflower (singular) in Eurovoc --- \url{http://eurovoc.europa.eu/4472}}.



The KOSs under analysis typically connect either to general knowledge graphs~\citep{peng2023knowledge}, such as Wikidata\footnote{Wikidata --- \url{https://www.wikidata.org}} or DBpedia\footnote{DBpedia --- \url{https://www.dbpedia.org/}}, or to other KOSs.
Such interconnections are generally beneficial, as they create a network of related resources, providing diverse perspectives on research topics and facilitating the creation of novel resources.

Wikidata stands out as the most externally linked knowledge graph since it is reached by concepts from \kos{Agrovoc Thesaurus}, \kos{Computer Science Ontology}, \kos{STW Thesaurus for Economics}, \kos{EuroVoc}, \kos{Open Biological and Biomedical Ontology}, and the \kos{Library of Congress Class. (and Subj. Head.)}. Conversely, \kos{Agrovoc Thesaurus}, \kos{ChemOnt}, \kos{STW Thesaurus for Economics}, \kos{EuroVoc} are the KOSs with the highest number of externally connected knowledge bases.
For instance, the \kos{AGROVOC Thesaurus} is highly interconnected, linking its concepts to \kos{EuroVoc}, Wikidata, DBpedia, the \kos{UNESCO Thesaurus}, and the \kos{Library of Congress Subject Headings (LCSH)}. Similarly, \kos{EuroVoc} is linked not only with the \kos{AGROVOC Thesaurus}, but also with \kos{TheSoz}, \kos{LCSH}, \kos{UNESCO}, \kos{STW}, \kos{MeSH}, Wikidata, and several other knowledge organization systems.

Thirteen out of the initial nineteen earn a five-star rating within the Linked Open Data deployment scheme proposed by Sir Tim Berners-Lee in 2010\footnote{Linked Data --- \url{https://www.w3.org/DesignIssues/LinkedData.html}}. Their extensive interconnections, facilitated by RDF standards such as OWL and SKOS, establish a rich network of information with related KOSs. Adhering to the 5-star scheme\footnote{Linked Open Data 5 Star --- \url{https://www.w3.org/DesignIssues/LinkedData.html}} optimises discoverability, reusability, and the potential for integration into the broader knowledge web, fostering innovative applications and the development of more comprehensive knowledge organization systems.






\subsection{Usage}
A KOS can be widely adopted across various applications, such as organising digital libraries, enhancing metadata, and ensuring interoperability between different data systems. For example, the \kos{Fields of Research (ANZSRC)} is utilised by Dimensions.ai to organise research metadata, by Figshare to manage research repositories, and by the Australian and New Zealand governments to measure and analyse research and experimental development.





Our analysis showed that KOSs are being employed to organise four main categories of resources:
\begin{inlist}
    \item \textit{digital libraries}, which contain research articles, policy documents, grant proposals, patents, and other kind of digital documents,
    \item \textit{research repositories}, which contain research data, code, research protocols, models and any other kind of research artefacts,
    \item \textit{bibliographic databases}, which organise metadata about publications and research artefacts,
    \item \textit{physical libraries}, which contain printed books and periodicals, as well as other media.
\end{inlist}

Twenty-nine KOSs are currently being employed to organise digital libraries including ACM Digital Library, IEEE Digital Library, MEDLINE/PubMed, Scopus, Nature, EconLit, Physical Review, and Mathematical Reviews. 
Ten KOSs are employed to organise research outputs like \kos{Modern Science Ontology} for the Open Research Knowledge Graph, \kos{ChemOnt} for DrugBank, T3DB, ChEBI, LIPID MAPS, \kos{TheSoz} for the Social Science Research Project Information System in Germany, \kos{EDAM} for bio.tools and Training eSupport System (TeSS), \kos{OpenAIRE's Field of Science Taxonomy} for OpenAIRE, and \kos{Subject Resource Application Ontology} for FAIRsharing. 
Ten KOSs are employed to organise bibliographic databases including Web of Science, Dimension.ai, Microsoft Academic Graph, APA PsycInfo database, the International System for Agricultural Science and Technology, and GeoRef database. Finally, the \kos{National Library of Medicine classification}, \kos{Dewey Decimal Classification}, and \kos{Library of Congress Classification} are mainly being employed for physical libraries.

Various KOSs are also utilised across a wide array of initiatives to directly advance research and create additional knowledge and tools. For instance, the \kos{Computer Science Ontology} has been employed for exploring and analysing scholarly data through platforms like Rexplore~\citep{osborne2013}, ScholarLensViz~\citep{ScholarLensViz2020}, ConceptScope~\citep{zhang2021conceptscope}, and VeTo~\citep{vergoulis2020veto}. It has also been instrumental in generating knowledge graphs, such as AIDA KG~\citep{angioni2022} and CS KG~\citep{dessi2022cskg}, as well as in recommending video lessons~\citep{borges2019semantic}. However, thoroughly analysing these specific applications would require a dedicated survey, as exemplified by Jing's usage analysis of the \kos{Unified Medical Language System}~\citep{xia2021}. 

\color{black}

%% file: tables/bigtable.tex
\begin{table*}[!h]
\caption{\label{tab:matrix}An overview on the 45 knowledge organization systems, reporting their main discipline, number of concepts, depth, type of system (\textbf{Sy} = {\textbf{O}ntology, \textbf{T}axonomy, Thesa\textbf{U}rus, and Term \textbf{L}ist}), kind of hierarchy (\textbf{Hr} = {\textbf{P}oly-hierarchical or \textbf{M}ono-hierarchical}), and related terms (\textbf{RT} = {\textbf{Y}es or \textbf{N}o}).}
\scriptsize
\rowcolors{2}{white}{orange!20}
\centering
\begin{tabular}{@{}L{5.5cm}L{1.9cm}rrrrr@{}}
\toprule
\textbf{Knowledge Organization System}                       & \textbf{Main Discipline} & \textbf{\#Concepts} & \textbf{Depth} & \textbf{Sy} & \textbf{Hr} & \textbf{RT}                          \\ \midrule
Agrovoc Thesaurus                                    & Agriculture              & 41K                 & 14             & O           & P           & Y           \\
Art and Architecture Thesaurus                       & Art \& Architecture      & 58K                 & 13             & O           & P           & Y           \\
EDAM                                                 & Bioinformatics           & 264                 & 7              & O           & P           & Y           \\
Open Biological and Biomedical Ontology              & Biology                  & 5.3M                & 39             & O           & P           & Y           \\
Biomedical Ontologies from BioPortal                 & Biomedicine              & 13M                 & -              & O           & P           & Y           \\
ChemOnt                                              & Chemistry                & 4825                & 11             & O           & M           & Y           \\
ACM Computing Classification Scheme                  & Computer Science         & 2114                & 6              & T           & P           & N           \\
Computer Science Ontology                            & Computer Science         & 14K                 & 13             & O           & P           & Y           \\
Computer Science Subject Headings from Wikipedia     & Computer Science         & 7354                & 20             & T           & M           & N           \\
Journal of Economic Literature                       & Economics                & 859                 & 3              & T           & M           & N           \\
STW Thesaurus for Economics                          & Economics                & 14K                 & 13             & O           & P           & Y           \\
IEEE Thesaurus                                       & Engineering              & 12K                 & 3              & U           & P           & Y           \\
GeoRef Thesaurus                                     & Geology                  & 33K                 & 5              & U           & M           & Y           \\
U.S. Geological Survey Library Classification System & Geology                  & 2k                  & 6              & T           & M           & N           \\
Mathematics Subject Classification                   & Mathematics              & 6598                & 3              & T           & M           & Y           \\
Medical Subject Headings                             & Medicine                 & 30K                 & 13             & O           & P           & Y           \\
National Library of Medicine classification          & Medicine                 & 4761                & 4              & T           & M           & N           \\
Unified Medical Language System                      & Medicine                 & 3.3M                & 30             & U           & P           & Y           \\
PhilPapers Taxonomy                                  & Philosophy               & 7447                & 5              & T           & P           & N           \\
Physics Subject Headings                            & Physics                  & 3749                & 6              & O           & P           & Y           \\
Physics and Astronomy Classif. Scheme                & Physics                  & 8260                & 5              & T           & M           & N           \\
PsycInfo and PsycTests Classification Systems        & Psychology               & 188                 & 3              & T           & M           & N           \\
TheSoz                                               & Social Science           & 8223                & 6              & O           & P           & Y           \\ \hline
All Science Journal Classification Codes             & Multi-field              & 334                 & 2              & T           & M           & N           \\
ArXiv Subjects                                       & Multi-field              & 176                 & 3              & T           & M           & N           \\
Dewey Decimal Classification                         & Multi-field              & 60K                 & 14             & T           & P           & Y           \\
DFG Classification                                   & Multi-field              & 278                 & 4              & T           & M           & N           \\
European Commission Taxonomy                         & Multi-field              & 629                 & 3              & T           & M           & N           \\
European Research Council Taxonomy                   & Multi-field              & 431                 & 3              & T           & M           & N           \\
EuroVoc                                              & Multi-field              & 7439                & 6              & O           & P           & Y           \\
Fields of Research (ANZSRC)                          & Multi-field              & 4406                & 3              & T           & M           & N           \\
Fields of research and development (OECD)            & Multi-field              & 48                  & 2              & T           & M           & N           \\
KNOWMAK                                              & Multi-field              & 72                  & 3              & O           & P           & Y           \\
Library of Congress Class. (and Subj. Head.)         & Multi-field              & 467K                & 29             & O           & P           & Y           \\
Microsoft's Fields of Study                          & Multi-field              & 704K                & 6              & T           & P           & N           \\
Modern Science Ontology                              & Multi-field              & 369                 & 6              & O           & P           & N           \\
Nature Subjects                                      & Multi-field              & 2852                & 8              & O           & P           & N           \\
OpenAIRE's Field of Science Taxonomy                 & Multi-field              & 50K                 & 6              & T           & P           & N           \\
OpenAlex Topics                                      & Multi-field              & 4798                & 4              & T           & M           & N           \\
Research Fields, Courses and Disciplines (ASRC)      & Multi-field              & 1061                & 3              & T           & M           & N           \\
Science Metrix Classification                        & Multi-field              & 199                 & 3              & T           & M           & N           \\
Socio-Economic Objective (ANZSRC)                    & Multi-field              & 1974                & 3              & T           & M           & N           \\
Subject Resource Application Ontology                & Multi-field              & 435                 & 8              & O           & P           & Y           \\
UNESCO Thesaurus                                     & Multi-field              & 4482                & 6              & O           & P           & Y           \\
Web of Science Categories                            & Multi-field              & 254                 & 1              & L           & -           & N           \\
\bottomrule
\end{tabular}
\end{table*}

%% file: tables/topic-distro-table.tex

\begin{table*}[!t]\centering
\caption{Coverage of the multi-field KOSs. 
In \textcolor[HTML]{6aa84f}{$\blacksquare$ green} the main fields, in \textcolor[HTML]{3d85c6}{$\blacksquare$ blue} the minor ones, only partially represented, and in \textcolor[HTML]{e69138}{$\blacksquare$ orange} the fields that are just mentioned. In \textbf{bold} are the KOSs that consistently cover all research fields.
}\label{tab:topic-spread}
\scriptsize
\begin{adjustwidth}{0cm}{}
\begin{tabular}{rlllllllllllllllllll}
\textbf{Multi-field Knowledge Organization Systems}&\Rot{Art} & \Rot{Biology} & \Rot{Business} & \Rot{Chemistry} & \Rot{Computer Science} & \Rot{Economics} & \Rot{Engineering} & \Rot{Environmental Science} & \Rot{Geography} & \Rot{Geology} & \Rot{History} & \Rot{Material Science} & \Rot{Mathematics} & \Rot{Medicine} & \Rot{Philosophy} & \Rot{Physics} & \Rot{Political Science} & \Rot{Psychology} & \Rot{Sociology}
\\
\midrule
\textbf{All Science Journal Classification Codes}        & \cellcolor[HTML]{6AA84F} & \cellcolor[HTML]{6AA84F} & \cellcolor[HTML]{6AA84F} & \cellcolor[HTML]{6AA84F} & \cellcolor[HTML]{6AA84F} & \cellcolor[HTML]{6AA84F} & \cellcolor[HTML]{6AA84F} & \cellcolor[HTML]{6AA84F} & \cellcolor[HTML]{6AA84F} & \cellcolor[HTML]{6AA84F} & \cellcolor[HTML]{6AA84F} & \cellcolor[HTML]{6AA84F} & \cellcolor[HTML]{6AA84F} & \cellcolor[HTML]{6AA84F} & \cellcolor[HTML]{6AA84F} & \cellcolor[HTML]{6AA84F} & \cellcolor[HTML]{6AA84F} & \cellcolor[HTML]{6AA84F} & \cellcolor[HTML]{6AA84F} \\
ArXiv Subjects                                  &                          & \cellcolor[HTML]{3D85C6} &                          &                          & \cellcolor[HTML]{6AA84F} & \cellcolor[HTML]{6AA84F} & \cellcolor[HTML]{3D85C6} &                          &                          &                          &                          &                          & \cellcolor[HTML]{6AA84F} &                          &                          & \cellcolor[HTML]{6AA84F} &                          &                          &                          \\
Dewey Decimal Classification                    & \cellcolor[HTML]{6AA84F} & \cellcolor[HTML]{6AA84F} & \cellcolor[HTML]{6AA84F} & \cellcolor[HTML]{6AA84F} & \cellcolor[HTML]{6AA84F} & \cellcolor[HTML]{6AA84F} & \cellcolor[HTML]{6AA84F} & \cellcolor[HTML]{6AA84F} & \cellcolor[HTML]{6AA84F} & \cellcolor[HTML]{6AA84F} & \cellcolor[HTML]{6AA84F} & \cellcolor[HTML]{3D85C6} & \cellcolor[HTML]{6AA84F} & \cellcolor[HTML]{6AA84F} & \cellcolor[HTML]{6AA84F} & \cellcolor[HTML]{6AA84F} & \cellcolor[HTML]{6AA84F} & \cellcolor[HTML]{6AA84F} & \cellcolor[HTML]{6AA84F} \\
DFG Classification                              & \cellcolor[HTML]{6AA84F} & \cellcolor[HTML]{6AA84F} & \cellcolor[HTML]{E69138} & \cellcolor[HTML]{6AA84F} & \cellcolor[HTML]{3D85C6} & \cellcolor[HTML]{3D85C6} & \cellcolor[HTML]{6AA84F} &                          & \cellcolor[HTML]{E69138} & \cellcolor[HTML]{6AA84F} & \cellcolor[HTML]{3D85C6} & \cellcolor[HTML]{3D85C6} & \cellcolor[HTML]{3D85C6} & \cellcolor[HTML]{6AA84F} & \cellcolor[HTML]{3D85C6} & \cellcolor[HTML]{6AA84F} & \cellcolor[HTML]{E69138} & \cellcolor[HTML]{3D85C6} & \cellcolor[HTML]{3D85C6} \\
European Commission Taxonomy                    & \cellcolor[HTML]{6AA84F} & \cellcolor[HTML]{6AA84F} & \cellcolor[HTML]{6AA84F} & \cellcolor[HTML]{6AA84F} & \cellcolor[HTML]{6AA84F} & \cellcolor[HTML]{6AA84F} & \cellcolor[HTML]{6AA84F} & \cellcolor[HTML]{6AA84F} & \cellcolor[HTML]{3D85C6} & \cellcolor[HTML]{6AA84F} & \cellcolor[HTML]{6AA84F} & \cellcolor[HTML]{3D85C6} & \cellcolor[HTML]{6AA84F} & \cellcolor[HTML]{6AA84F} & \cellcolor[HTML]{6AA84F} & \cellcolor[HTML]{6AA84F} & \cellcolor[HTML]{6AA84F} & \cellcolor[HTML]{6AA84F} & \cellcolor[HTML]{6AA84F} \\
European Research Council Taxonomy              & \cellcolor[HTML]{6AA84F} & \cellcolor[HTML]{6AA84F} & \cellcolor[HTML]{3D85C6} & \cellcolor[HTML]{6AA84F} & \cellcolor[HTML]{6AA84F} & \cellcolor[HTML]{6AA84F} & \cellcolor[HTML]{6AA84F} & \cellcolor[HTML]{6AA84F} & \cellcolor[HTML]{6AA84F} & \cellcolor[HTML]{6AA84F} & \cellcolor[HTML]{6AA84F} & \cellcolor[HTML]{6AA84F} & \cellcolor[HTML]{6AA84F} & \cellcolor[HTML]{6AA84F} & \cellcolor[HTML]{6AA84F} & \cellcolor[HTML]{6AA84F} & \cellcolor[HTML]{6AA84F} & \cellcolor[HTML]{6AA84F} & \cellcolor[HTML]{6AA84F} \\
EuroVoc                                         & \cellcolor[HTML]{6AA84F} & \cellcolor[HTML]{6AA84F} & \cellcolor[HTML]{6AA84F} & \cellcolor[HTML]{6AA84F} & \cellcolor[HTML]{6AA84F} & \cellcolor[HTML]{6AA84F} & \cellcolor[HTML]{6AA84F} & \cellcolor[HTML]{6AA84F} & \cellcolor[HTML]{6AA84F} & \cellcolor[HTML]{6AA84F} & \cellcolor[HTML]{6AA84F} & \cellcolor[HTML]{6AA84F} & \cellcolor[HTML]{3D85C6} & \cellcolor[HTML]{6AA84F} & \cellcolor[HTML]{3D85C6} & \cellcolor[HTML]{3D85C6} & \cellcolor[HTML]{6AA84F} & \cellcolor[HTML]{6AA84F} & \cellcolor[HTML]{6AA84F} \\
\textbf{Fields of Research (ANZSRC) }                    & \cellcolor[HTML]{6AA84F} & \cellcolor[HTML]{6AA84F} & \cellcolor[HTML]{6AA84F} & \cellcolor[HTML]{6AA84F} & \cellcolor[HTML]{6AA84F} & \cellcolor[HTML]{6AA84F} & \cellcolor[HTML]{6AA84F} & \cellcolor[HTML]{6AA84F} & \cellcolor[HTML]{6AA84F} & \cellcolor[HTML]{6AA84F} & \cellcolor[HTML]{6AA84F} & \cellcolor[HTML]{6AA84F} & \cellcolor[HTML]{6AA84F} & \cellcolor[HTML]{6AA84F} & \cellcolor[HTML]{6AA84F} & \cellcolor[HTML]{6AA84F} & \cellcolor[HTML]{6AA84F} & \cellcolor[HTML]{6AA84F} & \cellcolor[HTML]{6AA84F} \\
Fields of research and development              & \cellcolor[HTML]{6AA84F} & \cellcolor[HTML]{6AA84F} & \cellcolor[HTML]{6AA84F} & \cellcolor[HTML]{6AA84F} & \cellcolor[HTML]{6AA84F} & \cellcolor[HTML]{6AA84F} & \cellcolor[HTML]{6AA84F} & \cellcolor[HTML]{6AA84F} & \cellcolor[HTML]{3D85C6} & \cellcolor[HTML]{6AA84F} & \cellcolor[HTML]{3D85C6} & \cellcolor[HTML]{6AA84F} & \cellcolor[HTML]{6AA84F} & \cellcolor[HTML]{6AA84F} & \cellcolor[HTML]{6AA84F} & \cellcolor[HTML]{6AA84F} & \cellcolor[HTML]{6AA84F} & \cellcolor[HTML]{6AA84F} & \cellcolor[HTML]{6AA84F} \\
KNOWMAK                                         &                          & \cellcolor[HTML]{6AA84F} & \cellcolor[HTML]{3D85C6} &                          & \cellcolor[HTML]{3D85C6} &                          & \cellcolor[HTML]{6AA84F} & \cellcolor[HTML]{6AA84F} &                          &                          & \cellcolor[HTML]{3D85C6} & \cellcolor[HTML]{6AA84F} &                          &                          &                          &                          & \cellcolor[HTML]{3D85C6} &                          & \cellcolor[HTML]{6AA84F} \\
Library of Congress Class. (and Subj. Head.)    & \cellcolor[HTML]{6AA84F} & \cellcolor[HTML]{6AA84F} & \cellcolor[HTML]{6AA84F} & \cellcolor[HTML]{6AA84F} & \cellcolor[HTML]{6AA84F} & \cellcolor[HTML]{6AA84F} & \cellcolor[HTML]{6AA84F} & \cellcolor[HTML]{6AA84F} & \cellcolor[HTML]{6AA84F} & \cellcolor[HTML]{6AA84F} & \cellcolor[HTML]{6AA84F} & \cellcolor[HTML]{3D85C6} & \cellcolor[HTML]{6AA84F} & \cellcolor[HTML]{6AA84F} & \cellcolor[HTML]{6AA84F} & \cellcolor[HTML]{6AA84F} & \cellcolor[HTML]{6AA84F} & \cellcolor[HTML]{6AA84F} & \cellcolor[HTML]{6AA84F} \\
\textbf{Microsoft's Fields of Study }                    & \cellcolor[HTML]{6AA84F} & \cellcolor[HTML]{6AA84F} & \cellcolor[HTML]{6AA84F} & \cellcolor[HTML]{6AA84F} & \cellcolor[HTML]{6AA84F} & \cellcolor[HTML]{6AA84F} & \cellcolor[HTML]{6AA84F} & \cellcolor[HTML]{6AA84F} & \cellcolor[HTML]{6AA84F} & \cellcolor[HTML]{6AA84F} & \cellcolor[HTML]{6AA84F} & \cellcolor[HTML]{6AA84F} & \cellcolor[HTML]{6AA84F} & \cellcolor[HTML]{6AA84F} & \cellcolor[HTML]{6AA84F} & \cellcolor[HTML]{6AA84F} & \cellcolor[HTML]{6AA84F} & \cellcolor[HTML]{6AA84F} & \cellcolor[HTML]{6AA84F} \\
Modern Science Ontology                         &                          & \cellcolor[HTML]{6AA84F} &                          & \cellcolor[HTML]{6AA84F} & \cellcolor[HTML]{6AA84F} & \cellcolor[HTML]{3D85C6} & \cellcolor[HTML]{3D85C6} & \cellcolor[HTML]{3D85C6} & \cellcolor[HTML]{E69138} & \cellcolor[HTML]{6AA84F} &                          &                          & \cellcolor[HTML]{6AA84F} & \cellcolor[HTML]{E69138} &                          & \cellcolor[HTML]{6AA84F} & \cellcolor[HTML]{6AA84F} & \cellcolor[HTML]{E69138} & \cellcolor[HTML]{6AA84F} \\
Nature Subjects                                 & \cellcolor[HTML]{E69138} & \cellcolor[HTML]{6AA84F} & \cellcolor[HTML]{6AA84F} & \cellcolor[HTML]{6AA84F} & \cellcolor[HTML]{E69138} & \cellcolor[HTML]{6AA84F} & \cellcolor[HTML]{6AA84F} & \cellcolor[HTML]{6AA84F} & \cellcolor[HTML]{6AA84F} & \cellcolor[HTML]{3D85C6} & \cellcolor[HTML]{6AA84F} & \cellcolor[HTML]{6AA84F} & \cellcolor[HTML]{6AA84F} & \cellcolor[HTML]{6AA84F} & \cellcolor[HTML]{6AA84F} & \cellcolor[HTML]{6AA84F} &                          & \cellcolor[HTML]{6AA84F} & \cellcolor[HTML]{6AA84F} \\
OpenAIRE's Field of Science Taxonomy            & \cellcolor[HTML]{3D85C6} & \cellcolor[HTML]{6AA84F} & \cellcolor[HTML]{6AA84F} & \cellcolor[HTML]{6AA84F} & \cellcolor[HTML]{6AA84F} & \cellcolor[HTML]{6AA84F} & \cellcolor[HTML]{6AA84F} & \cellcolor[HTML]{6AA84F} & \cellcolor[HTML]{3D85C6} & \cellcolor[HTML]{3D85C6} & \cellcolor[HTML]{E69138} & \cellcolor[HTML]{6AA84F} & \cellcolor[HTML]{6AA84F} & \cellcolor[HTML]{6AA84F} & \cellcolor[HTML]{3D85C6} & \cellcolor[HTML]{6AA84F} & \cellcolor[HTML]{6AA84F} & \cellcolor[HTML]{6AA84F} & \cellcolor[HTML]{3D85C6} \\
OpenAlex Topics                                 & \cellcolor[HTML]{6AA84F} & \cellcolor[HTML]{6AA84F} & \cellcolor[HTML]{6AA84F} & \cellcolor[HTML]{6AA84F} & \cellcolor[HTML]{6AA84F} & \cellcolor[HTML]{6AA84F} & \cellcolor[HTML]{6AA84F} & \cellcolor[HTML]{6AA84F} & \cellcolor[HTML]{3D85C6} & \cellcolor[HTML]{3D85C6} & \cellcolor[HTML]{E69138} & \cellcolor[HTML]{6AA84F} & \cellcolor[HTML]{6AA84F} & \cellcolor[HTML]{6AA84F} & \cellcolor[HTML]{3D85C6} & \cellcolor[HTML]{6AA84F} & \cellcolor[HTML]{6AA84F} & \cellcolor[HTML]{6AA84F} & \cellcolor[HTML]{6AA84F} \\
\textbf{Research Fields, Courses and Disciplines (ASRC)} & \cellcolor[HTML]{6AA84F} & \cellcolor[HTML]{6AA84F} & \cellcolor[HTML]{6AA84F} & \cellcolor[HTML]{6AA84F} & \cellcolor[HTML]{6AA84F} & \cellcolor[HTML]{6AA84F} & \cellcolor[HTML]{6AA84F} & \cellcolor[HTML]{6AA84F} & \cellcolor[HTML]{6AA84F} & \cellcolor[HTML]{6AA84F} & \cellcolor[HTML]{6AA84F} & \cellcolor[HTML]{6AA84F} & \cellcolor[HTML]{6AA84F} & \cellcolor[HTML]{6AA84F} & \cellcolor[HTML]{6AA84F} & \cellcolor[HTML]{6AA84F} & \cellcolor[HTML]{6AA84F} & \cellcolor[HTML]{6AA84F} & \cellcolor[HTML]{6AA84F} \\
Science Metrix Classification                   & \cellcolor[HTML]{6AA84F} & \cellcolor[HTML]{3D85C6} & \cellcolor[HTML]{3D85C6} & \cellcolor[HTML]{3D85C6} & \cellcolor[HTML]{3D85C6} & \cellcolor[HTML]{3D85C6} & \cellcolor[HTML]{3D85C6} & \cellcolor[HTML]{E69138} & \cellcolor[HTML]{E69138} & \cellcolor[HTML]{E69138} & \cellcolor[HTML]{E69138} & \cellcolor[HTML]{E69138} & \cellcolor[HTML]{3D85C6} & \cellcolor[HTML]{6AA84F} & \cellcolor[HTML]{E69138} & \cellcolor[HTML]{3D85C6} & \cellcolor[HTML]{E69138} & \cellcolor[HTML]{3D85C6} & \cellcolor[HTML]{E69138} \\
Socio-Economic Objective (ANZSRC)               & \cellcolor[HTML]{6AA84F} &                          &                          &                          &                          & \cellcolor[HTML]{6AA84F} &                          & \cellcolor[HTML]{6AA84F} &                          & \cellcolor[HTML]{6AA84F} &                          & \cellcolor[HTML]{6AA84F} &                          & \cellcolor[HTML]{6AA84F} &                          &                          & \cellcolor[HTML]{6AA84F} &                          & \cellcolor[HTML]{6AA84F} \\
Subject Resource Application Ontology           & \cellcolor[HTML]{E69138} & \cellcolor[HTML]{6AA84F} & \cellcolor[HTML]{E69138} & \cellcolor[HTML]{6AA84F} & \cellcolor[HTML]{3D85C6} & \cellcolor[HTML]{6AA84F} & \cellcolor[HTML]{6AA84F} & \cellcolor[HTML]{E69138} & \cellcolor[HTML]{3D85C6} & \cellcolor[HTML]{6AA84F} & \cellcolor[HTML]{6AA84F} & \cellcolor[HTML]{6AA84F} & \cellcolor[HTML]{6AA84F} & \cellcolor[HTML]{6AA84F} & \cellcolor[HTML]{3D85C6} & \cellcolor[HTML]{6AA84F} & \cellcolor[HTML]{E69138} & \cellcolor[HTML]{6AA84F} & \cellcolor[HTML]{3D85C6} \\
Unesco Thesaurus                                & \cellcolor[HTML]{6AA84F} & \cellcolor[HTML]{6AA84F} & \cellcolor[HTML]{6AA84F} & \cellcolor[HTML]{6AA84F} & \cellcolor[HTML]{6AA84F} & \cellcolor[HTML]{6AA84F} & \cellcolor[HTML]{6AA84F} & \cellcolor[HTML]{6AA84F} & \cellcolor[HTML]{6AA84F} & \cellcolor[HTML]{6AA84F} & \cellcolor[HTML]{6AA84F} & \cellcolor[HTML]{E69138} & \cellcolor[HTML]{6AA84F} & \cellcolor[HTML]{6AA84F} & \cellcolor[HTML]{6AA84F} & \cellcolor[HTML]{6AA84F} & \cellcolor[HTML]{6AA84F} & \cellcolor[HTML]{6AA84F} & \cellcolor[HTML]{6AA84F} \\
\textbf{Web of Science Categories }                      & \cellcolor[HTML]{6AA84F} & \cellcolor[HTML]{6AA84F} & \cellcolor[HTML]{6AA84F} & \cellcolor[HTML]{6AA84F} & \cellcolor[HTML]{6AA84F} & \cellcolor[HTML]{6AA84F} & \cellcolor[HTML]{6AA84F} & \cellcolor[HTML]{6AA84F} & \cellcolor[HTML]{6AA84F} & \cellcolor[HTML]{6AA84F} & \cellcolor[HTML]{6AA84F} & \cellcolor[HTML]{6AA84F} & \cellcolor[HTML]{6AA84F} & \cellcolor[HTML]{6AA84F} & \cellcolor[HTML]{6AA84F} & \cellcolor[HTML]{6AA84F} & \cellcolor[HTML]{6AA84F} & \cellcolor[HTML]{6AA84F} & \cellcolor[HTML]{6AA84F} \\
\bottomrule
\end{tabular}
\end{adjustwidth}
\end{table*}

%% file: tables/formats.tex
\begin{table}[!h]
\caption{Analysis of the formats in which the different KOSs are currently released. Single-field KOSs in the upper part and multi-field KOSs in the lower part.}\label{tab:formats}
\centering
\scriptsize
\rowcolors{2}{white}{orange!20}
\begin{tabular}{L{5.5cm}|ccccr}
     \textbf{Knowledge Organization System}                       & \Rot{\textbf{RDF}} & \Rot{\textbf{CSV/Excel}} & \Rot{\textbf{HTML}} & \Rot{\textbf{PDF}} & \textbf{Other formats}     \\ \hline
Agrovoc Thesaurus                                    & x   &     & x    &     &               \\
Art and Architecture Thesaurus                       & x   &     & x    &     &               \\
EDAM                                                 & x   & x   & x    &     & TSV           \\
Open Biological and Biomedical Ontology              & x   &     & x    &     & OBO           \\
Biomedical Ontologies from BioPortal                 & x   & x   &      &     & OBO           \\
ChemOnt                                              &     &     & x    &     & OBO, JSON     \\
ACM Computing Classification Scheme                  & x   &     & x    & x   &               \\
Computer Science Ontology                            & x   &     & x    &     &               \\
Computer Science Subject Headings from Wikipedia     &     & x   &      &     &               \\
Journal of Economic Literature                       &     &     & x    &     & XML           \\
STW Thesaurus for Economics                          & x   &     &      &     &               \\
IEEE Thesaurus                                       &     &     &      & x   &               \\
GeoRef Thesaurus                                     &     &     &      & x   &               \\
U.S. Geological Survey Library Classification System &     &     &      & x   &               \\
Mathematics Subject Classification                   &     &     & x    & x   &               \\
Medical Subject Headings                             & x   &     & x    &     & MARC          \\
National Library of Medicine classification          &     &     & x    & x   &               \\
Unified Medical Language System                      &     &     & x    &     & RRF, ORF, SQL \\
PhilPapers Taxonomy                                  &     &     & x    &     &               \\
Physics Subject Headings                             & x   &     & x    &     &               \\
Physics and Astronomy Classif. Scheme                &     &     & x    &     &               \\
PsycInfo and PsycTests Classification Systems        &     &     & x    & x   &               \\
TheSoz                                               & x   &     &      &     &               \\ \hline
All Science Journal Classification Codes             &     &     & x    &     &               \\
ArXiv Subjects                                       &     &     & x    &     &               \\
Dewey Decimal Classification                         &     &     & x    & x   &               \\
DFG Classification                                   &     &     & x    & x   &               \\
European Commission Taxonomy                         &     &     &      & x   &               \\
European Research Council Taxonomy                   &     &     &      & x   &               \\
EuroVoc                                              & x   & x   & x    &     & MARC          \\
Fields of Research (ANZSRC)                          &     & x   & x    &     &               \\
Fields of research and development                   &     &     &      & x   &               \\
KNOWMAK                                              & x   &     &      &     & JSON          \\
Library of Congress Class. (and Subj. Head.)         & x   &     &      &     & MARC          \\
Microsoft's Fields of Study                          &     &     &      &     & TSV           \\
Modern Science Ontology                              & x   &     &      &     &               \\
Nature Subjects                                      &     & x   &      &     &               \\
OpenAIRE's Field of Science Taxonomy                 &     &     & x    &     & JSON          \\
OpenAlex Topics                                      &     & x   &      &     & JSON          \\
Research Fields, Courses and Disciplines (ASRC)      &     & x   & x    &     &               \\
Science Metrix Classification                        &     & x   &      &     &               \\
Socio-Economic Objective (ANZSRC)                    &     & x   & x    &     &               \\
Subject Resource Application Ontology                & x   &     &      &     &               \\
Unesco Thesaurus                                     & x   &     & x    &     &               \\
Web of Science Categories                            &     &     & x    &     &               \\
\hline
\end{tabular}%
\end{table}

%% file: new_challenges.tex
\section{Challenges and Future Directions}\label{challenges}
In this section, we present the main challenges and future directions of this field.
In Sec.~\ref{comprehensivescheme}, we focus on what we consider the most important challenge in this field: the creation of a comprehensive and granular KOS representing all academic disciplines. 
Sec.~\ref{integration} discusses the most important limitations and challenges regarding the integration of KOSs. 
We then present several other important challenges characterising this space, including: 
\begin{inlist}
\item expanding the coverage of different languages (Sec.~\ref{languagecoverage}), 
\item reconciling expert disagreements (Sec.~\ref{disagreement}), 
\item assessing the quality of KOSs (Sec.~\ref{qualityvsquantity}), 
\item handling ambiguous labels (Sec.~\ref{ambiguouslabels}), 
\item improving the generation and frequency of updates (Sec.~\ref{updatingscheme}), and 
\item developing scalable and accurate approaches for item classification (Sec.~\ref{classification}).
\end{inlist}

\subsection{Towards a comprehensive and granular representation of all academic disciplines}\label{comprehensivescheme}
An important limitation of the KOSs described in this work lies in their high fragmentation.
On one side, we have a good number of multi-field systems that cover multiple academic areas, but tend to be quite shallow and miss many research topics. 
On the other, we have a plethora of single-field KOSs, typically offering a much more granular representation of scholarly knowledge.  
However, there is no KOS that simultaneously is:
\begin{enumerate*}[label=\roman*)]
\item \textbf{comprehensive}, i.e., it encompasses all the main academic fields,
\item \textbf{granular}, i.e., it represents all the specific research areas that are typically used by researchers to refer to their work,
\item \textbf{maintained}, i.e., it is constantly updated to reflect the latest developments, and
\item \textbf{open}, i.e., it can be used freely, without restrictions.
\end{enumerate*}


We argue that such a resource would be transformative in this field and allow: 
\begin{enumerate*}[label=\roman*)]
\item organising content across various digital libraries, thereby enhancing interoperability; \item facilitating the retrieval and analysis of research outputs (e.g., articles, patents, project reports) and agents (e.g., researchers, organisations); \item enabling the monitoring of research, development, and innovation activities; \item ensuring data quality and compatibility across research institutions; and \item supporting evidence-informed policymaking. 
\end{enumerate*}

A potential approach to creating this resource is to adopt one or more multi-field KOSs to represent broad research areas and then integrate several single-field KOSs to effectively cover specific disciplines. As a first step, this section will explore which KOSs could be integrated to develop a comprehensive system. In Section~\ref{integration}, we will discuss current methods for interlinking multiple KOSs.

\paragraph{Multi-field KOSs that may support the creation of a general KOS.}
It is useful to analyse to what extent the current multi-field KOSs fulfil the requirements defined in the previous session. 
Table~\ref{tab:analysistables} presents an evaluation of the 22 multi-field KOSs examined in this study, considering four key features: comprehensiveness, depth, frequency of updates, and openness. 
Each of these categories is characterised using a colour-coded system to indicate performance levels: strong (green), acceptable (orange), or poor (red).
The criteria for assigning these performance levels are as follows.

For the comprehensiveness aspect, colours are assigned based on how well the KOSs cover the various fields, as detailed by Table~\ref{tab:topic-spread}. Specifically, green indicates comprehensive coverage, orange denotes partial coverage, and red signifies that the KOSs cover only a limited set of fields.
In the depth column, the colour coding is as follows: red indicates a depth of 2 or less (low or non-existent), orange represents a depth between 3 and 5 (medium), and green signifies a depth of 6 or higher (high). 
In the frequency of update column, red indicates that the KOS is discontinued with no future updates planned, orange is used for slow updates (i.e., between 5–10 years), and green denotes more frequent updates (i.e., within 5 years). If such information is unavailable, it is marked in white. 
Lastly, in the openness column, red indicates that the KOS is copyrighted and requires a special license for use, orange denotes that the KOS is available online and can be used with certain restrictions, and green signifies a fully open resource with no usage limitations.

\input{tables/definition-table}

Only five KOSs include all 19 broad fields of study: \kos{All Science Journal Classification Codes}, \kos{Fields of Research (ANZSRC)}, \kos{Research Fields, Courses and Disciplines (ASRC)}, \kos{Micorsoft's Fields of Study}, and \kos{Web of Science Categories}. However, none of these KOSs fulfils all the other requirements.
\kos{All Science Journal Classification Codes} and \kos{Web of Science Categories} are very shallow and not open. The \kos{Fields of Research, Research Fields, Courses and Disciplines (ASRC)}, and the \kos{Microsoft's Fields of Study} offer more depth compared to the previous two, but they were discontinued. 
The \kos{Fields of Research (ANZSRC)} is limited in depth, with many specific categories still overly broad, and updates are infrequent, often taking several years to implement. 
The \kos{Dewey Decimal Classification} and \kos{Library of Congress Subject Headings} offer extensive coverage across nearly all areas, along with substantial depth and frequent updates.  
However, their openness remains an issue, as users need to pay subscription fees to use them. 

In conclusion, the current state-of-the-art multi-field and open KOSs do not yet provide all the essential features necessary to serve as a robust foundation for a comprehensive general KOS. 
However, there are a few promising candidates that, if integrated effectively, could provide a solid starting point. In particular, one potential research path could begin with \kos{Eurovoc} as a multi-field KOSs, incorporating missing concepts from specialised single-field KOSs like \kos{Mathematics Subject Classification}, \kos{PhilPapers Taxonomy}, and \kos{Physics Subject Headings}. 

\paragraph{Academic fields that are not currently covered by specific KOSs.}


Several disciplines lack dedicated KOSs and are only superficially addressed by multi-field systems, making it challenging to categorise documents with the necessary precision.
In particular, we were unable to identify any KOS that offers a good characterisation of seven research fields:
\topic{History}, \topic{Political Science}, \topic{Environmental Science}, \topic{Material Science}, \topic{Geography}, \topic{Sociology}, and \topic{Business}. 
In the fields of \topic{Political Science} and \topic{Sociology}, \kos{The Soz} could potentially be utilised. However, \kos{The Soz} mostly focuses on \topic{Social Science} and does not fully cover these two disciplines.
The absence of fine-grained representation for these seven major areas is notable and underscores the need for further development in this space. Automated methods for generating KOSs can offer a valuable solution in this regard, a topic we will further discuss in Section~\ref{updatingscheme}.




\subsection{Integration of multiple KOSs}\label{integration}

In the following, we discuss future directions regarding the integration of multiple KOSs, which may lead to creating a more comprehensive and granular representation of research fields.  

A few of the KOSs that we described in this survey are already interlinked. 
For instance, some concepts in the \kos{Subject Resource Application Ontology} are mapped to concepts within \kos{AgroVoc}, \kos{EDAM}, as well as to a number of ontologies available in \kos{OBO}.
\kos{AgroVoc}, on its turn, has some concepts mapped to the \kos{Unesco Thesaurus}, and the \kos{Library of Congress Subject Headings}. 
However, several KOSs, like the \kos{Mathematical Subject Classification} and the \kos{Physics Subject Headings}, are not (yet) connected to any other KOS. Moreover, the existing mappings between KOSs are often incomplete. The integration of KOSs presents several challenges, which can be categorised into the following sub-challenges:

\begin{enumerate}[label={\textbf{Sub-challenge \arabic*:}},align=left]
\item Generating mappings between KOSs;

\item Adopting standard formats;

\item Developing tools for facilitating the integration of KOSs.
\end{enumerate}

\paragraph{Generating mappings between KOSs.}

Generating links between KOSs is a complex task. In the academic domain, this is usually done by identifying that two subjects from different systems refer to the same concept and linking them with a relation, such as \semrel{owl:sameAs} and \semrel{skos:exactMatch}. This process can be either manual, automatic, or semi-automatic~\citep{kalfoglou2003ontology}.

\textit{Manual approaches} are convenient for small KOSs, but they typically require a lot of effort and high-level expertise and may suffer from scalability issues. In addition, manual integration can lead to the introduction of inconsistencies, especially for large KOSs~\citep{halper2011auditing, erdogan2010exploiting, Solimando2014}.

\textit{Automatic approaches} typically use a combination of similarity metrics, natural language processing, and machine learning~\citep{salatino2020cso,zapilko2013,declerck2013integration}.
For instance, \cite{salatino2020cso} associated research topics to the corresponding DBpedia entity with the DBpedia Spotlight API~\citep{Daiber2013}. 
They fed the tool with artificial sentences listing the labels of the topic and of its direct sub- and super-topics, and then it returned the related DBpedia entities alongside the similarity score.
\cite{zapilko2013} mapped \kos{TheSoz} to \kos{Agrovoc} and DBpedia, using string similarity between terms and retaining the matches that have Levenshtein distance lower than a threshold.
In the domain of KOSs of academic fields, we still typically rely on simple approaches, which mainly use lexical heuristics, and may lead to three potentially unintended consequences~\citep{Shvaiko2008,Solimando2014,Slater2020}. 
First, the mapping might introduce new internal relationships between the entities of one system, and therefore modify inadvertently the description of the domain. 
Instead, the mapping should just enable the interaction across KOSs. 
Second, the automatic integration can introduce logical inconsistencies.
Finally, the mapping might connect entities belonging to different contexts, indicating a potential mapping error~\citep{jimenez2011logic}.
Some recent approaches address these limitations by relying on description logic~\citep{Dhombres2016} or deep learning~\citep{yip2019construction}.
For instance, \cite{Dhombres2016} developed an approach for mapping \kos{Human Phenotype Ontology} (HPO) and the \kos{Standardized Nomenclature of Medicine Clinical Terms} (SNOMED CT) using both lexical and logical approaches. The latter approach consists of using a representation based on description logic to compare the concepts. This method can mitigate the limitations of the lexical approach, however, not all KOSs are developed through description logic.
\cite{yip2019construction} developed a deep learning model with a Siamese recurrent architecture to identify synonyms across \kos{UMLS} concepts. 
The model provides good results; however, it still requires more research and fine-tuning because it presents several false positives (matches together non-synonyms) and false negatives (fails to identify synonyms).
The community still needs to further investigate and develop these new solutions in practical settings.

Finally, in \textit{Semi-automatic approaches}, domain experts analyse, correct, and give feedback on candidate mappings produced by automatic approaches~\citep{salvadores2013bioportal,yip2019construction}.
It is worth mentioning two big endeavours in this space: \kos{Unified Medical Language System} and the \kos{Biomedical Ontologies from BioPortal}.
The Unified Medical Language System is maintained by the US National Library of Medicine and currently integrates more than 200 vocabularies across different languages in the field of Medicine~\citep{Bodenreider2004}.
The US National Library of Medicine is willing to include new additional vocabularies as long as they meet the criteria for inclusion\footnote{Unified Medical Language System inclusion evaluation criteria --- \url{https://www.nlm.nih.gov/research/umls/knowledge_sources/metathesaurus/source_evaluation.html}}. 
These include whether the vocabulary brings new or unique content, is actively maintained, and is available in a machine-readable format.
The \kos{Biomedical Ontologies from BioPortal 
integrates more than 1,100 ontologies in the field of \topic{Biomedicine}.} Differently from \kos{UMLS}, registered users can submit new ontologies to the BioPortal without constraints. 

In conclusion, generating new interconnections between KOSs---either manually or through automatic approaches---is still an open challenge.

\paragraph{Adopting standard formats.}
The format of a KOS has a great impact on our ability to interconnect it with other knowledge bases. 
As shown in Table~\ref{tab:formats}, some KOSs are published in RDF\footnote{Resource Description Framework --- \url{https://www.w3.org/RDF}} (i.e., Resource Description Framework), which is a World Wide Web Consortium\footnote{World Wide Web Consortium --- \url{https://www.w3.org}} (W3C) standard and used for representing highly interconnected data, such as the \kos{Unesco Thesaurus}, and the \kos{Medical Subject Headings}. Some other KOSs are published in CSV, PDF, or browsable through web pages (HTML).

Based on the 5-star deployment scheme for open data\footnote{5-star Open Data --- \url{https://www.w3.org/DesignIssues/LinkedData.html}}, among the possible publishing formats, RDF is the one that fosters better integration. 
A system published according to this standard consists of a set of RDF statements.
Each statement is a three-part structure (also known as triple) which is the smallest irreducible representation for binary relationships, and it is expressed in the form of \triple{subject}{predicate}{object}~\citep{berners2001semantic}. 
For instance, \triple{Social Sciences}{narrower}{Politics} indicates that Social Sciences is a broader area of Politics. 
In this way, two entities (both subject and object) are linked via a predicate or verb. 
In addition, every part of a triple is individually addressable through unique URIs, such as: \triple{\nolinkurl{http://www.nature.com/subjects/social-sciences}}{ \nolinkurl{http://www.w3.org/2004/02/skos/core\#narrower}}{\nolinkurl{http://www.nature.com/subjects/politics}}. Such representation allows AI systems to interconnect, identify, disambiguate, and integrate data effectively.

On the contrary, KOSs published in CSV or HTML are more challenging to treat since the relevant knowledge is not described in a structured format, and relations may be implicit or interpretable only in a wider context. Generating a representation of these KOSs in a standard format such as RDF is not a trivial task.
For example, when working with CSV files, it is essential to understand the structure and data types of the columns (e.g., string, integer, float, date). Tools such as RML~\citep{dimou2014rml} can be used to automate the conversion process from CSV to RDF. However, this process still requires careful consideration of the data's format and structure to ensure accurate transformation.
Similarly, KOSs published in HTML often have arbitrary structures, making it necessary to develop custom parsers to convert them into RDF. 
When KOSs are available only as PDF files, parsing and extracting the information becomes even more challenging. 

In conclusion, the RDF format is adopted by only half (11 out of 23) of the single-field KOSs across the different disciplines, 
as shown in Table~\ref{tab:formats}. Several significant academic fields (e.g.,\topic{Psychology}, \topic{Chemistry}, \topic{Geology}, \topic{Engineering}, \topic{Philosophy}) do not yet rely on standard machine-readable formats, hindering the reuse and integration of their KOSs. In the literature, we can find a few research efforts to RDFy\footnote{\textit{RDFication} refers to the process of converting data into Resource Description Framework (RDF) format.} KOSs. Examples include the RDFification of the \kos{Mathematics Subject Classification}\footnote{MSC2020\_SKOS (TIBHannover) --- \url{https://github.com/TIBHannover/MSC2020\_SKOS}} and \kos{IEEE Thesaurus}\footnote{ieee-taxonomy-thesaurus-rdf (The Open University) --- \url{https://github.com/angelosalatino/ieee-taxonomy-thesaurus-rdf}}.  However, these RDF versions are not maintained by the original curators of the KOSs, making them difficult to update and reuse.

\paragraph{Developing tools for facilitating the integration of KOSs.}
The first KOSs were of relatively small size, such as the \kos{Library Classification for Environmental Science}\footnote{We excluded the Library Classification for Environmental Science from our analysis because it was developed over five decades ago and is no longer actively maintained or used by the scientific community.}, developed by~\cite{plate1966}. 
Typically, their creation would involve sketching ideas on paper or writing down topics on sticky notes, which would then be arranged on a table to create a hierarchical structure~\citep{motta1999reusable}. Today, we have access to a wide range of tools that streamline this process, enabling the creation of more structured and complex KOSs. For instance, the German Centre for Higher Education Research and Science Studies\footnote{German Centre for Higher Education Research and Science Studies (DZHW) --- \url{www.dzhw.eu}} (DZHW) used Trello to build the \kos{Research Core Dataset}\footnote{Research Core Dataset --- \url{https://w3id.org/kdsf-ffk}.  Despite being a KOS of academic disciplines, RCD was not included in our analysis in Sec.~\ref{results} because it is available only in German (see selection criteria in Sec.~\ref{inclusion_criteria}).} (RCD), a classification of interdisciplinary research fields~\citep{stiller2021}. Trello\footnote{Trello --- \url{https://trello.com}} is a web-based project management tool implementing the Kanban system, and it allows multiple users to collaborate on the same board. 
To build \kos{RCD}, the experts created a card for each subject and then arranged them over the board to create the hierarchical structure.
However, building complex KOSs using Trello presents some limitations, since it does not allow users to nest cards on more than two levels. In addition, as Trello is not intended for this task, it does not provide any tool to handle disagreement among the experts due to their different backgrounds.

More advanced tools for building KOSs are Protégé\footnote{Protégé --- \url{https://protege.stanford.edu}}, Semantic MediaWiki\footnote{Semantic MediaWiki --- \url{https://www.semantic-mediawiki.org}}, VocBench3\footnote{VocBench --- \url{https://vocbench.uniroma2.it}}, and PoolParty\footnote{PoolParty --- \url{https://www.poolparty.biz}}.
Protégé is an open-source software developed by Stanford University to support developers in creating reusable ontologies and building knowledge-based systems~\citep{musen2015}. 
Its graphical interface offers a range of functionalities for browsing and editing ontologies. While the original Protégé software lacks native collaborative features, its cloud-based counterpart, WebProtégé, enables multiple users to work simultaneously on ontology development~\citep{tudorache2013}.

Semantic MediaWiki is an open-source extension of MediaWiki, the engine that runs underneath Wikipedia~\citep{vrandecic2009}. Semantic MediaWiki provides a stable, powerful, and scalable environment, enabling users to browse and collaboratively edit ontologies. 
It also provides facilities for rating pages and users.

VocBench3 is an open-source web application that allows users to create, manage, and share ontologies and thesauri~\citep{stellato2020}. It provides a user-friendly interface for editing and managing RDF data, supports collaborative editing and version control, enabling tracking of changes and the ability to revert to prior versions.

PoolParty is a suite that supports the creation and maintenance of taxonomies, ontologies, knowledge graphs, and semantic search applications~\citep{schandl2010}. Similarly to WebProtégé, it allows users to browse and work collaboratively.
Other commercial solutions include Data Harmony\footnote{Data Harmony --- \url{https://www.accessinn.com/data-harmony}} and Synaptica\footnote{Synaptica --- \url{https://www.synaptica.com}} providing similar functionalities. 

In brief, a wide collection of tools facilitating the integration of the different KOSs is available. However, all these tools are general-purpose and built to support a multitude of use cases. 
As a result, users unfamiliar with the relevant technologies and the semantic web might find them quite complex and difficult to learn. 
We still lack user-friendly tools that are able to effectively support the creation and curation of KOSs.

\subsection{Improving the language coverage}\label{languagecoverage}
English is the \textit{de-facto} language of Science nowadays~\citep{sugimoto2018}. Most international conferences are held in English, and the world’s top scientific journals are published in English.
However, some countries, like China, Russia, and Japan, have several dedicated journals and conferences publishing scientific papers in their own language. It is crucial then to have KOSs in other languages as well so as to facilitate interoperability and the cross-language exploration and exploitation of digital artefacts~\citep{zeng2004}. This would also benefit students in non-English speaking countries, allowing them to understand and navigate KOSs without requiring fluency in English.

In Section~\ref{sec:language}, we acknowledged that some KOSs already have partial translations in other languages, such as the \kos{Dewey Decimal Classification}, the \kos{Agrovoc Thesaurus} and the \kos{Unesco Thesaurus}. However, the process is far from being complete. The \kos{Unesco Thesaurus} is available only in five languages (English, Arabic, Russian, French, and Spanish). 
The \kos{Agrovoc Thesaurus} has a very wide variety of languages; however, not all the forty-two languages are equally represented, as reported in Figure~\ref{fig:agrovoc-distro}. Notably, the Food and Agriculture Organization, who coordinates \kos{Agrovoc}, allows international institutions to contribute to \kos{Agrovoc} by authoring translations. This is an interesting solution that may be beneficial for several other KOSs.

Overall, the support for languages different from English is still very poor for most of the KOSs analysed in this survey. We need further work to produce resources able to support multiple languages~\citep{tudhope2006}. In this context, large language models offer a promising solution as they have proved to be highly effective on automatic language translation~\citep{lu2024llamax}.

\subsection{Reconciliating expert disagreements}\label{disagreement}
One of the major challenges appearing when working in collaborative environments is conflict management, and the process of building and integrating KOSs is not immune to this. Indeed, \cite{chilton2013cascade} argue that the main characteristic of taxonomies, and by extension KOSs, is that they are subjective. Experts have different backgrounds, based mainly on the paradigms they inhabit~\citep{kuhn_1962}, and therefore, they often tend to disagree on the properties and the structure of the system. 
Such disagreement also has an impact on the frequency of updates because it requires time to be addressed,  with a consequent delay in the new release.

In the literature, we can find different works studying disagreement among experts in the context of building KOSs.
\cite{gu2007evaluation} audited the semantic types assigned to \kos{UMLS} concepts with the support of four experts. In particular, the experts assessed whether the current semantic types assigned to the concepts were correct. In some cases, the disagreement prevailed even after multiple rounds and was eventually resolved with a subsequent discussion.
This experience indicates that the process of mitigating disagreement between experts is quite challenging and certainly time-consuming. 
\cite{osborne2019reducing} analysed the agreement of experts in characterising articles according to research areas in Software Architecture and highlighted that, in this domain, most of the disagreement was between domain experts of different seniority (e.g., Full Professor vs. PhD students).
\cite{fan2007combining} also explored the issue of disagreement among experts on research topics. They developed a classifier to annotate research documents with concepts derived from UMLS and then asked experts to categorise UMLS concepts into nine different types. While the inter-annotator agreement was relatively high (0.82), their ablation study revealed that experts often disagreed on vague concepts. For instance, the concept of \concept{promotion} was classified both as a behaviour and as a biological function. The authors emphasise that unresolved disagreements can introduce errors into gold standards, which in turn can propagate into downstream applications.



In conclusion, creating and integrating different KOSs is a collaborative effort, leading to experts' disagreement. 
We thus need to develop new tools and methodologies for handling disagreement and reconciling different suggestions.

\subsection{Assessing the quality of KOSs}\label{qualityvsquantity}
A high-quality KOS must be coherent at two different levels: \textit{structural} and \textit{conceptual}~\citep{MORREY2009468,Ayele2012,Raad2015}.

\begin{figure}[!h]
    \centering
      \includegraphics[width=0.6\linewidth]{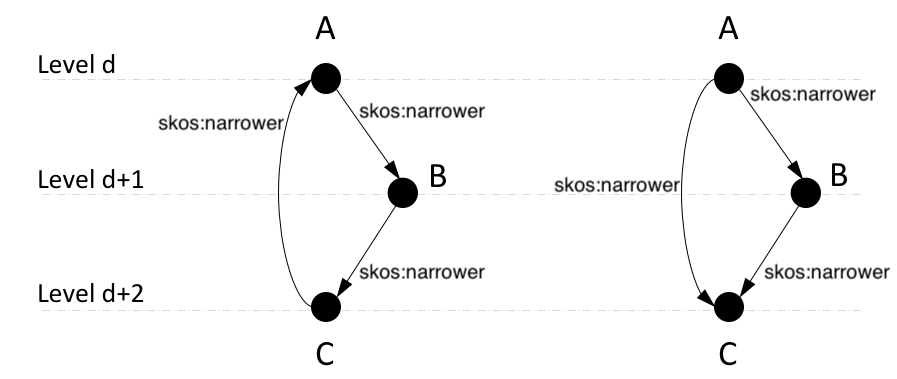}
    \caption{Example of inconsistencies within the KOSs. On the left the cyclic inconsistency and on the right the redundant inconsistency.}
    \label{fig:quality}
\end{figure}

At the structural level, the KOS can present logical inconsistencies, i.e., contradictions hindering the integrity of the system~\citep{Ayele2012}.
Here, we illustrate two of the most frequent cases of logical inconsistencies: cyclic and transitive. Figure~\ref{fig:quality} shows the two examples, and specifically the arrows identify the \semrel{skos:narrower}\footnote{\semrel{skos:narrower} --- \url{https://www.w3.org/TR/skos-reference/\#semantic-relations}} relationship, for instance \concept{A}$\rightarrow$\concept{B}  equals \triple{A}{skos:narrower}{B} meaning \concept{B} is a narrower concept of \concept{A}, and hence \concept{A} is broader than \concept{B}.
The first case, shown on the left, is the \textit{cyclic} inconsistency (also known as \textit{loop}): \concept{A} is broader than \concept{B}, \concept{B} is broader than \concept{C} and \concept{C} is broader than \concept{A}. This is inconsistent because the concept \textit{C} cannot be simultaneously narrower (via \concept{B}) and broader than \concept{A}. 
\cite{mougin2005approaches} highlight that automatically addressing this problem is challenging because identifying the appropriate connection to break is difficult, and performing this operation may introduce additional errors.


The second case of inconsistency is when there is a \textit{transitive hierarchy}\footnote{Transitive hierarchies --- \url{https://www.w3.org/TR/skos-primer/\#sectransitivebroader}}: \concept{A} is broader than \concept{B}, \concept{B} is broader than \concept{C}, but \concept{A} is also broader than \concept{C}. Although, in both cases \concept{A} is broader than \concept{C}, hence it may not be considered as a real mistake, however it adds confusion to the hierarchical structure which necessitates to be addressed. Besides, by convention, \semrel{skos:narrower} and \semrel{skos:broader} must be used to assert a direct or immediate hierarchical link between concepts and are not declared as transitive properties.


These two cases have been simplified just for the sake of clarity, but such inconsistencies might occur across longer chains of concepts~\citep{bodenreider2001circular}. However, ontology reasoners can help to identify these logical inconsistencies~\citep{jimenez2011logic}.

At a conceptual level, several challenges can hinder the quality of a KOS. These are related to the \textit{accuracy}~\citep{lambe2014organising}, i.e., how the concepts are related to each other, such as:
\begin{inlist}
    \item ensuring correct hierarchical relationships between two subjects, and
    \item addressing ambiguities in preserving multiple semantic views.
\end{inlist}
Additional challenges are instead related to the \textit{completeness} of the concepts~\citep{lambe2014organising}, such as: 
\begin{inlist}
    \item whether an area needs to be refined or enriched with sub-areas, 
    \item whether the description of concepts, including definition, scope, and editorial information, is appropriate, and
    \item if all related terms are correct and sufficient. 
\end{inlist}

The literature presents various approaches for assessing the quality of KOSs~\citep {MORREY2009468}. 
For instance, \cite{Raad2015} discusses four categories of ontology evaluation approaches: 
\begin{inlist}
    \item gold standard-based, 
    \item corpus-based, 
    \item task-based, and
    \item criteria-based. 
\end{inlist}
Gold standard-based evaluation approaches compare the KOS with a reference system. Corpus-based evaluation approaches assess the extent of coverage of the system in a given domain. Task-based evaluation approaches measure to what extent the system helps to improve a task. Finally, criteria-based evaluation approaches measure the extent of a system in adhering to certain criteria. 
The wide range of evaluation methods highlights the complexity of assessing the quality of a KOS.
Indeed, for gold standard-based evaluations,~\cite{Raad2015} point out that it is hard to find a suitable gold standard because it needs to be created with the same conditions as the evaluated system. 
Corpus-based evaluation approaches also have similar issues. On the other hand, task-based and criteria-based evaluation approaches have fewer challenges to tackle, but they fall short in assessing KOSs at the conceptual level (i.e., accuracy and completeness). 

In conclusion, when creating or integrating KOSs it is crucial to continuously assess the quality of the resulting knowledge representation~\citep{Shvaiko2008}. The community needs to further advance tools, metrics, and methodologies in this space.

\subsection{Handling ambiguous labels}\label{ambiguouslabels}

Polysemy is a linguistic feature whereby a term has multiple meanings in different contexts. For instance, \topic{Java} is both a programming language and an island in Indonesia. 
Polysemy has a negative effect during the process of building and aligning KOSs because it introduces ambiguity~\citep{johnson2007fault,Djeddi2014}. 

In the literature, we can find several approaches attempting to disambiguate the various senses of a term.
\cite{BELLA20171} mapped \kos{Eurovoc} with \kos{Universal Decimal Classification} (UDC)\footnote{We did not include UDC in our analysis because it has restricted access and requires paying a license.}, and they observed that indeed polysemy induces the appearance of false positives, matching together terms that are syntactically similar but have different semantics. To address this problem, they created a Word Sense Disambiguation (WSD) method. 
Other methods developing WSD techniques are applied to \kos{UMLS}~\citep{widdows2003unsupervised,mcinnes2007using}; however, these methods necessitate the surrounding context of a term and struggle to infer its correct sense when applied directly to concept labels.
In the context of ontologies, \cite{pisanelli2004coping} suggest to formally represent the several meanings of a word, specifying their types. 
For instance, the \topic{Java} entity, in the sense of programming language, can be defined both as a \semrel{skos:Concept} as well as the type that defines it as a programming language. In this way, the two \topic{Java} entities will be properly disambiguated through their specific type. 
\cite{GU200429} audited \kos{UMLS} and found out several polysemic concepts, and in a shared view with \cite{pisanelli2004coping}, they also suggest replacing the polysemous concept with new different concepts according to the different intersecting semantic types. 

In general, disambiguating subjects according to an explicit representation of their meaning is the \textit{de facto} solution to tackle ambiguous labels. 
However, it can also lead to some issues, since the resulting KOSs are more complex~\citep{shi2011comparison}, harder to maintain, and the type assignment may cause disagreements among domain experts~\citep{fan2007combining}.

\subsection{Automatic generation and updating KOSs}\label{updatingscheme}

The current landscape of KOSs reveals gaps in coverage for certain areas (see Sec.~\ref{sec:scope}) and highlights the vulnerability of some KOSs to becoming outdated (see Sec.~\ref{sec:frequency}).
Given the critical role KOSs play in various downstream services, the scientific community should prioritise developing more automatic tools that facilitate their creation, maintenance, and continuous updating. 

The literature presents various automated solutions for creating KOSs. These can either be employed to generate KOSs from scratch, e.g., \citep{osborne2015klink}, or to scan the recent literature and extend existing ones, e.g., \citep{huang2020corel,pisu2024w}.
Klink-2~\citep{osborne2015klink} demonstrated high effectiveness in automatically generating large-scale and granular ontologies of research areas. On a similar note, \cite{shang2020nettaxo} introduced NetTaxo, which is an automated topic taxonomy construction framework that leverages both text data and network structures to build hierarchical topic representations. Whereas, \cite{zhang2018taxogen} presented TaxoGen, an unsupervised method that employs term embeddings and hierarchical clustering to recursively build a topic taxonomy.
On the other hand, \cite{huang2020corel} introduced a technique for building topic taxonomies guided by seed concepts. Given a text corpus and an initial taxonomy of concepts, their approach develops a more comprehensive taxonomy. This approach is potentially valuable in the context of updating existing KOSs. 

In recent years, we witnessed the emergence of new automatic approaches that leverage advanced language models. For instance,~\cite{pisu2024w} introduced an AI-driven pipeline that extracts research concepts from articles and then determines their semantic relationships (hierarchical or synonymous) using SciBERT~\citep{beltagy2019scibertpretrainedlanguagemodel}. Other approaches rely on Large Language Models, which possess a deeper understanding of language, enabling them to identify semantic relationships between concepts more accurately~\citep{Revenko2024,aggarwal2024}. This challenge is attracting growing attention from the community due to the new possibilities offered by AI.

\subsection{Automatic classification of items}\label{classification}

The main objective of KOSs is to categorise a vast amount of items, such as articles, books, courses, patents, software, experiment materials, and so on. Classifying them manually may be unfeasible on a large scale. 
In recent years, various methods have been developed for the automatic classification of these items. The literature typically categorises these approaches into two main types: supervised and unsupervised methods.

\textit{Supervised methods} need to be trained according to a set of labelled samples, each associated with one or more research topics to learn from. The resulting model can be used to automatically classify new items according to the relevant research topics.
For instance, \cite{mai2018} employed deep learning techniques to develop a classifier applied to a set of papers annotated with the \kos{STW Thesaurus for Economics} and \kos{MeSH}. Similarly,~\cite{chernyak2015} presented a classifier in \topic{Computer Science} with topics drawn from the \kos{ACM Computing Classification System}.
\cite{caragea2015} trained their classifiers on a corpus of 3,186 papers distributed over 6 classes (subjects): \textit{agents}, \textit{artificial intelligence}, \textit{information retrieval}, \textit{human computer interaction}, \textit{machine learning}, and \textit{databases}. 
\cite{kandimalla2020} developed a deep attentive neural network for classifying papers according to 104 (out of 254) \kos{Web of Science subject categories}. Their classifier was trained on 9M abstracts from Web of Science.

Recently, the OpenAlex team developed a large deep learning model that leverages the research paper's title, abstract, citations, and journal name to assign classifications drawn from the 4,500 topics within the \kos{OpenAlex Topics}~\citep{openalex2024}. This model demonstrates modest accuracy, with the correct label appearing as the top prediction for 53\% of papers and within the top 10 predictions for 73\% of papers. To date, it represents a potentially unique example of a deep learning model deployed with such a large number of categories. 
Supervised approaches often face two significant limitations. First, they typically struggle with handling a large number of categories, which can lead to underperformance when applied to very large KOSs. Second, they are impractical for KOSs that lack a sufficiently large dataset of labelled items.


\textit{Unsupervised methods} in this space typically aim to associate research areas described in the KOSs to specific text segments using semantic similarity metrics, word embeddings, and a variety of NLP techniques. Their main advantage is that they do not require a training set. 

For instance, the CSO Classifier~\citep{salatino2021} is a tool designed to classify research documents within the field of Computer Science. It processes the textual elements of a research document, such as the title, abstract, and keywords, and outputs the relevant topics based on the Computer Science Ontology.
This classifier consists of two main modules. The first aims at identifying the \kos{CSO} concepts that are explicitly mentioned in the document. The second module leverages word embeddings to infer semantically related topics. Finally, an additional component post-processes the identified topics and removes outliers. 
\cite{soldaini2016quickumls} developed QuickUMLS, a scalable approach for annotating documents with concepts drawn from \kos{UMLS}. This approach takes the document, extracts parts of speech, then preprocesses the text to identify valid tokens and finally matches them against the labels in UMLS. 
It is able to get comparable results with state-of-the-art solutions, however it tends to confuse polysemic terms. 
MetaMap \citep{aronson2010overview} is a similar approach for annotating documents with \kos{UMLS} concepts. MetaMap extracts tokens and their part of speech, then it selects the candidate and matches them against \kos{UMLS} labels. It also implements word sense disambiguation to identify the right sense of the concept in the case of polysemic terms. 
However, it suffers from scalability issues and only identifies concepts whose labels are syntactically available in the text.
\cite{savova2010} developed cTAKES, which is an approach for extracting \kos{UMLS} concepts from medical records. cTAKES implements a named-entity recognition approach, identifying \kos{UMLS} terms within a noun-phrase look-up window. However, cTAKES currently does not resolve ambiguities.

A well-known limitation of unsupervised approaches is that they rely on language models, which typically need to be trained on a specific field. Therefore, they are typically unable to work cross-discipline. 
The emergence of large language models has opened up exciting new possibilities due to their ability to generalise across diverse research areas~\citep{kojima2023largelanguagemodelszeroshot}. However, further investigation is necessary before they can be confidently deployed in real-world applications.

In conclusion, only a few fields (e.g., \topic{Medicine}, \topic{Biology}, \topic{Chemistry}, \topic{Computer Science}, \topic{Economics}) currently have access to high-quality classifiers based on a granular representation of the discipline. Furthermore, these classifiers still face several technical limitations. As a result, it is essential to develop new solutions that not only expand the range of disciplines involved, but also improve accuracy and scalability.

%% file: tables/definition-table.tex

\begin{table*}[!h]\centering
\caption{Analysis of KOSs according to their comprehensiveness, depth, frequency of update, and openness. Features of each KOSs are evaluated using a colour-coded system: \textcolor[HTML]{6aa84f}{$\blacksquare$ green} indicates a strong performance, \textcolor[HTML]{e69138}{$\blacksquare$ orange} signifies an acceptable performance, and \textcolor[HTML]{cc0000}{$\blacksquare$ red} reflects poor performance.
A feature is blank when the information is not available.}\label{tab:analysistables}
\scriptsize
\begin{tabular}{rrrrr}
&\Rot{Comprehensiveness}&\Rot{Depth}&\Rot{Frequency of update}&\Rot{Openness}\\
\midrule
All Science Journal Classification Codes        & \cellcolor[HTML]{6AA84F} & \cellcolor[HTML]{CC0000} &                          & \cellcolor[HTML]{CC0000} \\
ArXiv Subjects                                  & \cellcolor[HTML]{CC0000} & \cellcolor[HTML]{E69138} &                          & \cellcolor[HTML]{6AA84F} \\
Dewey Decimal Classification                    & \cellcolor[HTML]{E69138} & \cellcolor[HTML]{6AA84F} & \cellcolor[HTML]{6AA84F} & \cellcolor[HTML]{CC0000} \\
DFG Classification                              & \cellcolor[HTML]{CC0000} & \cellcolor[HTML]{E69138} &                          & \cellcolor[HTML]{6AA84F} \\
European Commission Taxonomy                    & \cellcolor[HTML]{E69138} & \cellcolor[HTML]{E69138} &                          & \cellcolor[HTML]{6AA84F} \\
European Research Council Taxonomy              & \cellcolor[HTML]{E69138} & \cellcolor[HTML]{E69138} &                          & \cellcolor[HTML]{6AA84F} \\
EuroVoc                                         & \cellcolor[HTML]{E69138} & \cellcolor[HTML]{6AA84F} & \cellcolor[HTML]{6AA84F} & \cellcolor[HTML]{6AA84F} \\
Fields of Research (ANZSRC)                     & \cellcolor[HTML]{6AA84F} & \cellcolor[HTML]{E69138} & \cellcolor[HTML]{E69138} & \cellcolor[HTML]{6AA84F} \\
Fields of research and development              & \cellcolor[HTML]{E69138} & \cellcolor[HTML]{CC0000} &                          & \cellcolor[HTML]{CC0000} \\
KNOWMAK                                         & \cellcolor[HTML]{CC0000} & \cellcolor[HTML]{E69138} &                          & \cellcolor[HTML]{E69138} \\
Library of Congress Class. (and Subj. Head.)    & \cellcolor[HTML]{E69138} & \cellcolor[HTML]{6AA84F} & \cellcolor[HTML]{6AA84F} & \cellcolor[HTML]{CC0000} \\
Microsoft's Fields of Study                     & \cellcolor[HTML]{6AA84F} & \cellcolor[HTML]{6AA84F} & \cellcolor[HTML]{CC0000} & \cellcolor[HTML]{6AA84F} \\
Modern Science Ontology                         & \cellcolor[HTML]{CC0000} & \cellcolor[HTML]{6AA84F} &                          & \cellcolor[HTML]{6AA84F} \\
Nature Subjects                                 & \cellcolor[HTML]{E69138} & \cellcolor[HTML]{6AA84F} &                          & \cellcolor[HTML]{CC0000} \\
OpenAIRE's Field of Science Taxonomy            & \cellcolor[HTML]{E69138} & \cellcolor[HTML]{6AA84F} & \cellcolor[HTML]{6AA84F} & \cellcolor[HTML]{6AA84F} \\
OpenAlex Topics                                 & \cellcolor[HTML]{E69138} & \cellcolor[HTML]{E69138} & \cellcolor[HTML]{6AA84F} & \cellcolor[HTML]{6AA84F} \\
Research Fields, Courses and Disciplines (ASRC) & \cellcolor[HTML]{6AA84F} & \cellcolor[HTML]{E69138} & \cellcolor[HTML]{CC0000} & \cellcolor[HTML]{6AA84F} \\
Science Metrix Classification                   & \cellcolor[HTML]{E69138} & \cellcolor[HTML]{E69138} & \cellcolor[HTML]{E69138} & \cellcolor[HTML]{6AA84F} \\
Socio-Economic Objective (ANZSRC)               & \cellcolor[HTML]{CC0000} & \cellcolor[HTML]{E69138} & \cellcolor[HTML]{E69138} & \cellcolor[HTML]{6AA84F} \\
Subject Resource Application Ontology           & \cellcolor[HTML]{E69138} & \cellcolor[HTML]{6AA84F} & \cellcolor[HTML]{6AA84F} & \cellcolor[HTML]{6AA84F} \\
Unesco Thesaurus                                & \cellcolor[HTML]{E69138} & \cellcolor[HTML]{6AA84F} & \cellcolor[HTML]{6AA84F} & \cellcolor[HTML]{E69138} \\
Web of Science Categories                       & \cellcolor[HTML]{6AA84F} & \cellcolor[HTML]{CC0000} &                          & \cellcolor[HTML]{CC0000} \\
\bottomrule
\end{tabular}
\end{table*}

%% file: treats_to_validity.tex
\section{Threats to validity}\label{sec:ttv}

This section examines potential threats to validity of our study. We identify four key areas where the validity of our study could be challenged: internal validity, external validity, construct validity, and conclusion validity, as discussed in~\cite{wohlin2012experimentation}. In the following discussion, we evaluate these potential threats and detail the measures we have taken to minimise their impact.

\paragraph{Internal Validity.} 
Internal validity in surveys pertains to the rigour and accuracy of the adopted methodology. To guarantee the replicability of our survey, we carefully devised a methodologically sound protocol that included systematic and transparent phases for selecting knowledge organisation systems.
The initial protocol was developed by the first author and subsequently reviewed and refined by co-authors to establish consensus before initiating the survey process. We utilised well-known search engines (e.g., Google Scholar, Google) and various sources such as publishers, Wikipedia, and interconnections among the collected KOSs. Additionally, we consulted domain experts in relevant research fields to expand our result set further.

We performed a multi-stage selection process to ensure a rigorous evaluation and minimise selection bias. 
Initially, the first author analysed and selected all tools based on their description. Subsequently, all authors collaboratively conducted a comprehensive review of the shortlisted KOSs. In cases where information was unclear or unavailable, the first author directly contacted the curators of the respective KOSs for clarification.

Although we employed a systematic approach, biases could still arise from subjective decisions made during the application of inclusion and exclusion criteria. To address this, we conducted collaborative reviews of the shortlisted KOSs, which helped to minimise the impact of individual biases on the selection process.

In summary, although another research team replicating this study might find minor differences in the specific KOSs, the rigorous and systematic methodology used, combined with the collaborative nature of the process, strongly supports the internal validity of our results.

\paragraph{External Validity.} 
External validity refers to how well the results of this survey can be generalised and applied to other contexts and areas of study. To maximise the applicability of our findings, we utilised a variety of sources when choosing the KOSs for conducting this analysis.
While we aimed for a comprehensive identification of tools, some relevant KOSs may have been inadvertently excluded due to limitations in the search engines or query terms used. This could occur if KOSs were not adequately described or indexed using appropriate keywords. To address this, we continuously refined our search terms and consulted domain experts to ensure a wider range of potential KOSs were captured. Additionally, we explored the external links of identified KOSs to discover further relevant tools.

With regard to the inclusion and exclusion criteria, we identified two main potential threats to external validity. The first concerns the exclusion of KOSs that do not offer the English version. This was set due to the predominance of English as the language of scientific communication~\citep{sugimoto2018}. Additionally, the absence of English versions would have hindered our ability to conduct in-depth analyses, such as the one presented in Table~\ref{tab:topic-spread}. 
The second threat arises from the exclusion of KOSs that are designed specifically for individual digital libraries and have not been widely adopted by the broader community. This exclusion was necessary to avoid an unmanageable increase in the number of KOS candidates, which would have made the analysis impractical. Additionally, these KOSs are often customised to fit the unique content of their respective libraries, which could result in a skewed portrayal of the broader scientific landscape. 

\paragraph{Construct Validity.} 

Construct validity refers to the extent to which the operational measures used in a study accurately represent the concepts under investigation. In our survey, the primary concern is whether the 15 analysed features comprehensively cover all relevant aspects. 


To address any potential omissions in our analysis, all authors collaboratively and iteratively defined the five feature categories for evaluation (i.e., scope, structure, curation, external links, and usage) and the fifteen specific features.



We recognise that our analysis may not have fully addressed all relevant aspects. For instance, evaluating the quality of each KOS could offer valuable insights. However, there is no universally accepted definition of ``quality'' in the context of KOSs, which could lead to potential bias if a specific definition were adopted. Moreover, such an evaluation would be time-consuming, expensive, and require specialised expertise across various scientific disciplines.

\paragraph{Conclusion Validity.} 

Conclusion validity in surveys refers to how well the conclusions drawn are supported by the evidence and are reproducible. In our analysis, we placed great emphasis on minimising threats to conclusion validity by using a systematic approach to identify relevant KOSs and extract the relevant features.


To ensure precise and reliable data collection, we developed a data extraction form based on the 15 selected features and a protocol for identifying relevant information. 

Each author independently analysed a sample of KOSs using this standardised form and protocol. Furthermore, for KOSs available in machine-readable formats, we created custom Python scripts to extract structural features and external links (available on our GitHub repository: \githubrepo). Our protocol also included cross-checking our analyses to guarantee accuracy and consistency.


A continuous challenge to the validity of our conclusions is the dynamic nature of KOSs. Many of them are updated annually, acquiring new concepts. Consequently, it is anticipated that many KOSs will evolve in the near future. While our findings provide a snapshot of the current landscape, they may not fully capture the ongoing developments in this field.

\color{black}

%% file: main.bbl
\begin{thebibliography}{}

\bibitem [\protect \citeauthoryear {%
Aggarwal%
, Salatino%
, Osborne%
\BCBL {}\ \BBA {} Motta%
}{%
Aggarwal%
\ \protect \BOthers {.}}{%
{\protect \APACyear {2024}}%
}]{%
aggarwal2024}
\APACinsertmetastar {%
aggarwal2024}%
\begin{APACrefauthors}%
Aggarwal, T.%
, Salatino, A.%
, Osborne, F.%
\BCBL {}\ \BBA {} Motta, E.%
\end{APACrefauthors}%
\unskip\
\newblock
\APACrefYearMonthDay{2024}{}{}.
\newblock
{\BBOQ}\APACrefatitle {Identifying Semantic Relationships Between Research
  Topics Using Large Language Models in a Zero-Shot Learning Setting}
  {Identifying semantic relationships between research topics using large
  language models in a zero-shot learning setting}.{\BBCQ}
\newblock
\BIn{} \APACrefbtitle {Scientific Knowledge: Representation, Discovery, and
  Assessment 2024 @ ISWC 2024.} {Scientific knowledge: Representation,
  discovery, and assessment 2024 @ iswc 2024.}
\newblock
\APACaddressPublisher{}{CEUR}.
\newblock
\begin{APACrefURL} \url{https://ceur-ws.org/Vol-3780/paper3.pdf}
  \end{APACrefURL}
\PrintBackRefs{\CurrentBib}

\bibitem [\protect \citeauthoryear {%
Angioni%
, Salatino%
, Osborne%
, Birukou%
\BCBL {}\ \protect \BOthers {.}}{%
Angioni%
, Salatino%
, Osborne%
, Birukou%
\BCBL {}\ \protect \BOthers {.}}{%
{\protect \APACyear {2022}}%
}]{%
angioni2022leveraging}
\APACinsertmetastar {%
angioni2022leveraging}%
\begin{APACrefauthors}%
Angioni, S.%
, Salatino, A.%
, Osborne, F.%
, Birukou, A.%
, Recupero, D\BPBI R.%
\BCBL {}\ \BBA {} Motta, E.%
\end{APACrefauthors}%
\unskip\
\newblock
\APACrefYearMonthDay{2022}{}{}.
\newblock
{\BBOQ}\APACrefatitle {Leveraging knowledge graph technologies to assess
  journals and conferences at springer nature} {Leveraging knowledge graph
  technologies to assess journals and conferences at springer nature}.{\BBCQ}
\newblock
\BIn{} \APACrefbtitle {International Semantic Web Conference} {International
  semantic web conference}\ (\BPGS\ 735--752).
\newblock
\begin{APACrefDOI} \doi{10.1007/978-3-031-19433-7_42} \end{APACrefDOI}
\PrintBackRefs{\CurrentBib}

\bibitem [\protect \citeauthoryear {%
Angioni%
, Salatino%
, Osborne%
, Recupero%
\BCBL {}\ \BBA {} Motta%
}{%
Angioni%
, Salatino%
, Osborne%
, Recupero%
\BCBL {}\ \BBA {} Motta%
}{%
{\protect \APACyear {2022}}%
}]{%
angioni2022}
\APACinsertmetastar {%
angioni2022}%
\begin{APACrefauthors}%
Angioni, S.%
, Salatino, A.%
, Osborne, F.%
, Recupero, D\BPBI R.%
\BCBL {}\ \BBA {} Motta, E.%
\end{APACrefauthors}%
\unskip\
\newblock
\APACrefYearMonthDay{2022}{02}{}.
\newblock
{\BBOQ}\APACrefatitle {{AIDA: A knowledge graph about research dynamics in
  academia and industry}} {{AIDA: A knowledge graph about research dynamics in
  academia and industry}}.{\BBCQ}
\newblock
\APACjournalVolNumPages{Quantitative Science Studies}{2}{4}{1356-1398}.
\newblock
\begin{APACrefDOI} \doi{10.1162/qss_a_00162} \end{APACrefDOI}
\PrintBackRefs{\CurrentBib}

\bibitem [\protect \citeauthoryear {%
Aronson%
\ \BBA {} Lang%
}{%
Aronson%
\ \BBA {} Lang%
}{%
{\protect \APACyear {2010}}%
}]{%
aronson2010overview}
\APACinsertmetastar {%
aronson2010overview}%
\begin{APACrefauthors}%
Aronson, A\BPBI R.%
\BCBT {}\ \BBA {} Lang, F\BHBI M.%
\end{APACrefauthors}%
\unskip\
\newblock
\APACrefYearMonthDay{2010}{}{}.
\newblock
{\BBOQ}\APACrefatitle {An overview of MetaMap: historical perspective and
  recent advances} {An overview of metamap: historical perspective and recent
  advances}.{\BBCQ}
\newblock
\APACjournalVolNumPages{Journal of the American Medical Informatics
  Association}{17}{3}{229--236}.
\newblock
\begin{APACrefDOI} \doi{10.1136/jamia.2009.002733} \end{APACrefDOI}
\PrintBackRefs{\CurrentBib}

\bibitem [\protect \citeauthoryear {%
Auer%
\ \protect \BOthers {.}}{%
Auer%
\ \protect \BOthers {.}}{%
{\protect \APACyear {2018}}%
}]{%
auer2018towards}
\APACinsertmetastar {%
auer2018towards}%
\begin{APACrefauthors}%
Auer, S.%
, Kovtun, V.%
, Prinz, M.%
, Kasprzik, A.%
, Stocker, M.%
\BCBL {}\ \BBA {} Vidal, M\BPBI E.%
\end{APACrefauthors}%
\unskip\
\newblock
\APACrefYearMonthDay{2018}{}{}.
\newblock
{\BBOQ}\APACrefatitle {Towards a knowledge graph for science} {Towards a
  knowledge graph for science}.{\BBCQ}
\newblock
\BIn{} \APACrefbtitle {Proceedings of the 8th International Conference on Web
  Intelligence, Mining and Semantics} {Proceedings of the 8th international
  conference on web intelligence, mining and semantics}\ (\BPGS\ 1--6).
\newblock
\begin{APACrefDOI} \doi{10.1145/3227609.3227689} \end{APACrefDOI}
\PrintBackRefs{\CurrentBib}

\bibitem [\protect \citeauthoryear {%
Ayala%
\ \BBA {} Bechard%
}{%
Ayala%
\ \BBA {} Bechard%
}{%
{\protect \APACyear {2024}}%
}]{%
bechard2024reducinghallucinationstructuredoutputs}
\APACinsertmetastar {%
bechard2024reducinghallucinationstructuredoutputs}%
\begin{APACrefauthors}%
Ayala, O.%
\BCBT {}\ \BBA {} Bechard, P.%
\end{APACrefauthors}%
\unskip\
\newblock
\APACrefYearMonthDay{2024}{{\APACmonth{06}}}{}.
\newblock
{\BBOQ}\APACrefatitle {Reducing hallucination in structured outputs via
  Retrieval-Augmented Generation} {Reducing hallucination in structured outputs
  via retrieval-augmented generation}.{\BBCQ}
\newblock
\BIn{} Y.~Yang, A.~Davani, A.~Sil\BCBL {}\ \BBA {} A.~Kumar\ (\BEDS),
  \APACrefbtitle {Proceedings of the 2024 Conference of the North American
  Chapter of the Association for Computational Linguistics: Human Language
  Technologies (Volume 6: Industry Track)} {Proceedings of the 2024 conference
  of the north american chapter of the association for computational
  linguistics: Human language technologies (volume 6: Industry track)}\ (\BPGS\
  228--238).
\newblock
\APACaddressPublisher{Mexico City, Mexico}{Association for Computational
  Linguistics}.
\newblock
\begin{APACrefURL} \url{https://aclanthology.org/2024.naacl-industry.19/}
  \end{APACrefURL}
\newblock
\begin{APACrefDOI} \doi{10.18653/v1/2024.naacl-industry.19} \end{APACrefDOI}
\PrintBackRefs{\CurrentBib}

\bibitem [\protect \citeauthoryear {%
Ayele%
, Chevallet%
, Kassie%
\BCBL {}\ \BBA {} Meshesha%
}{%
Ayele%
\ \protect \BOthers {.}}{%
{\protect \APACyear {2012}}%
}]{%
Ayele2012}
\APACinsertmetastar {%
Ayele2012}%
\begin{APACrefauthors}%
Ayele, D.%
, Chevallet, J\BHBI P.%
, Kassie, G.%
\BCBL {}\ \BBA {} Meshesha, M.%
\end{APACrefauthors}%
\unskip\
\newblock
\APACrefYearMonthDay{2012}{}{}.
\newblock
{\BBOQ}\APACrefatitle {Enhancing Semantic Relation Quality of UMLS Knowledge
  Sources} {Enhancing semantic relation quality of umls knowledge
  sources}.{\BBCQ}
\newblock
\BIn{} \APACrefbtitle {Proceedings of the International Conference on
  Management of Emergent Digital EcoSystems} {Proceedings of the international
  conference on management of emergent digital ecosystems}\ (\BPG~59–66).
\newblock
\APACaddressPublisher{New York, NY, USA}{Association for Computing Machinery}.
\newblock
\begin{APACrefURL} \url{https://doi.org/10.1145/2457276.2457289}
  \end{APACrefURL}
\newblock
\begin{APACrefDOI} \doi{10.1145/2457276.2457289} \end{APACrefDOI}
\PrintBackRefs{\CurrentBib}

\bibitem [\protect \citeauthoryear {%
Bella%
, Giunchiglia%
\BCBL {}\ \BBA {} McNeill%
}{%
Bella%
\ \protect \BOthers {.}}{%
{\protect \APACyear {2017}}%
}]{%
BELLA20171}
\APACinsertmetastar {%
BELLA20171}%
\begin{APACrefauthors}%
Bella, G.%
, Giunchiglia, F.%
\BCBL {}\ \BBA {} McNeill, F.%
\end{APACrefauthors}%
\unskip\
\newblock
\APACrefYearMonthDay{2017}{}{}.
\newblock
{\BBOQ}\APACrefatitle {Language and domain aware lightweight ontology matching}
  {Language and domain aware lightweight ontology matching}.{\BBCQ}
\newblock
\APACjournalVolNumPages{Journal of Web Semantics}{43}{}{1-17}.
\newblock
\begin{APACrefURL}
  \url{https://www.sciencedirect.com/science/article/pii/S1570826817300161}
  \end{APACrefURL}
\newblock
\begin{APACrefDOI} \doi{10.1016/j.websem.2017.03.003} \end{APACrefDOI}
\PrintBackRefs{\CurrentBib}

\bibitem [\protect \citeauthoryear {%
Beltagy%
, Lo%
\BCBL {}\ \BBA {} Cohan%
}{%
Beltagy%
\ \protect \BOthers {.}}{%
{\protect \APACyear {2019}}%
}]{%
beltagy2019scibertpretrainedlanguagemodel}
\APACinsertmetastar {%
beltagy2019scibertpretrainedlanguagemodel}%
\begin{APACrefauthors}%
Beltagy, I.%
, Lo, K.%
\BCBL {}\ \BBA {} Cohan, A.%
\end{APACrefauthors}%
\unskip\
\newblock
\APACrefYearMonthDay{2019}{{\APACmonth{11}}}{}.
\newblock
{\BBOQ}\APACrefatitle {{S}ci{BERT}: A Pretrained Language Model for Scientific
  Text} {{S}ci{BERT}: A pretrained language model for scientific text}.{\BBCQ}
\newblock
\BIn{} K.~Inui, J.~Jiang, V.~Ng\BCBL {}\ \BBA {} X.~Wan\ (\BEDS),
  \APACrefbtitle {Proceedings of the 2019 Conference on Empirical Methods in
  Natural Language Processing and the 9th International Joint Conference on
  Natural Language Processing (EMNLP-IJCNLP)} {Proceedings of the 2019
  conference on empirical methods in natural language processing and the 9th
  international joint conference on natural language processing
  (emnlp-ijcnlp)}\ (\BPGS\ 3615--3620).
\newblock
\APACaddressPublisher{Hong Kong, China}{Association for Computational
  Linguistics}.
\newblock
\begin{APACrefURL} \url{https://aclanthology.org/D19-1371/} \end{APACrefURL}
\newblock
\begin{APACrefDOI} \doi{10.18653/v1/D19-1371} \end{APACrefDOI}
\PrintBackRefs{\CurrentBib}

\bibitem [\protect \citeauthoryear {%
Berners-Lee%
, Hendler%
\BCBL {}\ \BBA {} Lassila%
}{%
Berners-Lee%
\ \protect \BOthers {.}}{%
{\protect \APACyear {2001}}%
}]{%
berners2001semantic}
\APACinsertmetastar {%
berners2001semantic}%
\begin{APACrefauthors}%
Berners-Lee, T.%
, Hendler, J.%
\BCBL {}\ \BBA {} Lassila, O.%
\end{APACrefauthors}%
\unskip\
\newblock
\APACrefYearMonthDay{2001}{}{}.
\newblock
{\BBOQ}\APACrefatitle {The semantic web} {The semantic web}.{\BBCQ}
\newblock
\APACjournalVolNumPages{Scientific american}{284}{5}{34--43}.
\newblock
\begin{APACrefDOI} \doi{10.1038/scientificamerican052001-yL7Vw7HIOZ4iSjlnEeVsJ}
  \end{APACrefDOI}
\PrintBackRefs{\CurrentBib}

\bibitem [\protect \citeauthoryear {%
Bodenreider%
}{%
Bodenreider%
}{%
{\protect \APACyear {2001}}%
}]{%
bodenreider2001circular}
\APACinsertmetastar {%
bodenreider2001circular}%
\begin{APACrefauthors}%
Bodenreider, O.%
\end{APACrefauthors}%
\unskip\
\newblock
\APACrefYearMonthDay{2001}{}{}.
\newblock
{\BBOQ}\APACrefatitle {Circular hierarchical relationships in the UMLS:
  etiology, diagnosis, treatment, complications and prevention.} {Circular
  hierarchical relationships in the umls: etiology, diagnosis, treatment,
  complications and prevention.}{\BBCQ}
\newblock
\BIn{} \APACrefbtitle {Proceedings of the AMIA Symposium} {Proceedings of the
  amia symposium}\ (\BPG~57).
\PrintBackRefs{\CurrentBib}

\bibitem [\protect \citeauthoryear {%
Bodenreider%
}{%
Bodenreider%
}{%
{\protect \APACyear {2004}}%
}]{%
Bodenreider2004}
\APACinsertmetastar {%
Bodenreider2004}%
\begin{APACrefauthors}%
Bodenreider, O.%
\end{APACrefauthors}%
\unskip\
\newblock
\APACrefYearMonthDay{2004}{01}{}.
\newblock
{\BBOQ}\APACrefatitle {The Unified Medical Language System (UMLS): integrating
  biomedical terminology} {The unified medical language system (umls):
  integrating biomedical terminology}.{\BBCQ}
\newblock
\APACjournalVolNumPages{Nucleic Acids Research}{32}{suppl\_1}{D267-D270}.
\newblock
\begin{APACrefDOI} \doi{10.1093/nar/gkh061} \end{APACrefDOI}
\PrintBackRefs{\CurrentBib}

\bibitem [\protect \citeauthoryear {%
Bola{\~{n}}os%
, Salatino%
, Osborne%
\BCBL {}\ \BBA {} Motta%
}{%
Bola{\~{n}}os%
\ \protect \BOthers {.}}{%
{\protect \APACyear {2024}}%
}]{%
bolanos2024artificialintelligenceliteraturereviews}
\APACinsertmetastar {%
bolanos2024artificialintelligenceliteraturereviews}%
\begin{APACrefauthors}%
Bola{\~{n}}os, F.%
, Salatino, A.%
, Osborne, F.%
\BCBL {}\ \BBA {} Motta, E.%
\end{APACrefauthors}%
\unskip\
\newblock
\APACrefYearMonthDay{2024}{Aug}{17}.
\newblock
{\BBOQ}\APACrefatitle {Artificial intelligence for literature reviews:
  opportunities and challenges} {Artificial intelligence for literature
  reviews: opportunities and challenges}.{\BBCQ}
\newblock
\APACjournalVolNumPages{Artificial Intelligence Review}{57}{10}{259}.
\newblock
\begin{APACrefDOI} \doi{10.1007/s10462-024-10902-3} \end{APACrefDOI}
\PrintBackRefs{\CurrentBib}

\bibitem [\protect \citeauthoryear {%
Borges%
\ \BBA {} dos Reis%
}{%
Borges%
\ \BBA {} dos Reis%
}{%
{\protect \APACyear {2019}}%
}]{%
borges2019semantic}
\APACinsertmetastar {%
borges2019semantic}%
\begin{APACrefauthors}%
Borges, M\BPBI V\BPBI M.%
\BCBT {}\ \BBA {} dos Reis, J\BPBI C.%
\end{APACrefauthors}%
\unskip\
\newblock
\APACrefYearMonthDay{2019}{}{}.
\newblock
{\BBOQ}\APACrefatitle {Semantic-Enhanced Recommendation of Video Lectures}
  {Semantic-enhanced recommendation of video lectures}.{\BBCQ}
\newblock
\BIn{} \APACrefbtitle {2019 IEEE 19th International Conference on Advanced
  Learning Technologies (ICALT)} {2019 ieee 19th international conference on
  advanced learning technologies (icalt)}\ (\BVOL\ 2161, \BPGS\ 42--46).
\newblock
\begin{APACrefDOI} \doi{10.1109/ICALT.2019.00013} \end{APACrefDOI}
\PrintBackRefs{\CurrentBib}

\bibitem [\protect \citeauthoryear {%
Brainard%
}{%
Brainard%
}{%
{\protect \APACyear {2020}}%
}]{%
brainard2020scientists}
\APACinsertmetastar {%
brainard2020scientists}%
\begin{APACrefauthors}%
Brainard, J.%
\end{APACrefauthors}%
\unskip\
\newblock
\APACrefYearMonthDay{2020}{}{}.
\newblock
{\BBOQ}\APACrefatitle {Scientists are drowning in COVID-19 papers. Can new
  tools keep them afloat} {Scientists are drowning in covid-19 papers. can new
  tools keep them afloat}.{\BBCQ}
\newblock
\APACjournalVolNumPages{Science}{13}{10.1126}{}.
\newblock
\begin{APACrefDOI} \doi{10.1126/science.abc7839} \end{APACrefDOI}
\PrintBackRefs{\CurrentBib}

\bibitem [\protect \citeauthoryear {%
Caragea%
, Bulgarov%
\BCBL {}\ \BBA {} Mihalcea%
}{%
Caragea%
\ \protect \BOthers {.}}{%
{\protect \APACyear {2015}}%
}]{%
caragea2015}
\APACinsertmetastar {%
caragea2015}%
\begin{APACrefauthors}%
Caragea, C.%
, Bulgarov, F.%
\BCBL {}\ \BBA {} Mihalcea, R.%
\end{APACrefauthors}%
\unskip\
\newblock
\APACrefYearMonthDay{2015}{{\APACmonth{09}}}{}.
\newblock
{\BBOQ}\APACrefatitle {Co-Training for Topic Classification of Scholarly Data}
  {Co-training for topic classification of scholarly data}.{\BBCQ}
\newblock
\BIn{} L.~M{\`a}rquez, C.~Callison-Burch\BCBL {}\ \BBA {} J.~Su\ (\BEDS),
  \APACrefbtitle {Proceedings of the 2015 Conference on Empirical Methods in
  Natural Language Processing} {Proceedings of the 2015 conference on empirical
  methods in natural language processing}\ (\BPGS\ 2357--2366).
\newblock
\APACaddressPublisher{Lisbon, Portugal}{Association for Computational
  Linguistics}.
\newblock
\begin{APACrefDOI} \doi{10.18653/v1/D15-1283} \end{APACrefDOI}
\PrintBackRefs{\CurrentBib}

\bibitem [\protect \citeauthoryear {%
Chatzopoulos%
, Vergoulis%
, Kanellos%
, Dalamagas%
\BCBL {}\ \BBA {} Tryfonopoulos%
}{%
Chatzopoulos%
\ \protect \BOthers {.}}{%
{\protect \APACyear {2020}}%
}]{%
artsim2020}
\APACinsertmetastar {%
artsim2020}%
\begin{APACrefauthors}%
Chatzopoulos, S.%
, Vergoulis, T.%
, Kanellos, I.%
, Dalamagas, T.%
\BCBL {}\ \BBA {} Tryfonopoulos, C.%
\end{APACrefauthors}%
\unskip\
\newblock
\APACrefYearMonthDay{2020}{}{}.
\newblock
{\BBOQ}\APACrefatitle {ArtSim: Improved Estimation of Current Impact for Recent
  Articles} {Artsim: Improved estimation of current impact for recent
  articles}.{\BBCQ}
\newblock
\BIn{} L.~Bellatreche\ \BOthers {.}\ (\BEDS), \APACrefbtitle {ADBIS, TPDL and
  EDA 2020 Common Workshops and Doctoral Consortium} {Adbis, tpdl and eda 2020
  common workshops and doctoral consortium}\ (\BPGS\ 323--334).
\newblock
\APACaddressPublisher{Cham}{Springer International Publishing}.
\newblock
\begin{APACrefDOI} \doi{10.1007/978-3-030-55814-7_27} \end{APACrefDOI}
\PrintBackRefs{\CurrentBib}

\bibitem [\protect \citeauthoryear {%
Chernyak%
}{%
Chernyak%
}{%
{\protect \APACyear {2015}}%
}]{%
chernyak2015}
\APACinsertmetastar {%
chernyak2015}%
\begin{APACrefauthors}%
Chernyak, E.%
\end{APACrefauthors}%
\unskip\
\newblock
\APACrefYearMonthDay{2015}{}{}.
\newblock
{\BBOQ}\APACrefatitle {An Approach to the Problem of Annotation of Research
  Publications} {An approach to the problem of annotation of research
  publications}.{\BBCQ}
\newblock
\BIn{} \APACrefbtitle {Proceedings of the Eighth ACM International Conference
  on Web Search and Data Mining} {Proceedings of the eighth acm international
  conference on web search and data mining}\ (\BPG~429–434).
\newblock
\APACaddressPublisher{New York, NY, USA}{Association for Computing Machinery}.
\newblock
\begin{APACrefURL} \url{https://doi.org/10.1145/2684822.2697032}
  \end{APACrefURL}
\newblock
\begin{APACrefDOI} \doi{10.1145/2684822.2697032} \end{APACrefDOI}
\PrintBackRefs{\CurrentBib}

\bibitem [\protect \citeauthoryear {%
Chilton%
, Little%
, Edge%
, Weld%
\BCBL {}\ \BBA {} Landay%
}{%
Chilton%
\ \protect \BOthers {.}}{%
{\protect \APACyear {2013}}%
}]{%
chilton2013cascade}
\APACinsertmetastar {%
chilton2013cascade}%
\begin{APACrefauthors}%
Chilton, L\BPBI B.%
, Little, G.%
, Edge, D.%
, Weld, D\BPBI S.%
\BCBL {}\ \BBA {} Landay, J\BPBI A.%
\end{APACrefauthors}%
\unskip\
\newblock
\APACrefYearMonthDay{2013}{}{}.
\newblock
{\BBOQ}\APACrefatitle {Cascade: Crowdsourcing taxonomy creation} {Cascade:
  Crowdsourcing taxonomy creation}.{\BBCQ}
\newblock
\BIn{} \APACrefbtitle {Proceedings of the SIGCHI Conference on Human Factors in
  Computing Systems} {Proceedings of the sigchi conference on human factors in
  computing systems}\ (\BPGS\ 1999--2008).
\newblock
\begin{APACrefDOI} \doi{10.1145/2470654.2466265} \end{APACrefDOI}
\PrintBackRefs{\CurrentBib}

\bibitem [\protect \citeauthoryear {%
Chowdhury%
}{%
Chowdhury%
}{%
{\protect \APACyear {2010}}%
}]{%
chowdhury2010digital}
\APACinsertmetastar {%
chowdhury2010digital}%
\begin{APACrefauthors}%
Chowdhury, G.%
\end{APACrefauthors}%
\unskip\
\newblock
\APACrefYearMonthDay{2010}{}{}.
\newblock
{\BBOQ}\APACrefatitle {From digital libraries to digital preservation research:
  the importance of users and context} {From digital libraries to digital
  preservation research: the importance of users and context}.{\BBCQ}
\newblock
\APACjournalVolNumPages{Journal of documentation}{66}{2}{207--223}.
\newblock
\begin{APACrefDOI} \doi{10.1108/00220411011023625} \end{APACrefDOI}
\PrintBackRefs{\CurrentBib}

\bibitem [\protect \citeauthoryear {%
Cleverley%
\ \BBA {} Burnett%
}{%
Cleverley%
\ \BBA {} Burnett%
}{%
{\protect \APACyear {2015}}%
}]{%
cleverley2015best}
\APACinsertmetastar {%
cleverley2015best}%
\begin{APACrefauthors}%
Cleverley, P\BPBI H.%
\BCBT {}\ \BBA {} Burnett, S.%
\end{APACrefauthors}%
\unskip\
\newblock
\APACrefYearMonthDay{2015}{}{}.
\newblock
{\BBOQ}\APACrefatitle {The best of both worlds: highlighting the synergies of
  combining manual and automatic knowledge organization methods to improve
  information search and discovery.} {The best of both worlds: highlighting the
  synergies of combining manual and automatic knowledge organization methods to
  improve information search and discovery.}{\BBCQ}
\newblock
\APACjournalVolNumPages{Knowledge Organization}{42}{6}{}.
\newblock
\begin{APACrefDOI} \doi{10.5771/0943-7444-2015-6-428} \end{APACrefDOI}
\PrintBackRefs{\CurrentBib}

\bibitem [\protect \citeauthoryear {%
Dai%
\ \protect \BOthers {.}}{%
Dai%
\ \protect \BOthers {.}}{%
{\protect \APACyear {2020}}%
}]{%
dai2020fullmesh}
\APACinsertmetastar {%
dai2020fullmesh}%
\begin{APACrefauthors}%
Dai, S.%
, You, R.%
, Lu, Z.%
, Huang, X.%
, Mamitsuka, H.%
\BCBL {}\ \BBA {} Zhu, S.%
\end{APACrefauthors}%
\unskip\
\newblock
\APACrefYearMonthDay{2020}{}{}.
\newblock
{\BBOQ}\APACrefatitle {FullMeSH: improving large-scale MeSH indexing with full
  text} {Fullmesh: improving large-scale mesh indexing with full text}.{\BBCQ}
\newblock
\APACjournalVolNumPages{Bioinformatics}{36}{5}{1533--1541}.
\newblock
\begin{APACrefDOI} \doi{10.1093/bioinformatics/btz756} \end{APACrefDOI}
\PrintBackRefs{\CurrentBib}

\bibitem [\protect \citeauthoryear {%
Daiber%
, Jakob%
, Hokamp%
\BCBL {}\ \BBA {} Mendes%
}{%
Daiber%
\ \protect \BOthers {.}}{%
{\protect \APACyear {2013}}%
}]{%
Daiber2013}
\APACinsertmetastar {%
Daiber2013}%
\begin{APACrefauthors}%
Daiber, J.%
, Jakob, M.%
, Hokamp, C.%
\BCBL {}\ \BBA {} Mendes, P\BPBI N.%
\end{APACrefauthors}%
\unskip\
\newblock
\APACrefYearMonthDay{2013}{}{}.
\newblock
{\BBOQ}\APACrefatitle {Improving Efficiency and Accuracy in Multilingual Entity
  Extraction} {Improving efficiency and accuracy in multilingual entity
  extraction}.{\BBCQ}
\newblock
\BIn{} \APACrefbtitle {Proceedings of the 9th International Conference on
  Semantic Systems} {Proceedings of the 9th international conference on
  semantic systems}\ (\BPG~121–124).
\newblock
\APACaddressPublisher{New York, NY, USA}{Association for Computing Machinery}.
\newblock
\begin{APACrefURL} \url{https://doi.org/10.1145/2506182.2506198}
  \end{APACrefURL}
\newblock
\begin{APACrefDOI} \doi{10.1145/2506182.2506198} \end{APACrefDOI}
\PrintBackRefs{\CurrentBib}

\bibitem [\protect \citeauthoryear {%
Declerck%
}{%
Declerck%
}{%
{\protect \APACyear {2013}}%
}]{%
declerck2013integration}
\APACinsertmetastar {%
declerck2013integration}%
\begin{APACrefauthors}%
Declerck, T.%
\end{APACrefauthors}%
\unskip\
\newblock
\APACrefYearMonthDay{2013}{{\APACmonth{08}}}{}.
\newblock
{\BBOQ}\APACrefatitle {Integration of the Thesaurus for the Social Sciences
  ({T}he{S}oz) in an Information Extraction System} {Integration of the
  thesaurus for the social sciences ({T}he{S}oz) in an information extraction
  system}.{\BBCQ}
\newblock
\BIn{} P.~Lendvai\ \BBA {} K.~Zervanou\ (\BEDS), \APACrefbtitle {Proceedings of
  the 7th Workshop on Language Technology for Cultural Heritage, Social
  Sciences, and Humanities} {Proceedings of the 7th workshop on language
  technology for cultural heritage, social sciences, and humanities}\ (\BPGS\
  90--95).
\newblock
\APACaddressPublisher{Sofia, Bulgaria}{Association for Computational
  Linguistics}.
\newblock
\begin{APACrefURL} \url{https://aclanthology.org/W13-2712/} \end{APACrefURL}
\PrintBackRefs{\CurrentBib}

\bibitem [\protect \citeauthoryear {%
Dess{\`\i}%
, Osborne%
, Reforgiato~Recupero%
, Buscaldi%
\BCBL {}\ \BBA {} Motta%
}{%
Dess{\`\i}%
\ \protect \BOthers {.}}{%
{\protect \APACyear {2022}}%
}]{%
dessi2022cskg}
\APACinsertmetastar {%
dessi2022cskg}%
\begin{APACrefauthors}%
Dess{\`\i}, D.%
, Osborne, F.%
, Reforgiato~Recupero, D.%
, Buscaldi, D.%
\BCBL {}\ \BBA {} Motta, E.%
\end{APACrefauthors}%
\unskip\
\newblock
\APACrefYearMonthDay{2022}{}{}.
\newblock
{\BBOQ}\APACrefatitle {CS-KG: A Large-Scale Knowledge Graph of Research
  Entities and Claims in Computer Science} {Cs-kg: A large-scale knowledge
  graph of research entities and claims in computer science}.{\BBCQ}
\newblock
\BIn{} \APACrefbtitle {International Semantic Web Conference (ISWC).}
  {International semantic web conference (iswc).}
\newblock
\begin{APACrefDOI} \doi{10.1007/978-3-031-19433-7_39} \end{APACrefDOI}
\PrintBackRefs{\CurrentBib}

\bibitem [\protect \citeauthoryear {%
Dhombres%
\ \BBA {} Bodenreider%
}{%
Dhombres%
\ \BBA {} Bodenreider%
}{%
{\protect \APACyear {2016}}%
}]{%
Dhombres2016}
\APACinsertmetastar {%
Dhombres2016}%
\begin{APACrefauthors}%
Dhombres, F.%
\BCBT {}\ \BBA {} Bodenreider, O.%
\end{APACrefauthors}%
\unskip\
\newblock
\APACrefYearMonthDay{2016}{}{}.
\newblock
{\BBOQ}\APACrefatitle {Interoperability between phenotypes in research and
  healthcare terminologies-Investigating partial mappings between HPO and
  SNOMED CT} {Interoperability between phenotypes in research and healthcare
  terminologies-investigating partial mappings between hpo and snomed
  ct}.{\BBCQ}
\newblock
\APACjournalVolNumPages{Journal of Biomedical Semantics}{7}{1}{}.
\newblock
\begin{APACrefDOI} \doi{10.1186/s13326-016-0047-3} \end{APACrefDOI}
\PrintBackRefs{\CurrentBib}

\bibitem [\protect \citeauthoryear {%
Dimou%
\ \protect \BOthers {.}}{%
Dimou%
\ \protect \BOthers {.}}{%
{\protect \APACyear {2014}}%
}]{%
dimou2014rml}
\APACinsertmetastar {%
dimou2014rml}%
\begin{APACrefauthors}%
Dimou, A.%
, Vander~Sande, M.%
, Colpaert, P.%
, Verborgh, R.%
, Mannens, E.%
\BCBL {}\ \BBA {} Van~de Walle, R.%
\end{APACrefauthors}%
\unskip\
\newblock
\APACrefYearMonthDay{2014}{}{}.
\newblock
{\BBOQ}\APACrefatitle {RML: a generic language for integrated RDF mappings of
  heterogeneous data} {Rml: a generic language for integrated rdf mappings of
  heterogeneous data}.{\BBCQ}
\newblock
\BIn{} \APACrefbtitle {Ldow.} {Ldow.}
\newblock
\begin{APACrefURL} \url{http://ceur-ws.org/Vol-1184/ldow2014_paper_01.pdf}
  \end{APACrefURL}
\PrintBackRefs{\CurrentBib}

\bibitem [\protect \citeauthoryear {%
Ding%
, Rousseau%
\BCBL {}\ \BBA {} Wolfram%
}{%
Ding%
\ \protect \BOthers {.}}{%
{\protect \APACyear {2014}}%
}]{%
ding2014}
\APACinsertmetastar {%
ding2014}%
\begin{APACrefauthors}%
Ding, Y.%
, Rousseau, R.%
\BCBL {}\ \BBA {} Wolfram, D.%
\end{APACrefauthors}%
\ (\BEDS).
\unskip\
\newblock
\APACrefYear{2014}.
\newblock
\APACrefbtitle {Measuring {{Scholarly Impact}}} {Measuring {{Scholarly
  Impact}}}.
\newblock
\APACaddressPublisher{{Cham}}{{Springer International Publishing}}.
\newblock
\begin{APACrefDOI} \doi{10.1007/978-3-319-10377-8} \end{APACrefDOI}
\PrintBackRefs{\CurrentBib}

\bibitem [\protect \citeauthoryear {%
Djeddi%
\ \BBA {} Khadir%
}{%
Djeddi%
\ \BBA {} Khadir%
}{%
{\protect \APACyear {2014}}%
}]{%
Djeddi2014}
\APACinsertmetastar {%
Djeddi2014}%
\begin{APACrefauthors}%
Djeddi, W\BPBI E.%
\BCBT {}\ \BBA {} Khadir, M\BPBI T.%
\end{APACrefauthors}%
\unskip\
\newblock
\APACrefYearMonthDay{2014}{}{}.
\newblock
{\BBOQ}\APACrefatitle {A Novel Approach Using Context-Based Measure for
  Matching Large Scale Ontologies} {A novel approach using context-based
  measure for matching large scale ontologies}.{\BBCQ}
\newblock
\BIn{} L.~Bellatreche\ \BBA {} M\BPBI K.~Mohania\ (\BEDS), \APACrefbtitle {Data
  Warehousing and Knowledge Discovery} {Data warehousing and knowledge
  discovery}\ (\BPGS\ 320--331).
\newblock
\APACaddressPublisher{Cham}{Springer International Publishing}.
\newblock
\begin{APACrefDOI} \doi{10.1007/978-3-319-10160-6_29} \end{APACrefDOI}
\PrintBackRefs{\CurrentBib}

\bibitem [\protect \citeauthoryear {%
Dunne%
\ \BBA {} Hulek%
}{%
Dunne%
\ \BBA {} Hulek%
}{%
{\protect \APACyear {2020}}%
}]{%
dunne2020}
\APACinsertmetastar {%
dunne2020}%
\begin{APACrefauthors}%
Dunne, E.%
\BCBT {}\ \BBA {} Hulek, K.%
\end{APACrefauthors}%
\unskip\
\newblock
\APACrefYearMonthDay{2020}{}{}.
\newblock
{\BBOQ}\APACrefatitle {Mathematics subject classification 2020} {Mathematics
  subject classification 2020}.{\BBCQ}
\newblock
\APACjournalVolNumPages{Not. Am. Math. Soc}{67}{3}{410--411}.
\newblock
\begin{APACrefDOI} \doi{10.1090/noti2052} \end{APACrefDOI}
\PrintBackRefs{\CurrentBib}

\bibitem [\protect \citeauthoryear {%
Erdogan%
, Erdem%
\BCBL {}\ \BBA {} Bodenreider%
}{%
Erdogan%
\ \protect \BOthers {.}}{%
{\protect \APACyear {2010}}%
}]{%
erdogan2010exploiting}
\APACinsertmetastar {%
erdogan2010exploiting}%
\begin{APACrefauthors}%
Erdogan, H.%
, Erdem, E.%
\BCBL {}\ \BBA {} Bodenreider, O.%
\end{APACrefauthors}%
\unskip\
\newblock
\APACrefYearMonthDay{2010}{}{}.
\newblock
{\BBOQ}\APACrefatitle {Exploiting UMLS semantics for checking semantic
  consistency among UMLS concepts} {Exploiting umls semantics for checking
  semantic consistency among umls concepts}.{\BBCQ}
\newblock
\APACjournalVolNumPages{Studies in health technology and informatics}{160}{0
  1}{749}.
\PrintBackRefs{\CurrentBib}

\bibitem [\protect \citeauthoryear {%
Fan%
\ \BBA {} Friedman%
}{%
Fan%
\ \BBA {} Friedman%
}{%
{\protect \APACyear {2007}}%
}]{%
fan2007combining}
\APACinsertmetastar {%
fan2007combining}%
\begin{APACrefauthors}%
Fan, J\BHBI W.%
\BCBT {}\ \BBA {} Friedman, C.%
\end{APACrefauthors}%
\unskip\
\newblock
\APACrefYearMonthDay{2007}{}{}.
\newblock
{\BBOQ}\APACrefatitle {Combining contextual and lexical features to classify
  UMLS concepts} {Combining contextual and lexical features to classify umls
  concepts}.{\BBCQ}
\newblock
\BIn{} \APACrefbtitle {AMIA Annual Symposium Proceedings} {Amia annual
  symposium proceedings}\ (\BVOL\ 2007, \BPG~231).
\PrintBackRefs{\CurrentBib}

\bibitem [\protect \citeauthoryear {%
Genesereth%
\ \BBA {} Nilsson%
}{%
Genesereth%
\ \BBA {} Nilsson%
}{%
{\protect \APACyear {2012}}%
}]{%
genesereth2012}
\APACinsertmetastar {%
genesereth2012}%
\begin{APACrefauthors}%
Genesereth, M\BPBI R.%
\BCBT {}\ \BBA {} Nilsson, N\BPBI J.%
\end{APACrefauthors}%
\unskip\
\newblock
\APACrefYear{2012}.
\newblock
\APACrefbtitle {Logical foundations of artificial intelligence} {Logical
  foundations of artificial intelligence}.
\newblock
\APACaddressPublisher{}{Morgan Kaufmann}.
\newblock
\begin{APACrefDOI} \doi{10.1016/C2009-0-27551-9} \end{APACrefDOI}
\PrintBackRefs{\CurrentBib}

\bibitem [\protect \citeauthoryear {%
Gnoli%
\ \protect \BOthers {.}}{%
Gnoli%
\ \protect \BOthers {.}}{%
{\protect \APACyear {2024}}%
}]{%
gnoli2024library}
\APACinsertmetastar {%
gnoli2024library}%
\begin{APACrefauthors}%
Gnoli, C.%
, Golub, K.%
, Haynes, D.%
, Salaba, A.%
, Shiri, A.%
\BCBL {}\ \BBA {} Slavic, A.%
\end{APACrefauthors}%
\unskip\
\newblock
\APACrefYearMonthDay{2024}{}{}.
\newblock
{\BBOQ}\APACrefatitle {Library Catalog’s Search Interface: Making the Most of
  Subject Metadata} {Library catalog’s search interface: Making the most of
  subject metadata}.{\BBCQ}
\newblock
\APACjournalVolNumPages{KO KNOWLEDGE ORGANIZATION}{51}{3}{169--186}.
\newblock
\begin{APACrefDOI} \doi{10.5771/0943-7444-2024-3-169} \end{APACrefDOI}
\PrintBackRefs{\CurrentBib}

\bibitem [\protect \citeauthoryear {%
Gruber%
}{%
Gruber%
}{%
{\protect \APACyear {1993}}%
}]{%
gruber1993}
\APACinsertmetastar {%
gruber1993}%
\begin{APACrefauthors}%
Gruber, T\BPBI R.%
\end{APACrefauthors}%
\unskip\
\newblock
\APACrefYearMonthDay{1993}{}{}.
\newblock
{\BBOQ}\APACrefatitle {A translation approach to portable ontology
  specifications} {A translation approach to portable ontology
  specifications}.{\BBCQ}
\newblock
\APACjournalVolNumPages{Knowledge Acquisition}{5}{2}{199-220}.
\newblock
\begin{APACrefURL}
  \url{https://www.sciencedirect.com/science/article/pii/S1042814383710083}
  \end{APACrefURL}
\newblock
\begin{APACrefDOI} \doi{https://doi.org/10.1006/knac.1993.1008}
  \end{APACrefDOI}
\PrintBackRefs{\CurrentBib}

\bibitem [\protect \citeauthoryear {%
H.~Gu%
\ \protect \BOthers {.}}{%
H.~Gu%
\ \protect \BOthers {.}}{%
{\protect \APACyear {2007}}%
}]{%
gu2007evaluation}
\APACinsertmetastar {%
gu2007evaluation}%
\begin{APACrefauthors}%
Gu, H.%
, Hripcsak, G.%
, Chen, Y.%
, Morrey, C\BPBI P.%
, Elhanan, G.%
, Cimino, J\BPBI J.%
\BDBL {}Perl, Y.%
\end{APACrefauthors}%
\unskip\
\newblock
\APACrefYearMonthDay{2007}{}{}.
\newblock
{\BBOQ}\APACrefatitle {Evaluation of a UMLS auditing process of semantic type
  assignments} {Evaluation of a umls auditing process of semantic type
  assignments}.{\BBCQ}
\newblock
\BIn{} \APACrefbtitle {AMIA Annual Symposium Proceedings} {Amia annual
  symposium proceedings}\ (\BVOL\ 2007, \BPG~294).
\PrintBackRefs{\CurrentBib}

\bibitem [\protect \citeauthoryear {%
H\BPBI H.~Gu%
\ \protect \BOthers {.}}{%
H\BPBI H.~Gu%
\ \protect \BOthers {.}}{%
{\protect \APACyear {2004}}%
}]{%
GU200429}
\APACinsertmetastar {%
GU200429}%
\begin{APACrefauthors}%
Gu, H\BPBI H.%
, Perl, Y.%
, Elhanan, G.%
, Min, H.%
, Zhang, L.%
\BCBL {}\ \BBA {} Peng, Y.%
\end{APACrefauthors}%
\unskip\
\newblock
\APACrefYearMonthDay{2004}{}{}.
\newblock
{\BBOQ}\APACrefatitle {Auditing concept categorizations in the UMLS} {Auditing
  concept categorizations in the umls}.{\BBCQ}
\newblock
\APACjournalVolNumPages{Artificial Intelligence in Medicine}{31}{1}{29-44}.
\newblock
\begin{APACrefURL}
  \url{https://www.sciencedirect.com/science/article/pii/S0933365704000387}
  \end{APACrefURL}
\newblock
\begin{APACrefDOI} \doi{https://doi.org/10.1016/j.artmed.2004.02.002}
  \end{APACrefDOI}
\PrintBackRefs{\CurrentBib}

\bibitem [\protect \citeauthoryear {%
ANSI/NISO Z39.19-2005(R2010)}{%
ANSI/NISO Z39.19-2005(R2010)}{%
{\protect \APACyear {2010}}%
}]{%
niso2005}
\APACinsertmetastar {%
niso2005}%
\APACrefbtitle {Guidelines for the Construction, Format, and Management of
  Monolingual Controlled Vocabularies} {Guidelines for the construction,
  format, and management of monolingual controlled vocabularies}\ \APACbVolEdTR
  {}{Standard}.
\newblock
\APACrefYearMonthDay{2010}{{\APACmonth{07}}}{}.
\newblock
\APACaddressInstitution{Baltimore, Maryland}{National Information Standards
  Organization}.
\newblock
\begin{APACrefDOI} \doi{10.3789/ansi.niso.z39.19-2005R2010} \end{APACrefDOI}
\PrintBackRefs{\CurrentBib}

\bibitem [\protect \citeauthoryear {%
Halper%
\ \protect \BOthers {.}}{%
Halper%
\ \protect \BOthers {.}}{%
{\protect \APACyear {2011}}%
}]{%
halper2011auditing}
\APACinsertmetastar {%
halper2011auditing}%
\begin{APACrefauthors}%
Halper, M.%
, Morrey, C\BPBI P.%
, Chen, Y.%
, Elhanan, G.%
, Hripcsak, G.%
\BCBL {}\ \BBA {} Perl, Y.%
\end{APACrefauthors}%
\unskip\
\newblock
\APACrefYearMonthDay{2011}{}{}.
\newblock
{\BBOQ}\APACrefatitle {Auditing hierarchical cycles to locate other
  inconsistencies in the UMLS} {Auditing hierarchical cycles to locate other
  inconsistencies in the umls}.{\BBCQ}
\newblock
\BIn{} \APACrefbtitle {AMIA Annual Symposium Proceedings} {Amia annual
  symposium proceedings}\ (\BVOL\ 2011, \BPG~529).
\PrintBackRefs{\CurrentBib}

\bibitem [\protect \citeauthoryear {%
Han%
, Yang%
, Mishra%
\BCBL {}\ \BBA {} Diesner%
}{%
Han%
\ \protect \BOthers {.}}{%
{\protect \APACyear {2020}}%
}]{%
han2020}
\APACinsertmetastar {%
han2020}%
\begin{APACrefauthors}%
Han, K.%
, Yang, P.%
, Mishra, S.%
\BCBL {}\ \BBA {} Diesner, J.%
\end{APACrefauthors}%
\unskip\
\newblock
\APACrefYearMonthDay{2020}{}{}.
\newblock
{\BBOQ}\APACrefatitle {WikiCSSH: Extracting Computer Science Subject Headings
  from Wikipedia} {Wikicssh: Extracting computer science subject headings from
  wikipedia}.{\BBCQ}
\newblock
\BIn{} L.~Bellatreche\ \BOthers {.}\ (\BEDS), \APACrefbtitle {ADBIS, TPDL and
  EDA 2020 Common Workshops and Doctoral Consortium} {Adbis, tpdl and eda 2020
  common workshops and doctoral consortium}\ (\BPGS\ 207--218).
\newblock
\APACaddressPublisher{Cham}{Springer International Publishing}.
\newblock
\begin{APACrefDOI} \doi{10.1007/978-3-030-55814-7_17} \end{APACrefDOI}
\PrintBackRefs{\CurrentBib}

\bibitem [\protect \citeauthoryear {%
Hedden%
}{%
Hedden%
}{%
{\protect \APACyear {2010}}%
}]{%
hedden2010taxonomies}
\APACinsertmetastar {%
hedden2010taxonomies}%
\begin{APACrefauthors}%
Hedden, H.%
\end{APACrefauthors}%
\unskip\
\newblock
\APACrefYearMonthDay{2010}{}{}.
\newblock
{\BBOQ}\APACrefatitle {Taxonomies and controlled vocabularies best practices
  for metadata} {Taxonomies and controlled vocabularies best practices for
  metadata}.{\BBCQ}
\newblock
\APACjournalVolNumPages{Journal of Digital Asset Management}{6}{}{279--284}.
\newblock
\begin{APACrefDOI} \doi{10.1057/dam.2010.29} \end{APACrefDOI}
\PrintBackRefs{\CurrentBib}

\bibitem [\protect \citeauthoryear {%
Hodge%
}{%
Hodge%
}{%
{\protect \APACyear {2000}}%
}]{%
hodge2000systems}
\APACinsertmetastar {%
hodge2000systems}%
\begin{APACrefauthors}%
Hodge, G\BPBI M.%
\end{APACrefauthors}%
\unskip\
\newblock
\APACrefYear{2000}.
\newblock
\APACrefbtitle {Systems of knowledge organization for digital libraries: beyond
  traditional authority files} {Systems of knowledge organization for digital
  libraries: beyond traditional authority files}\ (\BNUM~91).
\newblock
\APACaddressPublisher{}{Digital Library Federation}.
\PrintBackRefs{\CurrentBib}

\bibitem [\protect \citeauthoryear {%
Huang%
, Xie%
, Meng%
, Zhang%
\BCBL {}\ \BBA {} Han%
}{%
Huang%
\ \protect \BOthers {.}}{%
{\protect \APACyear {2020}}%
}]{%
huang2020corel}
\APACinsertmetastar {%
huang2020corel}%
\begin{APACrefauthors}%
Huang, J.%
, Xie, Y.%
, Meng, Y.%
, Zhang, Y.%
\BCBL {}\ \BBA {} Han, J.%
\end{APACrefauthors}%
\unskip\
\newblock
\APACrefYearMonthDay{2020}{}{}.
\newblock
{\BBOQ}\APACrefatitle {Corel: Seed-guided topical taxonomy construction by
  concept learning and relation transferring} {Corel: Seed-guided topical
  taxonomy construction by concept learning and relation transferring}.{\BBCQ}
\newblock
\BIn{} \APACrefbtitle {Proceedings of the 26th ACM SIGKDD International
  Conference on Knowledge Discovery \& Data Mining} {Proceedings of the 26th
  acm sigkdd international conference on knowledge discovery \& data mining}\
  (\BPGS\ 1928--1936).
\newblock
\begin{APACrefDOI} \doi{10.1145/3394486.3403244} \end{APACrefDOI}
\PrintBackRefs{\CurrentBib}

\bibitem [\protect \citeauthoryear {%
Ilgisonis%
, Pyatnitskiy%
, Tarbeeva%
, Aldushin%
\BCBL {}\ \BBA {} Ponomarenko%
}{%
Ilgisonis%
\ \protect \BOthers {.}}{%
{\protect \APACyear {2022}}%
}]{%
ilgisonis2022catch}
\APACinsertmetastar {%
ilgisonis2022catch}%
\begin{APACrefauthors}%
Ilgisonis, E\BPBI V.%
, Pyatnitskiy, M\BPBI A.%
, Tarbeeva, S\BPBI N.%
, Aldushin, A\BPBI A.%
\BCBL {}\ \BBA {} Ponomarenko, E\BPBI A.%
\end{APACrefauthors}%
\unskip\
\newblock
\APACrefYearMonthDay{2022}{}{}.
\newblock
{\BBOQ}\APACrefatitle {How to catch trends using MeSH terms analysis?} {How to
  catch trends using mesh terms analysis?}{\BBCQ}
\newblock
\APACjournalVolNumPages{Scientometrics}{127}{4}{1953--1967}.
\newblock
\begin{APACrefDOI} \doi{10.1007/s11192-022-04292-y} \end{APACrefDOI}
\PrintBackRefs{\CurrentBib}

\bibitem [\protect \citeauthoryear {%
BS ISO 25964-2:2013}{%
BS ISO 25964-2:2013}{%
{\protect \APACyear {2013}}%
}]{%
ISO25964}
\APACinsertmetastar {%
ISO25964}%
\APACrefbtitle {Information and documentation. Thesauri and interoperability
  with othervocabularies. Interoperability with other vocabularies}
  {Information and documentation. thesauri and interoperability with
  othervocabularies. interoperability with other vocabularies}\ \APACbVolEdTR
  {}{Standard}.
\newblock
\APACrefYearMonthDay{2013}{{\APACmonth{03}}}{}.
\newblock
\APACaddressInstitution{}{International Organization for Standardization}.
\PrintBackRefs{\CurrentBib}

\bibitem [\protect \citeauthoryear {%
Jim{\'e}nez-Ruiz%
, Grau%
, Horrocks%
\BCBL {}\ \BBA {} Berlanga%
}{%
Jim{\'e}nez-Ruiz%
\ \protect \BOthers {.}}{%
{\protect \APACyear {2011}}%
}]{%
jimenez2011logic}
\APACinsertmetastar {%
jimenez2011logic}%
\begin{APACrefauthors}%
Jim{\'e}nez-Ruiz, E.%
, Grau, B\BPBI C.%
, Horrocks, I.%
\BCBL {}\ \BBA {} Berlanga, R.%
\end{APACrefauthors}%
\unskip\
\newblock
\APACrefYearMonthDay{2011}{}{}.
\newblock
{\BBOQ}\APACrefatitle {Logic-based assessment of the compatibility of UMLS
  ontology sources} {Logic-based assessment of the compatibility of umls
  ontology sources}.{\BBCQ}
\newblock
\APACjournalVolNumPages{Journal of biomedical semantics}{2}{1}{1--16}.
\newblock
\begin{APACrefDOI} \doi{10.1186/2041-1480-2-S1-S2} \end{APACrefDOI}
\PrintBackRefs{\CurrentBib}

\bibitem [\protect \citeauthoryear {%
Jing%
}{%
Jing%
}{%
{\protect \APACyear {2021}}%
}]{%
xia2021}
\APACinsertmetastar {%
xia2021}%
\begin{APACrefauthors}%
Jing, X.%
\end{APACrefauthors}%
\unskip\
\newblock
\APACrefYearMonthDay{2021}{Aug}{27}.
\newblock
{\BBOQ}\APACrefatitle {The Unified Medical Language System at 30 Years and How
  It Is Used and Published: Systematic Review and Content Analysis} {The
  unified medical language system at 30 years and how it is used and published:
  Systematic review and content analysis}.{\BBCQ}
\newblock
\APACjournalVolNumPages{JMIR Med Inform}{9}{8}{e20675}.
\newblock
\begin{APACrefURL} \url{https://medinform.jmir.org/2021/8/e20675}
  \end{APACrefURL}
\newblock
\begin{APACrefDOI} \doi{10.2196/20675} \end{APACrefDOI}
\PrintBackRefs{\CurrentBib}

\bibitem [\protect \citeauthoryear {%
Johnson%
, Bretonnel~Cohen%
\BCBL {}\ \BBA {} Hunter%
}{%
Johnson%
\ \protect \BOthers {.}}{%
{\protect \APACyear {2007}}%
}]{%
johnson2007fault}
\APACinsertmetastar {%
johnson2007fault}%
\begin{APACrefauthors}%
Johnson, H\BPBI L.%
, Bretonnel~Cohen, K.%
\BCBL {}\ \BBA {} Hunter, L.%
\end{APACrefauthors}%
\unskip\
\newblock
\APACrefYearMonthDay{2007}{}{}.
\newblock
{\BBOQ}\APACrefatitle {A fault model for ontology mapping, alignment, and
  linking systems} {A fault model for ontology mapping, alignment, and linking
  systems}.{\BBCQ}
\newblock
\BIn{} \APACrefbtitle {Biocomputing 2007} {Biocomputing 2007}\ (\BPGS\
  233--244).
\newblock
\APACaddressPublisher{}{World Scientific}.
\PrintBackRefs{\CurrentBib}

\bibitem [\protect \citeauthoryear {%
Kalfoglou%
\ \BBA {} Schorlemmer%
}{%
Kalfoglou%
\ \BBA {} Schorlemmer%
}{%
{\protect \APACyear {2003}}%
}]{%
kalfoglou2003ontology}
\APACinsertmetastar {%
kalfoglou2003ontology}%
\begin{APACrefauthors}%
Kalfoglou, Y.%
\BCBT {}\ \BBA {} Schorlemmer, M.%
\end{APACrefauthors}%
\unskip\
\newblock
\APACrefYearMonthDay{2003}{}{}.
\newblock
{\BBOQ}\APACrefatitle {Ontology mapping: the state of the art} {Ontology
  mapping: the state of the art}.{\BBCQ}
\newblock
\APACjournalVolNumPages{The knowledge engineering review}{18}{1}{1--31}.
\newblock
\begin{APACrefDOI} \doi{10.1017/S0269888903000651} \end{APACrefDOI}
\PrintBackRefs{\CurrentBib}

\bibitem [\protect \citeauthoryear {%
Kandimalla%
, Rohatgi%
, Wu%
\BCBL {}\ \BBA {} Giles%
}{%
Kandimalla%
\ \protect \BOthers {.}}{%
{\protect \APACyear {2021}}%
}]{%
kandimalla2020}
\APACinsertmetastar {%
kandimalla2020}%
\begin{APACrefauthors}%
Kandimalla, B.%
, Rohatgi, S.%
, Wu, J.%
\BCBL {}\ \BBA {} Giles, C\BPBI L.%
\end{APACrefauthors}%
\unskip\
\newblock
\APACrefYearMonthDay{2021}{}{}.
\newblock
{\BBOQ}\APACrefatitle {Large Scale Subject Category Classification of Scholarly
  Papers With Deep Attentive Neural Networks} {Large scale subject category
  classification of scholarly papers with deep attentive neural
  networks}.{\BBCQ}
\newblock
\APACjournalVolNumPages{Frontiers in Research Metrics and Analytics}{5}{}{31}.
\newblock
\begin{APACrefURL}
  \url{https://www.frontiersin.org/article/10.3389/frma.2020.600382}
  \end{APACrefURL}
\newblock
\begin{APACrefDOI} \doi{10.3389/frma.2020.600382} \end{APACrefDOI}
\PrintBackRefs{\CurrentBib}

\bibitem [\protect \citeauthoryear {%
Kang%
, Li%
\BCBL {}\ \BBA {} Coppel%
}{%
Kang%
\ \protect \BOthers {.}}{%
{\protect \APACyear {2015}}%
}]{%
Kang2015}
\APACinsertmetastar {%
Kang2015}%
\begin{APACrefauthors}%
Kang, Y\BHBI B.%
, Li, Y\BHBI F.%
\BCBL {}\ \BBA {} Coppel, R\BPBI L.%
\end{APACrefauthors}%
\unskip\
\newblock
\APACrefYearMonthDay{2015}{}{}.
\newblock
{\BBOQ}\APACrefatitle {Capturing Researcher Expertise through MeSH
  Classification} {Capturing researcher expertise through mesh
  classification}.{\BBCQ}
\newblock
\BIn{} \APACrefbtitle {Proceedings of the 8th International Conference on
  Knowledge Capture.} {Proceedings of the 8th international conference on
  knowledge capture.}
\newblock
\APACaddressPublisher{New York, NY, USA}{Association for Computing Machinery}.
\newblock
\begin{APACrefURL}
  \url{https://doi-org.libezproxy.open.ac.uk/10.1145/2815833.2815837}
  \end{APACrefURL}
\newblock
\begin{APACrefDOI} \doi{10.1145/2815833.2815837} \end{APACrefDOI}
\PrintBackRefs{\CurrentBib}

\bibitem [\protect \citeauthoryear {%
Kendall%
\ \BBA {} McGuinness%
}{%
Kendall%
\ \BBA {} McGuinness%
}{%
{\protect \APACyear {2019}}%
}]{%
kendall2019ontology}
\APACinsertmetastar {%
kendall2019ontology}%
\begin{APACrefauthors}%
Kendall, E\BPBI F.%
\BCBT {}\ \BBA {} McGuinness, D\BPBI L.%
\end{APACrefauthors}%
\unskip\
\newblock
\APACrefYear{2019}.
\newblock
\APACrefbtitle {Ontology engineering} {Ontology engineering}.
\newblock
\APACaddressPublisher{}{Morgan \& Claypool Publishers}.
\newblock
\begin{APACrefDOI} \doi{10.1007/978-3-031-79486-5} \end{APACrefDOI}
\PrintBackRefs{\CurrentBib}

\bibitem [\protect \citeauthoryear {%
Kojima%
, Gu%
, Reid%
, Matsuo%
\BCBL {}\ \BBA {} Iwasawa%
}{%
Kojima%
\ \protect \BOthers {.}}{%
{\protect \APACyear {2023}}%
}]{%
kojima2023largelanguagemodelszeroshot}
\APACinsertmetastar {%
kojima2023largelanguagemodelszeroshot}%
\begin{APACrefauthors}%
Kojima, T.%
, Gu, S\BPBI S.%
, Reid, M.%
, Matsuo, Y.%
\BCBL {}\ \BBA {} Iwasawa, Y.%
\end{APACrefauthors}%
\unskip\
\newblock
\APACrefYearMonthDay{2023}{}{}.
\newblock
\APACrefbtitle {Large Language Models are Zero-Shot Reasoners.} {Large language
  models are zero-shot reasoners.}
\newblock
\begin{APACrefURL} \url{https://arxiv.org/abs/2205.11916} \end{APACrefURL}
\PrintBackRefs{\CurrentBib}

\bibitem [\protect \citeauthoryear {%
Kopácsi%
, Hudak%
\BCBL {}\ \BBA {} Ganguly%
}{%
Kopácsi%
\ \protect \BOthers {.}}{%
{\protect \APACyear {2017}}%
}]{%
Ganguly_2017}
\APACinsertmetastar {%
Ganguly_2017}%
\begin{APACrefauthors}%
Kopácsi, S.%
, Hudak, R.%
\BCBL {}\ \BBA {} Ganguly, R.%
\end{APACrefauthors}%
\unskip\
\newblock
\APACrefYearMonthDay{2017}{Sep.}{}.
\newblock
{\BBOQ}\APACrefatitle {Implementation of a Classification Server to Support
  Metadata Organization for Long Term Preservation Systems} {Implementation of
  a classification server to support metadata organization for long term
  preservation systems}.{\BBCQ}
\newblock
\APACjournalVolNumPages{Communications of the Association of Austrian
  Librarians}{70}{2}{225–243}.
\newblock
\begin{APACrefURL}
  \url{https://journals.univie.ac.at/index.php/voebm/article/view/2075}
  \end{APACrefURL}
\newblock
\begin{APACrefDOI} \doi{10.31263/voebm.v70i2.1897} \end{APACrefDOI}
\PrintBackRefs{\CurrentBib}

\bibitem [\protect \citeauthoryear {%
Kuhn%
}{%
Kuhn%
}{%
{\protect \APACyear {1962}}%
}]{%
kuhn_1962}
\APACinsertmetastar {%
kuhn_1962}%
\begin{APACrefauthors}%
Kuhn, T\BPBI S.%
\end{APACrefauthors}%
\unskip\
\newblock
\APACrefYear{1962}.
\newblock
\APACrefbtitle {The structure of Scientific Revolutions} {The structure of
  scientific revolutions}.
\newblock
\APACaddressPublisher{}{The University of Chicago Press}.
\PrintBackRefs{\CurrentBib}

\bibitem [\protect \citeauthoryear {%
Lambe%
}{%
Lambe%
}{%
{\protect \APACyear {2014}}%
}]{%
lambe2014organising}
\APACinsertmetastar {%
lambe2014organising}%
\begin{APACrefauthors}%
Lambe, P.%
\end{APACrefauthors}%
\unskip\
\newblock
\APACrefYear{2014}.
\newblock
\APACrefbtitle {Organising Knowledge: Taxonomies, Knowledge and Organisational
  Effectiveness} {Organising knowledge: Taxonomies, knowledge and
  organisational effectiveness}.
\newblock
\APACaddressPublisher{}{Elsevier Science}.
\newblock
\begin{APACrefURL} \url{https://books.google.co.uk/books?id=z1mpAgAAQBAJ}
  \end{APACrefURL}
\PrintBackRefs{\CurrentBib}

\bibitem [\protect \citeauthoryear {%
Lei~Zeng%
\ \BBA {} Mai~Chan%
}{%
Lei~Zeng%
\ \BBA {} Mai~Chan%
}{%
{\protect \APACyear {2004}}%
}]{%
zeng2004}
\APACinsertmetastar {%
zeng2004}%
\begin{APACrefauthors}%
Lei~Zeng, M.%
\BCBT {}\ \BBA {} Mai~Chan, L.%
\end{APACrefauthors}%
\unskip\
\newblock
\APACrefYearMonthDay{2004}{}{}.
\newblock
{\BBOQ}\APACrefatitle {Trends and issues in establishing interoperability among
  knowledge organization systems} {Trends and issues in establishing
  interoperability among knowledge organization systems}.{\BBCQ}
\newblock
\APACjournalVolNumPages{Journal of the American Society for Information Science
  and Technology}{55}{5}{377-395}.
\newblock
\begin{APACrefURL}
  \url{https://onlinelibrary.wiley.com/doi/abs/10.1002/asi.10387}
  \end{APACrefURL}
\newblock
\begin{APACrefDOI} \doi{10.1002/asi.10387} \end{APACrefDOI}
\PrintBackRefs{\CurrentBib}

\bibitem [\protect \citeauthoryear {%
L\"offler%
\ \protect \BOthers {.}}{%
L\"offler%
\ \protect \BOthers {.}}{%
{\protect \APACyear {2020}}%
}]{%
ScholarLensViz2020}
\APACinsertmetastar {%
ScholarLensViz2020}%
\begin{APACrefauthors}%
L\"offler, F.%
, Wesp, V.%
, Babalou, S.%
, Kahn, P.%
, Lachmann, R.%
, Sateli, B.%
\BDBL {}K\"onig-Ries, B.%
\end{APACrefauthors}%
\unskip\
\newblock
\APACrefYearMonthDay{2020}{}{}.
\newblock
{\BBOQ}\APACrefatitle {ScholarLensViz: A Visualization Framework for
  Transparency in Semantic User Profiles} {Scholarlensviz: A visualization
  framework for transparency in semantic user profiles}.{\BBCQ}
\newblock
\BIn{} K.~Taylor, R.~Gonçalves, F.~Lecue\BCBL {}\ \BBA {} J.~Yan\ (\BEDS),
  \APACrefbtitle {Proceedings of the ISWC 2020 Demos and Industry Tracks: From
  Novel Ideas to Industrial Practice co-located with 19th International
  Semantic Web Conference (ISWC 2020), Globally online, November 1-6, 2020
  (UTC).} {Proceedings of the iswc 2020 demos and industry tracks: From novel
  ideas to industrial practice co-located with 19th international semantic web
  conference (iswc 2020), globally online, november 1-6, 2020 (utc).}
\newblock
\begin{APACrefURL} \url{https://ceur-ws.org/Vol-2721/paper485.pdf}
  \end{APACrefURL}
\PrintBackRefs{\CurrentBib}

\bibitem [\protect \citeauthoryear {%
Lu%
, Zhu%
, Li%
, Qiao%
\BCBL {}\ \BBA {} Yuan%
}{%
Lu%
\ \protect \BOthers {.}}{%
{\protect \APACyear {2024}}%
}]{%
lu2024llamax}
\APACinsertmetastar {%
lu2024llamax}%
\begin{APACrefauthors}%
Lu, Y.%
, Zhu, W.%
, Li, L.%
, Qiao, Y.%
\BCBL {}\ \BBA {} Yuan, F.%
\end{APACrefauthors}%
\unskip\
\newblock
\APACrefYearMonthDay{2024}{{\APACmonth{11}}}{}.
\newblock
{\BBOQ}\APACrefatitle {{LL}a{MAX}: Scaling Linguistic Horizons of {LLM} by
  Enhancing Translation Capabilities Beyond 100 Languages} {{LL}a{MAX}: Scaling
  linguistic horizons of {LLM} by enhancing translation capabilities beyond 100
  languages}.{\BBCQ}
\newblock
\BIn{} Y.~Al-Onaizan, M.~Bansal\BCBL {}\ \BBA {} Y\BHBI N.~Chen\ (\BEDS),
  \APACrefbtitle {Findings of the Association for Computational Linguistics:
  EMNLP 2024} {Findings of the association for computational linguistics: Emnlp
  2024}\ (\BPGS\ 10748--10772).
\newblock
\APACaddressPublisher{Miami, Florida, USA}{Association for Computational
  Linguistics}.
\newblock
\begin{APACrefURL} \url{https://aclanthology.org/2024.findings-emnlp.631/}
  \end{APACrefURL}
\newblock
\begin{APACrefDOI} \doi{10.18653/v1/2024.findings-emnlp.631} \end{APACrefDOI}
\PrintBackRefs{\CurrentBib}

\bibitem [\protect \citeauthoryear {%
Mai%
, Galke%
\BCBL {}\ \BBA {} Scherp%
}{%
Mai%
\ \protect \BOthers {.}}{%
{\protect \APACyear {2018}}%
}]{%
mai2018}
\APACinsertmetastar {%
mai2018}%
\begin{APACrefauthors}%
Mai, F.%
, Galke, L.%
\BCBL {}\ \BBA {} Scherp, A.%
\end{APACrefauthors}%
\unskip\
\newblock
\APACrefYearMonthDay{2018}{}{}.
\newblock
{\BBOQ}\APACrefatitle {Using Deep Learning for Title-Based Semantic Subject
  Indexing to Reach Competitive Performance to Full-Text} {Using deep learning
  for title-based semantic subject indexing to reach competitive performance to
  full-text}.{\BBCQ}
\newblock
\BIn{} \APACrefbtitle {Proceedings of the 18th ACM/IEEE on Joint Conference on
  Digital Libraries} {Proceedings of the 18th acm/ieee on joint conference on
  digital libraries}\ (\BPG~169–178).
\newblock
\APACaddressPublisher{New York, NY, USA}{Association for Computing Machinery}.
\newblock
\begin{APACrefURL} \url{https://doi.org/10.1145/3197026.3197039}
  \end{APACrefURL}
\newblock
\begin{APACrefDOI} \doi{10.1145/3197026.3197039} \end{APACrefDOI}
\PrintBackRefs{\CurrentBib}

\bibitem [\protect \citeauthoryear {%
Martínez-González%
\ \BBA {} Alvite-Díez%
}{%
Martínez-González%
\ \BBA {} Alvite-Díez%
}{%
{\protect \APACyear {2019}}%
}]{%
8873649}
\APACinsertmetastar {%
8873649}%
\begin{APACrefauthors}%
Martínez-González, M\BPBI M.%
\BCBT {}\ \BBA {} Alvite-Díez, M\BHBI L.%
\end{APACrefauthors}%
\unskip\
\newblock
\APACrefYearMonthDay{2019}{}{}.
\newblock
{\BBOQ}\APACrefatitle {Thesauri and Semantic Web: Discussion of the Evolution
  of Thesauri Toward Their Integration With the Semantic Web} {Thesauri and
  semantic web: Discussion of the evolution of thesauri toward their
  integration with the semantic web}.{\BBCQ}
\newblock
\APACjournalVolNumPages{IEEE Access}{7}{}{153151-153170}.
\newblock
\begin{APACrefDOI} \doi{10.1109/ACCESS.2019.2948028} \end{APACrefDOI}
\PrintBackRefs{\CurrentBib}

\bibitem [\protect \citeauthoryear {%
Mazzocchi%
}{%
Mazzocchi%
}{%
{\protect \APACyear {2018}}%
}]{%
mazzocchi2018}
\APACinsertmetastar {%
mazzocchi2018}%
\begin{APACrefauthors}%
Mazzocchi, F.%
\end{APACrefauthors}%
\unskip\
\newblock
\APACrefYearMonthDay{2018}{}{}.
\newblock
{\BBOQ}\APACrefatitle {{Knowledge Organization System (KOS): An Introductory
  Critical Account}} {{Knowledge Organization System (KOS): An Introductory
  Critical Account}}.{\BBCQ}
\newblock
\APACjournalVolNumPages{KO KNOWLEDGE ORGANIZATION}{45}{1}{54--78}.
\newblock
\begin{APACrefDOI} \doi{10.5771/0943-7444-2018-1-54} \end{APACrefDOI}
\PrintBackRefs{\CurrentBib}

\bibitem [\protect \citeauthoryear {%
McInnes%
, Pedersen%
\BCBL {}\ \BBA {} Carlis%
}{%
McInnes%
\ \protect \BOthers {.}}{%
{\protect \APACyear {2007}}%
}]{%
mcinnes2007using}
\APACinsertmetastar {%
mcinnes2007using}%
\begin{APACrefauthors}%
McInnes, B\BPBI T.%
, Pedersen, T.%
\BCBL {}\ \BBA {} Carlis, J.%
\end{APACrefauthors}%
\unskip\
\newblock
\APACrefYearMonthDay{2007}{}{}.
\newblock
{\BBOQ}\APACrefatitle {Using UMLS Concept Unique Identifiers (CUIs) for word
  sense disambiguation in the biomedical domain} {Using umls concept unique
  identifiers (cuis) for word sense disambiguation in the biomedical
  domain}.{\BBCQ}
\newblock
\BIn{} \APACrefbtitle {AMIA annual symposium proceedings} {Amia annual
  symposium proceedings}\ (\BVOL\ 2007, \BPG~533).
\PrintBackRefs{\CurrentBib}

\bibitem [\protect \citeauthoryear {%
Mitchell%
}{%
Mitchell%
}{%
{\protect \APACyear {2001}}%
}]{%
mitchell2001relationships}
\APACinsertmetastar {%
mitchell2001relationships}%
\begin{APACrefauthors}%
Mitchell, J\BPBI S.%
\end{APACrefauthors}%
\unskip\
\newblock
\APACrefYearMonthDay{2001}{}{}.
\newblock
{\BBOQ}\APACrefatitle {Relationships in the Dewey decimal classification
  system} {Relationships in the dewey decimal classification system}.{\BBCQ}
\newblock
\APACjournalVolNumPages{Relationships in the organization of
  knowledge}{}{}{211--226}.
\newblock
\begin{APACrefDOI} \doi{10.1007/978-94-015-9696-1_14} \end{APACrefDOI}
\PrintBackRefs{\CurrentBib}

\bibitem [\protect \citeauthoryear {%
Montiel-Ponsoda%
, De~Cea%
, G{\'o}mez-P{\'e}rez%
\BCBL {}\ \BBA {} Peters%
}{%
Montiel-Ponsoda%
\ \protect \BOthers {.}}{%
{\protect \APACyear {2011}}%
}]{%
montiel2011enriching}
\APACinsertmetastar {%
montiel2011enriching}%
\begin{APACrefauthors}%
Montiel-Ponsoda, E.%
, De~Cea, G\BPBI A.%
, G{\'o}mez-P{\'e}rez, A.%
\BCBL {}\ \BBA {} Peters, W.%
\end{APACrefauthors}%
\unskip\
\newblock
\APACrefYearMonthDay{2011}{}{}.
\newblock
{\BBOQ}\APACrefatitle {Enriching ontologies with multilingual information}
  {Enriching ontologies with multilingual information}.{\BBCQ}
\newblock
\APACjournalVolNumPages{Natural language engineering}{17}{3}{283--309}.
\newblock
\begin{APACrefDOI} \doi{10.1007/978-3-540-68234-9_26} \end{APACrefDOI}
\PrintBackRefs{\CurrentBib}

\bibitem [\protect \citeauthoryear {%
Morrey%
, Geller%
, Halper%
\BCBL {}\ \BBA {} Perl%
}{%
Morrey%
\ \protect \BOthers {.}}{%
{\protect \APACyear {2009}}%
}]{%
MORREY2009468}
\APACinsertmetastar {%
MORREY2009468}%
\begin{APACrefauthors}%
Morrey, C\BPBI P.%
, Geller, J.%
, Halper, M.%
\BCBL {}\ \BBA {} Perl, Y.%
\end{APACrefauthors}%
\unskip\
\newblock
\APACrefYearMonthDay{2009}{}{}.
\newblock
{\BBOQ}\APACrefatitle {The Neighborhood Auditing Tool: A hybrid interface for
  auditing the UMLS} {The neighborhood auditing tool: A hybrid interface for
  auditing the umls}.{\BBCQ}
\newblock
\APACjournalVolNumPages{Journal of Biomedical Informatics}{42}{3}{468-489}.
\newblock
\begin{APACrefURL}
  \url{https://www.sciencedirect.com/science/article/pii/S1532046409000197}
  \end{APACrefURL}
\newblock
\APACrefnote{Auditing of Terminologies}
\newblock
\begin{APACrefDOI} \doi{10.1016/j.jbi.2009.01.006} \end{APACrefDOI}
\PrintBackRefs{\CurrentBib}

\bibitem [\protect \citeauthoryear {%
Motta%
}{%
Motta%
}{%
{\protect \APACyear {1999}}%
}]{%
motta1999reusable}
\APACinsertmetastar {%
motta1999reusable}%
\begin{APACrefauthors}%
Motta, E.%
\end{APACrefauthors}%
\unskip\
\newblock
\APACrefYear{1999}.
\newblock
\APACrefbtitle {Reusable Components for Knowledge Modelling: Case Studies in
  Parametric Design Problem Solving} {Reusable components for knowledge
  modelling: Case studies in parametric design problem solving}.
\newblock
\APACaddressPublisher{}{IOS Press}.
\newblock
\begin{APACrefURL} \url{https://books.google.co.uk/books?id=S6g8FTcQAVYC}
  \end{APACrefURL}
\PrintBackRefs{\CurrentBib}

\bibitem [\protect \citeauthoryear {%
Mougin%
\ \BBA {} Bodenreider%
}{%
Mougin%
\ \BBA {} Bodenreider%
}{%
{\protect \APACyear {2005}}%
}]{%
mougin2005approaches}
\APACinsertmetastar {%
mougin2005approaches}%
\begin{APACrefauthors}%
Mougin, F.%
\BCBT {}\ \BBA {} Bodenreider, O.%
\end{APACrefauthors}%
\unskip\
\newblock
\APACrefYearMonthDay{2005}{}{}.
\newblock
{\BBOQ}\APACrefatitle {Approaches to eliminating cycles in the UMLS
  Metathesaurus: Na{\"\i}ve vs. formal} {Approaches to eliminating cycles in
  the umls metathesaurus: Na{\"\i}ve vs. formal}.{\BBCQ}
\newblock
\BIn{} \APACrefbtitle {AMIA Annual Symposium Proceedings} {Amia annual
  symposium proceedings}\ (\BVOL\ 2005, \BPG~550).
\PrintBackRefs{\CurrentBib}

\bibitem [\protect \citeauthoryear {%
Mu%
, Lu%
\BCBL {}\ \BBA {} Ryu%
}{%
Mu%
\ \protect \BOthers {.}}{%
{\protect \APACyear {2014}}%
}]{%
mu2014explicitly}
\APACinsertmetastar {%
mu2014explicitly}%
\begin{APACrefauthors}%
Mu, X.%
, Lu, K.%
\BCBL {}\ \BBA {} Ryu, H.%
\end{APACrefauthors}%
\unskip\
\newblock
\APACrefYearMonthDay{2014}{}{}.
\newblock
{\BBOQ}\APACrefatitle {Explicitly integrating MeSH thesaurus help into health
  information retrieval systems: An empirical user study} {Explicitly
  integrating mesh thesaurus help into health information retrieval systems: An
  empirical user study}.{\BBCQ}
\newblock
\APACjournalVolNumPages{Information Processing \& Management}{50}{1}{24--40}.
\newblock
\begin{APACrefDOI} \doi{10.1016/j.ipm.2013.03.005} \end{APACrefDOI}
\PrintBackRefs{\CurrentBib}

\bibitem [\protect \citeauthoryear {%
Musen%
}{%
Musen%
}{%
{\protect \APACyear {2015}}%
}]{%
musen2015}
\APACinsertmetastar {%
musen2015}%
\begin{APACrefauthors}%
Musen, M\BPBI A.%
\end{APACrefauthors}%
\unskip\
\newblock
\APACrefYearMonthDay{2015}{Jun}{}.
\newblock
{\BBOQ}\APACrefatitle {{{T}he {P}rotégé {P}roject: {A} {L}ook {B}ack and a
  {L}ook {F}orward}} {{{T}he {P}rotégé {P}roject: {A} {L}ook {B}ack and a
  {L}ook {F}orward}}.{\BBCQ}
\newblock
\APACjournalVolNumPages{AI Matters}{1}{4}{4--12}.
\newblock
\begin{APACrefDOI} \doi{10.1145/2757001.2757003} \end{APACrefDOI}
\PrintBackRefs{\CurrentBib}

\bibitem [\protect \citeauthoryear {%
Newman%
, Noh%
, Talley%
, Karimi%
\BCBL {}\ \BBA {} Baldwin%
}{%
Newman%
\ \protect \BOthers {.}}{%
{\protect \APACyear {2010}}%
}]{%
newman2010evaluating}
\APACinsertmetastar {%
newman2010evaluating}%
\begin{APACrefauthors}%
Newman, D.%
, Noh, Y.%
, Talley, E.%
, Karimi, S.%
\BCBL {}\ \BBA {} Baldwin, T.%
\end{APACrefauthors}%
\unskip\
\newblock
\APACrefYearMonthDay{2010}{}{}.
\newblock
{\BBOQ}\APACrefatitle {Evaluating topic models for digital libraries}
  {Evaluating topic models for digital libraries}.{\BBCQ}
\newblock
\BIn{} \APACrefbtitle {Proceedings of the 10th annual joint conference on
  Digital libraries} {Proceedings of the 10th annual joint conference on
  digital libraries}\ (\BPGS\ 215--224).
\newblock
\begin{APACrefDOI} \doi{10.1145/1816123.1816156} \end{APACrefDOI}
\PrintBackRefs{\CurrentBib}

\bibitem [\protect \citeauthoryear {%
OpenAlex%
}{%
OpenAlex%
}{%
{\protect \APACyear {2024}}%
}]{%
openalex2024}
\APACinsertmetastar {%
openalex2024}%
\begin{APACrefauthors}%
OpenAlex.%
\end{APACrefauthors}%
\unskip\
\newblock
\APACrefYearMonthDay{2024}{}{}.
\newblock
{\BBOQ}\APACrefatitle {OpenAlex: End-to-End Process for Topic Classification}
  {Openalex: End-to-end process for topic classification}.{\BBCQ}.
\newblock
\begin{APACrefURL}
  \url{https://docs.google.com/document/d/1bDopkhuGieQ4F8gGNj7sEc8WSE8mvLZS/edit}
  \end{APACrefURL}
\PrintBackRefs{\CurrentBib}

\bibitem [\protect \citeauthoryear {%
Osborne%
\ \BBA {} Motta%
}{%
Osborne%
\ \BBA {} Motta%
}{%
{\protect \APACyear {2015}}%
}]{%
osborne2015klink}
\APACinsertmetastar {%
osborne2015klink}%
\begin{APACrefauthors}%
Osborne, F.%
\BCBT {}\ \BBA {} Motta, E.%
\end{APACrefauthors}%
\unskip\
\newblock
\APACrefYearMonthDay{2015}{}{}.
\newblock
{\BBOQ}\APACrefatitle {Klink-2: integrating multiple web sources to generate
  semantic topic networks} {Klink-2: integrating multiple web sources to
  generate semantic topic networks}.{\BBCQ}
\newblock
\BIn{} \APACrefbtitle {The Semantic Web-ISWC 2015: 14th International Semantic
  Web Conference, Bethlehem, PA, USA, October 11-15, 2015, Proceedings, Part I
  14} {The semantic web-iswc 2015: 14th international semantic web conference,
  bethlehem, pa, usa, october 11-15, 2015, proceedings, part i 14}\ (\BPGS\
  408--424).
\newblock
\begin{APACrefDOI} \doi{10.1007/978-3-319-25007-6_24} \end{APACrefDOI}
\PrintBackRefs{\CurrentBib}

\bibitem [\protect \citeauthoryear {%
Osborne%
, Motta%
\BCBL {}\ \BBA {} Mulholland%
}{%
Osborne%
\ \protect \BOthers {.}}{%
{\protect \APACyear {2013}}%
}]{%
osborne2013}
\APACinsertmetastar {%
osborne2013}%
\begin{APACrefauthors}%
Osborne, F.%
, Motta, E.%
\BCBL {}\ \BBA {} Mulholland, P.%
\end{APACrefauthors}%
\unskip\
\newblock
\APACrefYearMonthDay{2013}{}{}.
\newblock
{\BBOQ}\APACrefatitle {Exploring Scholarly Data with Rexplore} {Exploring
  scholarly data with rexplore}.{\BBCQ}
\newblock
\BIn{} H.~Alani\ \BOthers {.}\ (\BEDS), \APACrefbtitle {The Semantic Web --
  ISWC 2013} {The semantic web -- iswc 2013}\ (\BPGS\ 460--477).
\newblock
\APACaddressPublisher{Berlin, Heidelberg}{Springer Berlin Heidelberg}.
\newblock
\begin{APACrefDOI} \doi{10.1007/978-3-642-41335-3_29} \end{APACrefDOI}
\PrintBackRefs{\CurrentBib}

\bibitem [\protect \citeauthoryear {%
Osborne%
, Muccini%
, Lago%
\BCBL {}\ \BBA {} Motta%
}{%
Osborne%
\ \protect \BOthers {.}}{%
{\protect \APACyear {2019}}%
}]{%
osborne2019reducing}
\APACinsertmetastar {%
osborne2019reducing}%
\begin{APACrefauthors}%
Osborne, F.%
, Muccini, H.%
, Lago, P.%
\BCBL {}\ \BBA {} Motta, E.%
\end{APACrefauthors}%
\unskip\
\newblock
\APACrefYearMonthDay{2019}{}{}.
\newblock
{\BBOQ}\APACrefatitle {Reducing the effort for systematic reviews in software
  engineering} {Reducing the effort for systematic reviews in software
  engineering}.{\BBCQ}
\newblock
\APACjournalVolNumPages{Data Science}{2}{1-2}{311--340}.
\newblock
\begin{APACrefDOI} \doi{10.3233/DS-190019} \end{APACrefDOI}
\PrintBackRefs{\CurrentBib}

\bibitem [\protect \citeauthoryear {%
Ovalle-Perandones%
, Gorraiz%
, Wieland%
, Gumpenberger%
\BCBL {}\ \BBA {} Olmeda-G{\'o}mez%
}{%
Ovalle-Perandones%
\ \protect \BOthers {.}}{%
{\protect \APACyear {2013}}%
}]{%
ovalle2013influence}
\APACinsertmetastar {%
ovalle2013influence}%
\begin{APACrefauthors}%
Ovalle-Perandones, M\BHBI A.%
, Gorraiz, J.%
, Wieland, M.%
, Gumpenberger, C.%
\BCBL {}\ \BBA {} Olmeda-G{\'o}mez, C.%
\end{APACrefauthors}%
\unskip\
\newblock
\APACrefYearMonthDay{2013}{}{}.
\newblock
{\BBOQ}\APACrefatitle {The influence of European Framework Programmes on
  scientific collaboration in nanotechnology} {The influence of european
  framework programmes on scientific collaboration in nanotechnology}.{\BBCQ}
\newblock
\APACjournalVolNumPages{Scientometrics}{97}{}{59--74}.
\newblock
\begin{APACrefDOI} \doi{10.1007/s11192-013-1028-2} \end{APACrefDOI}
\PrintBackRefs{\CurrentBib}

\bibitem [\protect \citeauthoryear {%
Peng%
, Xia%
, Naseriparsa%
\BCBL {}\ \BBA {} Osborne%
}{%
Peng%
\ \protect \BOthers {.}}{%
{\protect \APACyear {2023}}%
}]{%
peng2023knowledge}
\APACinsertmetastar {%
peng2023knowledge}%
\begin{APACrefauthors}%
Peng, C.%
, Xia, F.%
, Naseriparsa, M.%
\BCBL {}\ \BBA {} Osborne, F.%
\end{APACrefauthors}%
\unskip\
\newblock
\APACrefYearMonthDay{2023}{}{}.
\newblock
{\BBOQ}\APACrefatitle {Knowledge graphs: Opportunities and challenges}
  {Knowledge graphs: Opportunities and challenges}.{\BBCQ}
\newblock
\APACjournalVolNumPages{Artificial Intelligence Review}{56}{11}{13071--13102}.
\newblock
\begin{APACrefDOI} \doi{10.1007/s10462-023-10465-9} \end{APACrefDOI}
\PrintBackRefs{\CurrentBib}

\bibitem [\protect \citeauthoryear {%
Pisanelli%
, Gangemi%
, Battaglia%
\BCBL {}\ \BBA {} Catenacci%
}{%
Pisanelli%
\ \protect \BOthers {.}}{%
{\protect \APACyear {2004}}%
}]{%
pisanelli2004coping}
\APACinsertmetastar {%
pisanelli2004coping}%
\begin{APACrefauthors}%
Pisanelli, D\BPBI M.%
, Gangemi, A.%
, Battaglia, M.%
\BCBL {}\ \BBA {} Catenacci, C.%
\end{APACrefauthors}%
\unskip\
\newblock
\APACrefYearMonthDay{2004}{}{}.
\newblock
{\BBOQ}\APACrefatitle {Coping with medical polysemy in the semantic web: the
  role of ontologies} {Coping with medical polysemy in the semantic web: the
  role of ontologies}.{\BBCQ}
\newblock
\BIn{} \APACrefbtitle {MEDINFO 2004} {Medinfo 2004}\ (\BPGS\ 416--419).
\newblock
\begin{APACrefDOI} \doi{10.3233/978-1-60750-949-3-416} \end{APACrefDOI}
\PrintBackRefs{\CurrentBib}

\bibitem [\protect \citeauthoryear {%
Pisu%
\ \protect \BOthers {.}}{%
Pisu%
\ \protect \BOthers {.}}{%
{\protect \APACyear {2024}}%
}]{%
pisu2024w}
\APACinsertmetastar {%
pisu2024w}%
\begin{APACrefauthors}%
Pisu, A.%
, Pompianu, L.%
, Salatino, A.%
, Osborne, F.%
, Riboni, D.%
, Motta, E.%
\BCBL {}\ \BBA {} Reforgiato~Recupero, D.%
\end{APACrefauthors}%
\unskip\
\newblock
\APACrefYearMonthDay{2024}{}{}.
\newblock
{\BBOQ}\APACrefatitle {Leveraging Language Models for Generating Ontologies of
  Research Topics} {Leveraging language models for generating ontologies of
  research topics}.{\BBCQ}
\newblock
\BIn{} \APACrefbtitle {Text2KG @ ESWC 2024.} {Text2kg @ eswc 2024.}
\newblock
\APACaddressPublisher{}{CEUR}.
\newblock
\begin{APACrefURL} \url{https://ceur-ws.org/Vol-3747/text2kg_paper6.pdf}
  \end{APACrefURL}
\PrintBackRefs{\CurrentBib}

\bibitem [\protect \citeauthoryear {%
Plate%
}{%
Plate%
}{%
{\protect \APACyear {1966}}%
}]{%
plate1966}
\APACinsertmetastar {%
plate1966}%
\begin{APACrefauthors}%
Plate, K\BPBI H.%
\end{APACrefauthors}%
\unskip\
\newblock
\APACrefYearMonthDay{1966}{}{}.
\newblock
{\BBOQ}\APACrefatitle {Library Classification for Environmental Science}
  {Library classification for environmental science}.{\BBCQ}
\newblock
\APACjournalVolNumPages{Journal (Water Pollution Control
  Federation)}{38}{4}{580--584}.
\newblock
\begin{APACrefURL} \url{http://www.jstor.org/stable/25035533} \end{APACrefURL}
\PrintBackRefs{\CurrentBib}

\bibitem [\protect \citeauthoryear {%
Pranckutė%
}{%
Pranckutė%
}{%
{\protect \APACyear {2021}}%
}]{%
publications9010012}
\APACinsertmetastar {%
publications9010012}%
\begin{APACrefauthors}%
Pranckutė, R.%
\end{APACrefauthors}%
\unskip\
\newblock
\APACrefYearMonthDay{2021}{}{}.
\newblock
{\BBOQ}\APACrefatitle {Web of Science (WoS) and Scopus: The Titans of
  Bibliographic Information in Today’s Academic World} {Web of science (wos)
  and scopus: The titans of bibliographic information in today’s academic
  world}.{\BBCQ}
\newblock
\APACjournalVolNumPages{Publications}{9}{1}{}.
\newblock
\begin{APACrefURL} \url{https://www.mdpi.com/2304-6775/9/1/12} \end{APACrefURL}
\newblock
\begin{APACrefDOI} \doi{10.3390/publications9010012} \end{APACrefDOI}
\PrintBackRefs{\CurrentBib}

\bibitem [\protect \citeauthoryear {%
Qiu%
, Zhao%
, Yang%
\BCBL {}\ \BBA {} Dong%
}{%
Qiu%
\ \protect \BOthers {.}}{%
{\protect \APACyear {2017}}%
}]{%
qiu2017}
\APACinsertmetastar {%
qiu2017}%
\begin{APACrefauthors}%
Qiu, J.%
, Zhao, R.%
, Yang, S.%
\BCBL {}\ \BBA {} Dong, K.%
\end{APACrefauthors}%
\unskip\
\newblock
\APACrefYear{2017}.
\newblock
\APACrefbtitle {Informetrics} {Informetrics}.
\newblock
\APACaddressPublisher{{Singapore}}{{Springer Singapore}}.
\newblock
\begin{APACrefDOI} \doi{10.1007/978-981-10-4032-0} \end{APACrefDOI}
\PrintBackRefs{\CurrentBib}

\bibitem [\protect \citeauthoryear {%
Raad%
\ \BBA {} Cruz%
}{%
Raad%
\ \BBA {} Cruz%
}{%
{\protect \APACyear {2015}}%
}]{%
Raad2015}
\APACinsertmetastar {%
Raad2015}%
\begin{APACrefauthors}%
Raad, J.%
\BCBT {}\ \BBA {} Cruz, C.%
\end{APACrefauthors}%
\unskip\
\newblock
\APACrefYearMonthDay{2015}{}{}.
\newblock
{\BBOQ}\APACrefatitle {A Survey on Ontology Evaluation Methods} {A survey on
  ontology evaluation methods}.{\BBCQ}
\newblock
\BIn{} \APACrefbtitle {Proceedings of the 7th International Joint Conference on
  Knowledge Discovery, Knowledge Engineering and Knowledge Management - KEOD,
  (IC3K 2015)} {Proceedings of the 7th international joint conference on
  knowledge discovery, knowledge engineering and knowledge management - keod,
  (ic3k 2015)}\ (\BPG~179-186).
\newblock
\APACaddressPublisher{}{SciTePress}.
\newblock
\begin{APACrefDOI} \doi{10.5220/0005591001790186} \end{APACrefDOI}
\PrintBackRefs{\CurrentBib}

\bibitem [\protect \citeauthoryear {%
Rasch%
}{%
Rasch%
}{%
{\protect \APACyear {1987}}%
}]{%
rasch1987nature}
\APACinsertmetastar {%
rasch1987nature}%
\begin{APACrefauthors}%
Rasch, R\BPBI F.%
\end{APACrefauthors}%
\unskip\
\newblock
\APACrefYearMonthDay{1987}{}{}.
\newblock
{\BBOQ}\APACrefatitle {The nature of taxonomy} {The nature of taxonomy}.{\BBCQ}
\newblock
\APACjournalVolNumPages{Image: Journal of Nursing
  Scholarship}{19}{3}{147--149}.
\PrintBackRefs{\CurrentBib}

\bibitem [\protect \citeauthoryear {%
Revenko%
, Breit%
, Mahmoud%
, Szabo%
\BCBL {}\ \BBA {} Knap%
}{%
Revenko%
\ \protect \BOthers {.}}{%
{\protect \APACyear {2024}}%
}]{%
Revenko2024}
\APACinsertmetastar {%
Revenko2024}%
\begin{APACrefauthors}%
Revenko, A.%
, Breit, A.%
, Mahmoud, S.%
, Szabo, M.%
\BCBL {}\ \BBA {} Knap, T.%
\end{APACrefauthors}%
\unskip\
\newblock
\APACrefYearMonthDay{2024}{}{}.
\newblock
{\BBOQ}\APACrefatitle {Investigate the Impact of Contextual Information on LLMs
  for Taxonomy Expansion} {Investigate the impact of contextual information on
  llms for taxonomy expansion}.{\BBCQ}
\newblock
\BIn{} \APACrefbtitle {Knowledge Graphs in the Time of Language Models and
  Neuro-Symbolic AI.} {Knowledge graphs in the time of language models and
  neuro-symbolic ai.}
\newblock
\APACaddressPublisher{}{IOS Press}.
\newblock
\begin{APACrefDOI} \doi{10.3233/SSW240005} \end{APACrefDOI}
\PrintBackRefs{\CurrentBib}

\bibitem [\protect \citeauthoryear {%
Reymond%
}{%
Reymond%
}{%
{\protect \APACyear {2020}}%
}]{%
reymond2020patents}
\APACinsertmetastar {%
reymond2020patents}%
\begin{APACrefauthors}%
Reymond, D.%
\end{APACrefauthors}%
\unskip\
\newblock
\APACrefYearMonthDay{2020}{}{}.
\newblock
{\BBOQ}\APACrefatitle {Patents information for humanities research: Could there
  be something?} {Patents information for humanities research: Could there be
  something?}{\BBCQ}
\newblock
\APACjournalVolNumPages{Iberoamerican Journal of Science Measurement and
  Communication}{1}{1}{6}.
\newblock
\begin{APACrefDOI} \doi{10.47909/ijsmc.02} \end{APACrefDOI}
\PrintBackRefs{\CurrentBib}

\bibitem [\protect \citeauthoryear {%
A.~Salatino%
, Angioni%
, Osborne%
, Recupero%
\BCBL {}\ \BBA {} Motta%
}{%
A.~Salatino%
\ \protect \BOthers {.}}{%
{\protect \APACyear {2023}}%
}]{%
salatino2023}
\APACinsertmetastar {%
salatino2023}%
\begin{APACrefauthors}%
Salatino, A.%
, Angioni, S.%
, Osborne, F.%
, Recupero, D\BPBI R.%
\BCBL {}\ \BBA {} Motta, E.%
\end{APACrefauthors}%
\unskip\
\newblock
\APACrefYearMonthDay{2023}{}{}.
\newblock
{\BBOQ}\APACrefatitle {Diversity of Expertise is Key to Scientific Impact: a
  Large-Scale Analysis in the Field of Computer Science} {Diversity of
  expertise is key to scientific impact: a large-scale analysis in the field of
  computer science}.{\BBCQ}.
\newblock
\begin{APACrefURL}
  \url{https://dapp.orvium.io/deposits/6442f3fd947802668eee976c/view}
  \end{APACrefURL}
\newblock
\begin{APACrefDOI} \doi{10.55835/6442f3fd947802668eee976c} \end{APACrefDOI}
\PrintBackRefs{\CurrentBib}

\bibitem [\protect \citeauthoryear {%
A.~Salatino%
, Osborne%
\BCBL {}\ \BBA {} Motta%
}{%
A.~Salatino%
\ \protect \BOthers {.}}{%
{\protect \APACyear {2021}}%
}]{%
salatino2021}
\APACinsertmetastar {%
salatino2021}%
\begin{APACrefauthors}%
Salatino, A.%
, Osborne, F.%
\BCBL {}\ \BBA {} Motta, E.%
\end{APACrefauthors}%
\unskip\
\newblock
\APACrefYearMonthDay{2021}{}{}.
\newblock
{\BBOQ}\APACrefatitle {CSO Classifier 3.0: a scalable unsupervised method for
  classifying documents in terms of research topics} {Cso classifier 3.0: a
  scalable unsupervised method for classifying documents in terms of research
  topics}.{\BBCQ}
\newblock
\APACjournalVolNumPages{International Journal on Digital Libraries}{}{}{}.
\newblock
\begin{APACrefURL} \url{https://doi.org/10.1007/s00799-021-00305-y}
  \end{APACrefURL}
\newblock
\begin{APACrefDOI} \doi{10.1007/s00799-021-00305-y} \end{APACrefDOI}
\PrintBackRefs{\CurrentBib}

\bibitem [\protect \citeauthoryear {%
A\BPBI A.~Salatino%
}{%
A\BPBI A.~Salatino%
}{%
{\protect \APACyear {2019}}%
}]{%
salatino2019early}
\APACinsertmetastar {%
salatino2019early}%
\begin{APACrefauthors}%
Salatino, A\BPBI A.%
\end{APACrefauthors}%
\unskip\
\newblock
\APACrefYearMonthDay{2019}{}{}.
\newblock
\APACrefbtitle {Early Detection of Research Trends.} {Early detection of
  research trends.}
\newblock
\begin{APACrefURL} \url{http://oro.open.ac.uk/67224/} \end{APACrefURL}
\PrintBackRefs{\CurrentBib}

\bibitem [\protect \citeauthoryear {%
A\BPBI A.~Salatino%
, Osborne%
, Birukou%
\BCBL {}\ \BBA {} Motta%
}{%
A\BPBI A.~Salatino%
\ \protect \BOthers {.}}{%
{\protect \APACyear {2019}}%
}]{%
salatino2019improving}
\APACinsertmetastar {%
salatino2019improving}%
\begin{APACrefauthors}%
Salatino, A\BPBI A.%
, Osborne, F.%
, Birukou, A.%
\BCBL {}\ \BBA {} Motta, E.%
\end{APACrefauthors}%
\unskip\
\newblock
\APACrefYearMonthDay{2019}{}{}.
\newblock
{\BBOQ}\APACrefatitle {Improving Editorial Workflow and Metadata Quality at
  Springer Nature} {Improving editorial workflow and metadata quality at
  springer nature}.{\BBCQ}
\newblock
\BIn{} \APACrefbtitle {The Semantic Web -- ISWC 2019} {The semantic web -- iswc
  2019}\ (\BPGS\ 507--525).
\newblock
\APACaddressPublisher{Cham}{Springer International Publishing}.
\newblock
\begin{APACrefDOI} \doi{10.1007/978-3-030-30796-7_31} \end{APACrefDOI}
\PrintBackRefs{\CurrentBib}

\bibitem [\protect \citeauthoryear {%
A\BPBI A.~Salatino%
, Osborne%
\BCBL {}\ \BBA {} Motta%
}{%
A\BPBI A.~Salatino%
, Osborne%
\BCBL {}\ \BBA {} Motta%
}{%
{\protect \APACyear {2018}}%
}]{%
salatino2018augur}
\APACinsertmetastar {%
salatino2018augur}%
\begin{APACrefauthors}%
Salatino, A\BPBI A.%
, Osborne, F.%
\BCBL {}\ \BBA {} Motta, E.%
\end{APACrefauthors}%
\unskip\
\newblock
\APACrefYearMonthDay{2018}{}{}.
\newblock
{\BBOQ}\APACrefatitle {AUGUR: forecasting the emergence of new research topics}
  {Augur: forecasting the emergence of new research topics}.{\BBCQ}
\newblock
\BIn{} \APACrefbtitle {Proceedings of the 18th ACM/IEEE on joint conference on
  digital libraries} {Proceedings of the 18th acm/ieee on joint conference on
  digital libraries}\ (\BPGS\ 303--312).
\newblock
\begin{APACrefDOI} \doi{10.1145/3197026.3197052} \end{APACrefDOI}
\PrintBackRefs{\CurrentBib}

\bibitem [\protect \citeauthoryear {%
A\BPBI A.~Salatino%
, Osborne%
\BCBL {}\ \BBA {} Motta%
}{%
A\BPBI A.~Salatino%
, Osborne%
\BCBL {}\ \BBA {} Motta%
}{%
{\protect \APACyear {2020}}%
}]{%
salatino2020ontology}
\APACinsertmetastar {%
salatino2020ontology}%
\begin{APACrefauthors}%
Salatino, A\BPBI A.%
, Osborne, F.%
\BCBL {}\ \BBA {} Motta, E.%
\end{APACrefauthors}%
\unskip\
\newblock
\APACrefYearMonthDay{2020}{}{}.
\newblock
{\BBOQ}\APACrefatitle {Ontology Extraction and Usage in the Scholarly Knowledge
  Domain} {Ontology extraction and usage in the scholarly knowledge
  domain}.{\BBCQ}
\newblock
\BIn{} \APACrefbtitle {Applications and Practices in Ontology Design,
  Extraction, and Reasoning} {Applications and practices in ontology design,
  extraction, and reasoning}\ (\BPGS\ 91--106).
\newblock
\APACaddressPublisher{}{Ios Press}.
\newblock
\begin{APACrefDOI} \doi{10.3233/SSW200037} \end{APACrefDOI}
\PrintBackRefs{\CurrentBib}

\bibitem [\protect \citeauthoryear {%
A\BPBI A.~Salatino%
, Thanapalasingam%
\BCBL {}\ \protect \BOthers {.}}{%
A\BPBI A.~Salatino%
, Thanapalasingam%
\BCBL {}\ \protect \BOthers {.}}{%
{\protect \APACyear {2020}}%
}]{%
salatino2020cso}
\APACinsertmetastar {%
salatino2020cso}%
\begin{APACrefauthors}%
Salatino, A\BPBI A.%
, Thanapalasingam, T.%
, Mannocci, A.%
, Birukou, A.%
, Osborne, F.%
\BCBL {}\ \BBA {} Motta, E.%
\end{APACrefauthors}%
\unskip\
\newblock
\APACrefYearMonthDay{2020}{07}{}.
\newblock
{\BBOQ}\APACrefatitle {{The Computer Science Ontology: A Comprehensive
  Automatically-Generated Taxonomy of Research Areas}} {{The Computer Science
  Ontology: A Comprehensive Automatically-Generated Taxonomy of Research
  Areas}}.{\BBCQ}
\newblock
\APACjournalVolNumPages{Data Intelligence}{2}{3}{379-416}.
\newblock
\begin{APACrefURL} \url{https://doi.org/10.1162/dint\_a\_00055}
  \end{APACrefURL}
\newblock
\begin{APACrefDOI} \doi{10.1162/dint_a_00055} \end{APACrefDOI}
\PrintBackRefs{\CurrentBib}

\bibitem [\protect \citeauthoryear {%
A\BPBI A.~Salatino%
, Thanapalasingam%
, Mannocci%
, Osborne%
\BCBL {}\ \BBA {} Motta%
}{%
A\BPBI A.~Salatino%
, Thanapalasingam%
\BCBL {}\ \protect \BOthers {.}}{%
{\protect \APACyear {2018}}%
}]{%
salatino2018}
\APACinsertmetastar {%
salatino2018}%
\begin{APACrefauthors}%
Salatino, A\BPBI A.%
, Thanapalasingam, T.%
, Mannocci, A.%
, Osborne, F.%
\BCBL {}\ \BBA {} Motta, E.%
\end{APACrefauthors}%
\unskip\
\newblock
\APACrefYearMonthDay{2018}{}{}.
\newblock
{\BBOQ}\APACrefatitle {The Computer Science Ontology: A Large-Scale Taxonomy of
  Research Areas} {The computer science ontology: A large-scale taxonomy of
  research areas}.{\BBCQ}
\newblock
\BIn{} D.~Vrande{\v{c}}i{\'{c}}\ \BOthers {.}\ (\BEDS), \APACrefbtitle {The
  Semantic Web -- ISWC 2018} {The semantic web -- iswc 2018}\ (\BPGS\
  187--205).
\newblock
\APACaddressPublisher{Cham}{Springer International Publishing}.
\newblock
\begin{APACrefDOI} \doi{10.1007/978-3-030-00668-6_12} \end{APACrefDOI}
\PrintBackRefs{\CurrentBib}

\bibitem [\protect \citeauthoryear {%
Salvadores%
, Alexander%
, Musen%
\BCBL {}\ \BBA {} Noy%
}{%
Salvadores%
\ \protect \BOthers {.}}{%
{\protect \APACyear {2013}}%
}]{%
salvadores2013bioportal}
\APACinsertmetastar {%
salvadores2013bioportal}%
\begin{APACrefauthors}%
Salvadores, M.%
, Alexander, P\BPBI R.%
, Musen, M\BPBI A.%
\BCBL {}\ \BBA {} Noy, N\BPBI F.%
\end{APACrefauthors}%
\unskip\
\newblock
\APACrefYearMonthDay{2013}{}{}.
\newblock
{\BBOQ}\APACrefatitle {BioPortal as a dataset of linked biomedical ontologies
  and terminologies in RDF} {Bioportal as a dataset of linked biomedical
  ontologies and terminologies in rdf}.{\BBCQ}
\newblock
\APACjournalVolNumPages{Semantic web}{4}{3}{277--284}.
\PrintBackRefs{\CurrentBib}

\bibitem [\protect \citeauthoryear {%
Savova%
\ \protect \BOthers {.}}{%
Savova%
\ \protect \BOthers {.}}{%
{\protect \APACyear {2010}}%
}]{%
savova2010}
\APACinsertmetastar {%
savova2010}%
\begin{APACrefauthors}%
Savova, G\BPBI K.%
, Masanz, J\BPBI J.%
, Ogren, P\BPBI V.%
, Zheng, J.%
, Sohn, S.%
, Kipper-Schuler, K\BPBI C.%
\BCBL {}\ \BBA {} Chute, C\BPBI G.%
\end{APACrefauthors}%
\unskip\
\newblock
\APACrefYearMonthDay{2010}{09}{}.
\newblock
{\BBOQ}\APACrefatitle {{Mayo clinical Text Analysis and Knowledge Extraction
  System (cTAKES): architecture, component evaluation and applications}} {{Mayo
  clinical Text Analysis and Knowledge Extraction System (cTAKES):
  architecture, component evaluation and applications}}.{\BBCQ}
\newblock
\APACjournalVolNumPages{Journal of the American Medical Informatics
  Association}{17}{5}{507-513}.
\newblock
\begin{APACrefURL} \url{https://doi.org/10.1136/jamia.2009.001560}
  \end{APACrefURL}
\newblock
\begin{APACrefDOI} \doi{10.1136/jamia.2009.001560} \end{APACrefDOI}
\PrintBackRefs{\CurrentBib}

\bibitem [\protect \citeauthoryear {%
Schandl%
\ \BBA {} Blumauer%
}{%
Schandl%
\ \BBA {} Blumauer%
}{%
{\protect \APACyear {2010}}%
}]{%
schandl2010}
\APACinsertmetastar {%
schandl2010}%
\begin{APACrefauthors}%
Schandl, T.%
\BCBT {}\ \BBA {} Blumauer, A.%
\end{APACrefauthors}%
\unskip\
\newblock
\APACrefYearMonthDay{2010}{}{}.
\newblock
{\BBOQ}\APACrefatitle {PoolParty: SKOS Thesaurus Management Utilizing Linked
  Data} {Poolparty: Skos thesaurus management utilizing linked data}.{\BBCQ}
\newblock
\BIn{} L.~Aroyo\ \BOthers {.}\ (\BEDS), \APACrefbtitle {The Semantic Web:
  Research and Applications} {The semantic web: Research and applications}\
  (\BPGS\ 421--425).
\newblock
\APACaddressPublisher{Berlin, Heidelberg}{Springer Berlin Heidelberg}.
\newblock
\begin{APACrefDOI} \doi{10.1007/978-3-642-13489-0_36} \end{APACrefDOI}
\PrintBackRefs{\CurrentBib}

\bibitem [\protect \citeauthoryear {%
Scharnhorst%
, B{\"o}rner%
\BCBL {}\ \BBA {} {van den Besselaar}%
}{%
Scharnhorst%
\ \protect \BOthers {.}}{%
{\protect \APACyear {2012}}%
}]{%
scharnhorst2012}
\APACinsertmetastar {%
scharnhorst2012}%
\begin{APACrefauthors}%
Scharnhorst, A.%
, B{\"o}rner, K.%
\BCBL {}\ \BBA {} {van den Besselaar}, P.%
\end{APACrefauthors}%
\ (\BEDS).
\unskip\
\newblock
\APACrefYear{2012}.
\newblock
\APACrefbtitle {Models of {{Science Dynamics}}} {Models of {{Science
  Dynamics}}}\ (\BVOL~53).
\newblock
\APACaddressPublisher{{Berlin, Heidelberg}}{{Springer Berlin Heidelberg}}.
\newblock
\begin{APACrefDOI} \doi{10.1007/978-3-642-23068-4} \end{APACrefDOI}
\PrintBackRefs{\CurrentBib}

\bibitem [\protect \citeauthoryear {%
Shang%
, Zhang%
, Liu%
, Li%
\BCBL {}\ \BBA {} Han%
}{%
Shang%
\ \protect \BOthers {.}}{%
{\protect \APACyear {2020}}%
}]{%
shang2020nettaxo}
\APACinsertmetastar {%
shang2020nettaxo}%
\begin{APACrefauthors}%
Shang, J.%
, Zhang, X.%
, Liu, L.%
, Li, S.%
\BCBL {}\ \BBA {} Han, J.%
\end{APACrefauthors}%
\unskip\
\newblock
\APACrefYearMonthDay{2020}{}{}.
\newblock
{\BBOQ}\APACrefatitle {Nettaxo: Automated topic taxonomy construction from
  text-rich network} {Nettaxo: Automated topic taxonomy construction from
  text-rich network}.{\BBCQ}
\newblock
\BIn{} \APACrefbtitle {Proceedings of the web conference 2020} {Proceedings of
  the web conference 2020}\ (\BPGS\ 1908--1919).
\newblock
\begin{APACrefDOI} \doi{10.1145/3366423.3380259} \end{APACrefDOI}
\PrintBackRefs{\CurrentBib}

\bibitem [\protect \citeauthoryear {%
Shi%
, Maly%
, Zeil%
\BCBL {}\ \BBA {} Zubair%
}{%
Shi%
\ \protect \BOthers {.}}{%
{\protect \APACyear {2011}}%
}]{%
shi2011comparison}
\APACinsertmetastar {%
shi2011comparison}%
\begin{APACrefauthors}%
Shi, H.%
, Maly, K.%
, Zeil, S.%
\BCBL {}\ \BBA {} Zubair, M.%
\end{APACrefauthors}%
\unskip\
\newblock
\APACrefYearMonthDay{2011}{}{}.
\newblock
{\BBOQ}\APACrefatitle {Comparison of ontology reasoning systems using custom
  rules} {Comparison of ontology reasoning systems using custom rules}.{\BBCQ}
\newblock
\BIn{} \APACrefbtitle {Proceedings of the international conference on web
  intelligence, mining and semantics} {Proceedings of the international
  conference on web intelligence, mining and semantics}\ (\BPGS\ 1--9).
\newblock
\begin{APACrefDOI} \doi{10.1145/1988688.1988708} \end{APACrefDOI}
\PrintBackRefs{\CurrentBib}

\bibitem [\protect \citeauthoryear {%
Shiri%
, Revie%
\BCBL {}\ \BBA {} Chowdhury%
}{%
Shiri%
\ \protect \BOthers {.}}{%
{\protect \APACyear {2002}}%
}]{%
shiri2002thesaurus}
\APACinsertmetastar {%
shiri2002thesaurus}%
\begin{APACrefauthors}%
Shiri, A\BPBI A.%
, Revie, C\BPBI W.%
\BCBL {}\ \BBA {} Chowdhury, G.%
\end{APACrefauthors}%
\unskip\
\newblock
\APACrefYearMonthDay{2002}{}{}.
\newblock
{\BBOQ}\APACrefatitle {Thesaurus-assisted search term selection and query
  expansion: a review of user-centred studies} {Thesaurus-assisted search term
  selection and query expansion: a review of user-centred studies}.{\BBCQ}
\newblock
\APACjournalVolNumPages{Knowledge organization}{29}{1}{1--19}.
\PrintBackRefs{\CurrentBib}

\bibitem [\protect \citeauthoryear {%
Shvaiko%
\ \BBA {} Euzenat%
}{%
Shvaiko%
\ \BBA {} Euzenat%
}{%
{\protect \APACyear {2008}}%
}]{%
Shvaiko2008}
\APACinsertmetastar {%
Shvaiko2008}%
\begin{APACrefauthors}%
Shvaiko, P.%
\BCBT {}\ \BBA {} Euzenat, J.%
\end{APACrefauthors}%
\unskip\
\newblock
\APACrefYearMonthDay{2008}{}{}.
\newblock
{\BBOQ}\APACrefatitle {Ten Challenges for Ontology Matching} {Ten challenges
  for ontology matching}.{\BBCQ}
\newblock
\BIn{} R.~Meersman\ \BBA {} Z.~Tari\ (\BEDS), \APACrefbtitle {On the Move to
  Meaningful Internet Systems: OTM 2008} {On the move to meaningful internet
  systems: Otm 2008}\ (\BPGS\ 1164--1182).
\newblock
\APACaddressPublisher{Berlin, Heidelberg}{Springer Berlin Heidelberg}.
\newblock
\begin{APACrefDOI} \doi{10.1007/978-3-540-88873-4_18} \end{APACrefDOI}
\PrintBackRefs{\CurrentBib}

\bibitem [\protect \citeauthoryear {%
Sinha%
\ \protect \BOthers {.}}{%
Sinha%
\ \protect \BOthers {.}}{%
{\protect \APACyear {2015}}%
}]{%
sinha2015}
\APACinsertmetastar {%
sinha2015}%
\begin{APACrefauthors}%
Sinha, A.%
, Shen, Z.%
, Song, Y.%
, Ma, H.%
, Eide, D.%
, Hsu, B\BHBI J\BPBI P.%
\BCBL {}\ \BBA {} Wang, K.%
\end{APACrefauthors}%
\unskip\
\newblock
\APACrefYearMonthDay{2015}{}{}.
\newblock
{\BBOQ}\APACrefatitle {An Overview of Microsoft Academic Service (MAS) and
  Applications} {An overview of microsoft academic service (mas) and
  applications}.{\BBCQ}
\newblock
\BIn{} \APACrefbtitle {Proceedings of the 24th International Conference on
  World Wide Web} {Proceedings of the 24th international conference on world
  wide web}\ (\BPG~243–246).
\newblock
\APACaddressPublisher{New York, NY, USA}{Association for Computing Machinery}.
\newblock
\begin{APACrefURL} \url{https://doi.org/10.1145/2740908.2742839}
  \end{APACrefURL}
\newblock
\begin{APACrefDOI} \doi{10.1145/2740908.2742839} \end{APACrefDOI}
\PrintBackRefs{\CurrentBib}

\bibitem [\protect \citeauthoryear {%
Sj{\"o}g{\aa}rde%
\ \BBA {} Didegah%
}{%
Sj{\"o}g{\aa}rde%
\ \BBA {} Didegah%
}{%
{\protect \APACyear {2022}}%
}]{%
sjogaarde2022association}
\APACinsertmetastar {%
sjogaarde2022association}%
\begin{APACrefauthors}%
Sj{\"o}g{\aa}rde, P.%
\BCBT {}\ \BBA {} Didegah, F.%
\end{APACrefauthors}%
\unskip\
\newblock
\APACrefYearMonthDay{2022}{}{}.
\newblock
{\BBOQ}\APACrefatitle {The association between topic growth and citation impact
  of research publications} {The association between topic growth and citation
  impact of research publications}.{\BBCQ}
\newblock
\APACjournalVolNumPages{Scientometrics}{127}{4}{1903--1921}.
\newblock
\begin{APACrefDOI} \doi{10.1007/s11192-022-04293-x} \end{APACrefDOI}
\PrintBackRefs{\CurrentBib}

\bibitem [\protect \citeauthoryear {%
Sjögårde%
\ \BBA {} Ahlgren%
}{%
Sjögårde%
\ \BBA {} Ahlgren%
}{%
{\protect \APACyear {2020}}%
}]{%
Sjogarde2020}
\APACinsertmetastar {%
Sjogarde2020}%
\begin{APACrefauthors}%
Sjögårde, P.%
\BCBT {}\ \BBA {} Ahlgren, P.%
\end{APACrefauthors}%
\unskip\
\newblock
\APACrefYearMonthDay{2020}{02}{}.
\newblock
{\BBOQ}\APACrefatitle {{Granularity of algorithmically constructed
  publication-level classifications of research publications: Identification of
  specialties}} {{Granularity of algorithmically constructed publication-level
  classifications of research publications: Identification of
  specialties}}.{\BBCQ}
\newblock
\APACjournalVolNumPages{Quantitative Science Studies}{1}{1}{207-238}.
\newblock
\begin{APACrefURL} \url{https://doi.org/10.1162/qss\_a\_00004} \end{APACrefURL}
\newblock
\begin{APACrefDOI} \doi{10.1162/qss_a_00004} \end{APACrefDOI}
\PrintBackRefs{\CurrentBib}

\bibitem [\protect \citeauthoryear {%
Slater%
, Gkoutos%
\BCBL {}\ \BBA {} Hoehndorf%
}{%
Slater%
\ \protect \BOthers {.}}{%
{\protect \APACyear {2020}}%
}]{%
Slater2020}
\APACinsertmetastar {%
Slater2020}%
\begin{APACrefauthors}%
Slater, L\BPBI T.%
, Gkoutos, G\BPBI V.%
\BCBL {}\ \BBA {} Hoehndorf, R.%
\end{APACrefauthors}%
\unskip\
\newblock
\APACrefYearMonthDay{2020}{}{}.
\newblock
{\BBOQ}\APACrefatitle {{Towards semantic interoperability: finding and
  repairing hidden contradictions in biomedical ontologies}} {{Towards semantic
  interoperability: finding and repairing hidden contradictions in biomedical
  ontologies}}.{\BBCQ}
\newblock
\APACjournalVolNumPages{BMC Medical Informatics and Decision
  Making}{20}{10}{311}.
\newblock
\begin{APACrefURL} \url{https://doi.org/10.1186/s12911-020-01336-2}
  \end{APACrefURL}
\newblock
\begin{APACrefDOI} \doi{10.1186/s12911-020-01336-2} \end{APACrefDOI}
\PrintBackRefs{\CurrentBib}

\bibitem [\protect \citeauthoryear {%
Soldaini%
\ \BBA {} Goharian%
}{%
Soldaini%
\ \BBA {} Goharian%
}{%
{\protect \APACyear {2016}}%
}]{%
soldaini2016quickumls}
\APACinsertmetastar {%
soldaini2016quickumls}%
\begin{APACrefauthors}%
Soldaini, L.%
\BCBT {}\ \BBA {} Goharian, N.%
\end{APACrefauthors}%
\unskip\
\newblock
\APACrefYearMonthDay{2016}{}{}.
\newblock
{\BBOQ}\APACrefatitle {Quickumls: a fast, unsupervised approach for medical
  concept extraction} {Quickumls: a fast, unsupervised approach for medical
  concept extraction}.{\BBCQ}
\newblock
\BIn{} \APACrefbtitle {MedIR workshop, sigir} {Medir workshop, sigir}\ (\BPGS\
  1--4).
\PrintBackRefs{\CurrentBib}

\bibitem [\protect \citeauthoryear {%
Solimando%
, Jim{\'e}nez-Ruiz%
\BCBL {}\ \BBA {} Guerrini%
}{%
Solimando%
\ \protect \BOthers {.}}{%
{\protect \APACyear {2014}}%
}]{%
Solimando2014}
\APACinsertmetastar {%
Solimando2014}%
\begin{APACrefauthors}%
Solimando, A.%
, Jim{\'e}nez-Ruiz, E.%
\BCBL {}\ \BBA {} Guerrini, G.%
\end{APACrefauthors}%
\unskip\
\newblock
\APACrefYearMonthDay{2014}{}{}.
\newblock
{\BBOQ}\APACrefatitle {Detecting and Correcting Conservativity Principle
  Violations in Ontology-to-Ontology Mappings} {Detecting and correcting
  conservativity principle violations in ontology-to-ontology mappings}.{\BBCQ}
\newblock
\BIn{} P.~Mika\ \BOthers {.}\ (\BEDS), \APACrefbtitle {The Semantic Web -- ISWC
  2014} {The semantic web -- iswc 2014}\ (\BPGS\ 1--16).
\newblock
\APACaddressPublisher{Cham}{Springer International Publishing}.
\newblock
\begin{APACrefDOI} \doi{10.1007/978-3-319-11915-1_1} \end{APACrefDOI}
\PrintBackRefs{\CurrentBib}

\bibitem [\protect \citeauthoryear {%
Stellato%
\ \protect \BOthers {.}}{%
Stellato%
\ \protect \BOthers {.}}{%
{\protect \APACyear {2020}}%
}]{%
stellato2020}
\APACinsertmetastar {%
stellato2020}%
\begin{APACrefauthors}%
Stellato, A.%
, Fiorelli, M.%
, Turbati, A.%
, Lorenzetti, T.%
, van Gemert, W.%
, Dechandon, D.%
\BDBL {}Keizer, J.%
\end{APACrefauthors}%
\unskip\
\newblock
\APACrefYearMonthDay{2020}{{\APACmonth{01}}}{}.
\newblock
{\BBOQ}\APACrefatitle {VocBench 3: A collaborative Semantic Web editor for
  ontologies, thesauri and lexicons} {Vocbench 3: A collaborative semantic web
  editor for ontologies, thesauri and lexicons}.{\BBCQ}
\newblock
\APACjournalVolNumPages{Semant. Web}{11}{5}{855–881}.
\newblock
\begin{APACrefURL} \url{https://doi.org/10.3233/SW-200370} \end{APACrefURL}
\newblock
\begin{APACrefDOI} \doi{10.3233/SW-200370} \end{APACrefDOI}
\PrintBackRefs{\CurrentBib}

\bibitem [\protect \citeauthoryear {%
Stiller%
, Trkulja%
, Biesenbender%
\BCBL {}\ \BBA {} Petras%
}{%
Stiller%
\ \protect \BOthers {.}}{%
{\protect \APACyear {2021}}%
}]{%
stiller2021}
\APACinsertmetastar {%
stiller2021}%
\begin{APACrefauthors}%
Stiller, J.%
, Trkulja, V.%
, Biesenbender, S.%
\BCBL {}\ \BBA {} Petras, V.%
\end{APACrefauthors}%
\unskip\
\newblock
\APACrefYearMonthDay{2021}{}{}.
\newblock
\APACrefbtitle {Entwicklung einer Klassifikation für interdisziplinäre
  Forschungsfelder im Rahmen des Kerndatensatz Forschung: Dokumentation des
  Projekts und der Projektergebnisse.} {Entwicklung einer klassifikation für
  interdisziplinäre forschungsfelder im rahmen des kerndatensatz forschung:
  Dokumentation des projekts und der projektergebnisse.}\
  \APACbVolEdTR{}{\BTR{}}.
\newblock
\APACaddressInstitution{}{Deutsches Zentrum für Hochschul- und
  Wissenschaftsforschung}.
\newblock
\begin{APACrefURL}
  \url{http://web.archive.org/web/20211004184211/https://www.kerndatensatz-forschung.de/docs_ff/2021-04-30_ffk-projekt_dokumentation.pdf}
  \end{APACrefURL}
\PrintBackRefs{\CurrentBib}

\bibitem [\protect \citeauthoryear {%
Sugimoto%
\ \BBA {} Larivi{\`e}re%
}{%
Sugimoto%
\ \BBA {} Larivi{\`e}re%
}{%
{\protect \APACyear {2018}}%
}]{%
sugimoto2018}
\APACinsertmetastar {%
sugimoto2018}%
\begin{APACrefauthors}%
Sugimoto, C\BPBI R.%
\BCBT {}\ \BBA {} Larivi{\`e}re, V.%
\end{APACrefauthors}%
\unskip\
\newblock
\APACrefYear{2018}.
\newblock
\APACrefbtitle {Measuring {{Research}}: What {{Everyone Needs}} to {{Know}}}
  {Measuring {{Research}}: What {{Everyone Needs}} to {{Know}}}\ (\BVOL~68).
\newblock
\APACaddressPublisher{}{{Oxford University Press}}.
\PrintBackRefs{\CurrentBib}

\bibitem [\protect \citeauthoryear {%
Thanapalasingam%
, Osborne%
, Birukou%
\BCBL {}\ \BBA {} Motta%
}{%
Thanapalasingam%
\ \protect \BOthers {.}}{%
{\protect \APACyear {2018}}%
}]{%
thanapalasingam2018}
\APACinsertmetastar {%
thanapalasingam2018}%
\begin{APACrefauthors}%
Thanapalasingam, T.%
, Osborne, F.%
, Birukou, A.%
\BCBL {}\ \BBA {} Motta, E.%
\end{APACrefauthors}%
\unskip\
\newblock
\APACrefYearMonthDay{2018}{}{}.
\newblock
{\BBOQ}\APACrefatitle {Ontology-Based Recommendation of Editorial Products}
  {Ontology-based recommendation of editorial products}.{\BBCQ}
\newblock
\BIn{} D.~Vrande{\v{c}}i{\'{c}}\ \BOthers {.}\ (\BEDS), \APACrefbtitle {The
  Semantic Web -- ISWC 2018} {The semantic web -- iswc 2018}\ (\BPGS\
  341--358).
\newblock
\APACaddressPublisher{Cham}{Springer International Publishing}.
\newblock
\begin{APACrefDOI} \doi{10.1007/978-3-030-00668-6_21} \end{APACrefDOI}
\PrintBackRefs{\CurrentBib}

\bibitem [\protect \citeauthoryear {%
Tudhope%
\ \BBA {} Nielsen%
}{%
Tudhope%
\ \BBA {} Nielsen%
}{%
{\protect \APACyear {2006}}%
}]{%
tudhope2006}
\APACinsertmetastar {%
tudhope2006}%
\begin{APACrefauthors}%
Tudhope, D.%
\BCBT {}\ \BBA {} Nielsen, M\BPBI L.%
\end{APACrefauthors}%
\unskip\
\newblock
\APACrefYearMonthDay{2006}{}{}.
\newblock
{\BBOQ}\APACrefatitle {Introduction to Knowledge Organization Systems and
  Services} {Introduction to knowledge organization systems and
  services}.{\BBCQ}
\newblock
\APACjournalVolNumPages{New Review of Hypermedia and Multimedia}{12}{1}{3--9}.
\newblock
\begin{APACrefURL} \url{https://doi.org/10.1080/13614560600856433}
  \end{APACrefURL}
\newblock
\begin{APACrefDOI} \doi{10.1080/13614560600856433} \end{APACrefDOI}
\PrintBackRefs{\CurrentBib}

\bibitem [\protect \citeauthoryear {%
Tudorache%
, Nyulas%
, Noy%
\BCBL {}\ \BBA {} Musen%
}{%
Tudorache%
\ \protect \BOthers {.}}{%
{\protect \APACyear {2013}}%
}]{%
tudorache2013}
\APACinsertmetastar {%
tudorache2013}%
\begin{APACrefauthors}%
Tudorache, T.%
, Nyulas, C.%
, Noy, N\BPBI F.%
\BCBL {}\ \BBA {} Musen, M\BPBI A.%
\end{APACrefauthors}%
\unskip\
\newblock
\APACrefYearMonthDay{2013}{Jan}{}.
\newblock
{\BBOQ}\APACrefatitle {{{W}eb{P}rotégé: {A} {C}ollaborative {O}ntology
  {E}ditor and {K}nowledge {A}cquisition {T}ool for the {W}eb}}
  {{{W}eb{P}rotégé: {A} {C}ollaborative {O}ntology {E}ditor and {K}nowledge
  {A}cquisition {T}ool for the {W}eb}}.{\BBCQ}
\newblock
\APACjournalVolNumPages{Semant Web}{4}{1}{89--99}.
\PrintBackRefs{\CurrentBib}

\bibitem [\protect \citeauthoryear {%
Vergoulis%
, Chatzopoulos%
, Dalamagas%
\BCBL {}\ \BBA {} Tryfonopoulos%
}{%
Vergoulis%
\ \protect \BOthers {.}}{%
{\protect \APACyear {2020}}%
}]{%
vergoulis2020veto}
\APACinsertmetastar {%
vergoulis2020veto}%
\begin{APACrefauthors}%
Vergoulis, T.%
, Chatzopoulos, S.%
, Dalamagas, T.%
\BCBL {}\ \BBA {} Tryfonopoulos, C.%
\end{APACrefauthors}%
\unskip\
\newblock
\APACrefYearMonthDay{2020}{}{}.
\newblock
{\BBOQ}\APACrefatitle {VeTo: Expert Set Expansion in Academia} {Veto: Expert
  set expansion in academia}.{\BBCQ}
\newblock
\BIn{} M.~Hall, T.~Mer{\v{c}}un, T.~Risse\BCBL {}\ \BBA {} F.~Duchateau\
  (\BEDS), \APACrefbtitle {Digital Libraries for Open Knowledge} {Digital
  libraries for open knowledge}\ (\BPGS\ 48--61).
\newblock
\APACaddressPublisher{Cham}{Springer International Publishing}.
\newblock
\begin{APACrefDOI} \doi{10.1007/978-3-030-54956-5\_4} \end{APACrefDOI}
\PrintBackRefs{\CurrentBib}

\bibitem [\protect \citeauthoryear {%
Vrande{\v{c}}i{\'{c}}%
\ \BBA {} Kr{\"o}tzsch%
}{%
Vrande{\v{c}}i{\'{c}}%
\ \BBA {} Kr{\"o}tzsch%
}{%
{\protect \APACyear {2009}}%
}]{%
vrandecic2009}
\APACinsertmetastar {%
vrandecic2009}%
\begin{APACrefauthors}%
Vrande{\v{c}}i{\'{c}}, D.%
\BCBT {}\ \BBA {} Kr{\"o}tzsch, M.%
\end{APACrefauthors}%
\unskip\
\newblock
\APACrefYearMonthDay{2009}{}{}.
\newblock
{\BBOQ}\APACrefatitle {Semantic MediaWiki} {Semantic mediawiki}.{\BBCQ}
\newblock
\BIn{} J.~Davies, M.~Grobelnik\BCBL {}\ \BBA {} D.~Mladeni{\'{c}}\ (\BEDS),
  \APACrefbtitle {Semantic Knowledge Management: Integrating Ontology
  Management, Knowledge Discovery, and Human Language Technologies} {Semantic
  knowledge management: Integrating ontology management, knowledge discovery,
  and human language technologies}\ (\BPGS\ 171--179).
\newblock
\APACaddressPublisher{Berlin, Heidelberg}{Springer Berlin Heidelberg}.
\newblock
\begin{APACrefURL} \url{https://doi.org/10.1007/978-3-540-88845-1_13}
  \end{APACrefURL}
\newblock
\begin{APACrefDOI} \doi{10.1007/978-3-540-88845-1_13} \end{APACrefDOI}
\PrintBackRefs{\CurrentBib}

\bibitem [\protect \citeauthoryear {%
Widdows%
\ \protect \BOthers {.}}{%
Widdows%
\ \protect \BOthers {.}}{%
{\protect \APACyear {2003}}%
}]{%
widdows2003unsupervised}
\APACinsertmetastar {%
widdows2003unsupervised}%
\begin{APACrefauthors}%
Widdows, D.%
, Peters, S.%
, Cederberg, S.%
, Chan, C\BHBI K.%
, Steffen, D.%
\BCBL {}\ \BBA {} Buitelaar, P.%
\end{APACrefauthors}%
\unskip\
\newblock
\APACrefYearMonthDay{2003}{{\APACmonth{07}}}{}.
\newblock
{\BBOQ}\APACrefatitle {Unsupervised Monolingual and Bilingual Word-Sense
  Disambiguation of Medical Documents using {UMLS}} {Unsupervised monolingual
  and bilingual word-sense disambiguation of medical documents using
  {UMLS}}.{\BBCQ}
\newblock
\BIn{} \APACrefbtitle {Proceedings of the {ACL} 2003 Workshop on Natural
  Language Processing in Biomedicine} {Proceedings of the {ACL} 2003 workshop
  on natural language processing in biomedicine}\ (\BPGS\ 9--16).
\newblock
\APACaddressPublisher{Sapporo, Japan}{Association for Computational
  Linguistics}.
\newblock
\begin{APACrefURL} \url{https://aclanthology.org/W03-1302/} \end{APACrefURL}
\newblock
\begin{APACrefDOI} \doi{10.3115/1118958.1118960} \end{APACrefDOI}
\PrintBackRefs{\CurrentBib}

\bibitem [\protect \citeauthoryear {%
Wohlin%
\ \protect \BOthers {.}}{%
Wohlin%
\ \protect \BOthers {.}}{%
{\protect \APACyear {2012}}%
}]{%
wohlin2012experimentation}
\APACinsertmetastar {%
wohlin2012experimentation}%
\begin{APACrefauthors}%
Wohlin, C.%
, Runeson, P.%
, H{\"o}st, M.%
, Ohlsson, M\BPBI C.%
, Regnell, B.%
, Wessl{\'e}n, A.%
\BCBL {}\ \BOthersPeriod {.}\end{APACrefauthors}%
\unskip\
\newblock
\APACrefYear{2012}.
\newblock
\APACrefbtitle {Experimentation in software engineering} {Experimentation in
  software engineering}\ (\BVOL~236).
\newblock
\APACaddressPublisher{}{Springer}.
\newblock
\begin{APACrefDOI} \doi{10.1007/978-3-662-69306-3} \end{APACrefDOI}
\PrintBackRefs{\CurrentBib}

\bibitem [\protect \citeauthoryear {%
Yan%
}{%
Yan%
}{%
{\protect \APACyear {2014}}%
}]{%
yan2014research}
\APACinsertmetastar {%
yan2014research}%
\begin{APACrefauthors}%
Yan, E.%
\end{APACrefauthors}%
\unskip\
\newblock
\APACrefYearMonthDay{2014}{}{}.
\newblock
{\BBOQ}\APACrefatitle {Research dynamics: Measuring the continuity and
  popularity of research topics} {Research dynamics: Measuring the continuity
  and popularity of research topics}.{\BBCQ}
\newblock
\APACjournalVolNumPages{Journal of Informetrics}{8}{1}{98--110}.
\newblock
\begin{APACrefDOI} \doi{10.1016/j.ssresearch.2022.102772} \end{APACrefDOI}
\PrintBackRefs{\CurrentBib}

\bibitem [\protect \citeauthoryear {%
Yang%
\ \BBA {} Lee%
}{%
Yang%
\ \BBA {} Lee%
}{%
{\protect \APACyear {2018}}%
}]{%
ijerph15061113}
\APACinsertmetastar {%
ijerph15061113}%
\begin{APACrefauthors}%
Yang, H.%
\BCBT {}\ \BBA {} Lee, H\BPBI J.%
\end{APACrefauthors}%
\unskip\
\newblock
\APACrefYearMonthDay{2018}{}{}.
\newblock
{\BBOQ}\APACrefatitle {Research Trend Visualization by MeSH Terms from PubMed}
  {Research trend visualization by mesh terms from pubmed}.{\BBCQ}
\newblock
\APACjournalVolNumPages{International Journal of Environmental Research and
  Public Health}{15}{6}{}.
\newblock
\begin{APACrefURL} \url{https://www.mdpi.com/1660-4601/15/6/1113}
  \end{APACrefURL}
\newblock
\begin{APACrefDOI} \doi{10.3390/ijerph15061113} \end{APACrefDOI}
\PrintBackRefs{\CurrentBib}

\bibitem [\protect \citeauthoryear {%
Yip%
, Nguyen%
\BCBL {}\ \BBA {} Bodenreider%
}{%
Yip%
\ \protect \BOthers {.}}{%
{\protect \APACyear {2019}}%
}]{%
yip2019construction}
\APACinsertmetastar {%
yip2019construction}%
\begin{APACrefauthors}%
Yip, H\BPBI Y.%
, Nguyen, V.%
\BCBL {}\ \BBA {} Bodenreider, O.%
\end{APACrefauthors}%
\unskip\
\newblock
\APACrefYearMonthDay{2019}{}{}.
\newblock
{\BBOQ}\APACrefatitle {Construction of UMLS Metathesaurus with
  Knowledge-Infused Deep Learning.} {Construction of umls metathesaurus with
  knowledge-infused deep learning.}{\BBCQ}
\newblock
\BIn{} \APACrefbtitle {BlockSW/CKG@ ISWC.} {Blocksw/ckg@ iswc.}
\PrintBackRefs{\CurrentBib}

\bibitem [\protect \citeauthoryear {%
Zaharee%
}{%
Zaharee%
}{%
{\protect \APACyear {2013}}%
}]{%
Zaharee2013}
\APACinsertmetastar {%
Zaharee2013}%
\begin{APACrefauthors}%
Zaharee, M.%
\end{APACrefauthors}%
\unskip\
\newblock
\APACrefYearMonthDay{2013}{}{}.
\newblock
{\BBOQ}\APACrefatitle {Building controlled vocabularies for metadata
  harmonization} {Building controlled vocabularies for metadata
  harmonization}.{\BBCQ}
\newblock
\APACjournalVolNumPages{Bulletin of the American Society for Information
  Science and Technology}{39}{2}{39-42}.
\newblock
\begin{APACrefURL}
  \url{https://asistdl.onlinelibrary.wiley.com/doi/abs/10.1002/bult.2013.1720390211}
  \end{APACrefURL}
\newblock
\begin{APACrefDOI} \doi{10.1002/bult.2013.1720390211} \end{APACrefDOI}
\PrintBackRefs{\CurrentBib}

\bibitem [\protect \citeauthoryear {%
Zapilko%
, Schaible%
, Mayr%
\BCBL {}\ \BBA {} Mathiak%
}{%
Zapilko%
\ \protect \BOthers {.}}{%
{\protect \APACyear {2013}}%
}]{%
zapilko2013}
\APACinsertmetastar {%
zapilko2013}%
\begin{APACrefauthors}%
Zapilko, B.%
, Schaible, J.%
, Mayr, P.%
\BCBL {}\ \BBA {} Mathiak, B.%
\end{APACrefauthors}%
\unskip\
\newblock
\APACrefYearMonthDay{2013}{{\APACmonth{07}}}{}.
\newblock
{\BBOQ}\APACrefatitle {TheSoz: A SKOS Representation of the Thesaurus for the
  Social Sciences} {Thesoz: A skos representation of the thesaurus for the
  social sciences}.{\BBCQ}
\newblock
\APACjournalVolNumPages{Semant. Web}{4}{3}{257–263}.
\newblock
\begin{APACrefDOI} \doi{10.3233/SW-2012-0081} \end{APACrefDOI}
\PrintBackRefs{\CurrentBib}

\bibitem [\protect \citeauthoryear {%
Zeng%
}{%
Zeng%
}{%
{\protect \APACyear {2008}}%
}]{%
zeng2008knowledge}
\APACinsertmetastar {%
zeng2008knowledge}%
\begin{APACrefauthors}%
Zeng, M\BPBI L.%
\end{APACrefauthors}%
\unskip\
\newblock
\APACrefYearMonthDay{2008}{}{}.
\newblock
{\BBOQ}\APACrefatitle {Knowledge organization systems (KOS)} {Knowledge
  organization systems (kos)}.{\BBCQ}
\newblock
\APACjournalVolNumPages{KO KNOWLEDGE ORGANIZATION}{35}{2-3}{160--182}.
\newblock
\begin{APACrefDOI} \doi{10.5771/0943-7444-2008-2-3-160} \end{APACrefDOI}
\PrintBackRefs{\CurrentBib}

\bibitem [\protect \citeauthoryear {%
Zeng%
\ \BBA {} Mayr%
}{%
Zeng%
\ \BBA {} Mayr%
}{%
{\protect \APACyear {2019}}%
}]{%
zeng2019knowledge}
\APACinsertmetastar {%
zeng2019knowledge}%
\begin{APACrefauthors}%
Zeng, M\BPBI L.%
\BCBT {}\ \BBA {} Mayr, P.%
\end{APACrefauthors}%
\unskip\
\newblock
\APACrefYearMonthDay{2019}{}{}.
\newblock
{\BBOQ}\APACrefatitle {Knowledge Organization Systems (KOS) in the Semantic
  Web: a multi-dimensional review} {Knowledge organization systems (kos) in the
  semantic web: a multi-dimensional review}.{\BBCQ}
\newblock
\APACjournalVolNumPages{International Journal on Digital
  Libraries}{20}{3}{209--230}.
\newblock
\begin{APACrefDOI} \doi{10.1007/s00799-018-0241-2} \end{APACrefDOI}
\PrintBackRefs{\CurrentBib}

\bibitem [\protect \citeauthoryear {%
C.~Zhang%
\ \protect \BOthers {.}}{%
C.~Zhang%
\ \protect \BOthers {.}}{%
{\protect \APACyear {2018}}%
}]{%
zhang2018taxogen}
\APACinsertmetastar {%
zhang2018taxogen}%
\begin{APACrefauthors}%
Zhang, C.%
, Tao, F.%
, Chen, X.%
, Shen, J.%
, Jiang, M.%
, Sadler, B.%
\BDBL {}Han, J.%
\end{APACrefauthors}%
\unskip\
\newblock
\APACrefYearMonthDay{2018}{}{}.
\newblock
{\BBOQ}\APACrefatitle {Taxogen: Unsupervised topic taxonomy construction by
  adaptive term embedding and clustering} {Taxogen: Unsupervised topic taxonomy
  construction by adaptive term embedding and clustering}.{\BBCQ}
\newblock
\BIn{} \APACrefbtitle {Proceedings of the 24th ACM SIGKDD International
  Conference on Knowledge Discovery \& Data Mining} {Proceedings of the 24th
  acm sigkdd international conference on knowledge discovery \& data mining}\
  (\BPGS\ 2701--2709).
\newblock
\begin{APACrefDOI} \doi{10.1145/3219819.3220064} \end{APACrefDOI}
\PrintBackRefs{\CurrentBib}

\bibitem [\protect \citeauthoryear {%
X.~Zhang%
, Chandrasegaran%
\BCBL {}\ \BBA {} Ma%
}{%
X.~Zhang%
\ \protect \BOthers {.}}{%
{\protect \APACyear {2021}}%
}]{%
zhang2021conceptscope}
\APACinsertmetastar {%
zhang2021conceptscope}%
\begin{APACrefauthors}%
Zhang, X.%
, Chandrasegaran, S.%
\BCBL {}\ \BBA {} Ma, K\BHBI L.%
\end{APACrefauthors}%
\unskip\
\newblock
\APACrefYearMonthDay{2021}{}{}.
\newblock
{\BBOQ}\APACrefatitle {ConceptScope: Organizing and Visualizing Knowledge in
  Documents based on Domain Ontology} {Conceptscope: Organizing and visualizing
  knowledge in documents based on domain ontology}.{\BBCQ}
\newblock
\BIn{} \APACrefbtitle {Proceedings of the 2021 CHI Conference on Human Factors
  in Computing Systems} {Proceedings of the 2021 chi conference on human
  factors in computing systems}\ (\BPGS\ 1--13).
\newblock
\begin{APACrefDOI} \doi{10.1145/3411764.3445396} \end{APACrefDOI}
\PrintBackRefs{\CurrentBib}

\end{thebibliography}
